\documentclass[12pt]{article}
\usepackage{algpseudocode}
\usepackage{amsmath, amssymb, amsthm}
\usepackage{float, fullpage, graphicx, multirow,parskip, subcaption, setspace}
\usepackage{comment}
\usepackage{url}
\usepackage{enumitem}
\usepackage{tikz}
\usepackage{bm, bbm}
\usetikzlibrary{shapes.geometric, positioning}
\usetikzlibrary{quotes, angles}
\usepackage{rotating}
\usepackage{hyperref}
\usepackage{natbib}
\usepackage{placeins}
\usepackage{algorithm2e}
\RestyleAlgo{ruled}

\definecolor{SkyBlue}{RGB}{14, 118, 188}
\definecolor{BrightRed}{RGB}{223,82, 78}

\hypersetup{pdfborder = {0 0 0.5 [3 3]}, colorlinks = true, linkcolor = BrightRed, citecolor = SkyBlue}

\DeclareMathOperator{\tr}{tr}

\DeclareMathOperator{\diag}{diag}
\DeclareMathOperator{\vect}{vec}
\DeclareMathOperator{\GA}{Gamma}

\DeclareMathOperator{\Pois}{Poisson}
\DeclareMathOperator{\Multinomial}{Multinomial}

\newcommand{\ysrevision}{\textcolor{black}}

\def\keywordname{{\bfseries \emph Keywords}}%
\def\keywords#1{\par\addvspace\medskipamount{\rightskip=0pt plus1cm
\def\and{\ifhmode\unskip\nobreak\fi\ $\cdot$
}\noindent\keywordname\enspace\ignorespaces#1\par}}

\title{Bayesian \ysrevision{Chain Graph} LASSO Models to Learn Sparse Biological Networks with Predictors}
\author{Yunyi Shen\thanks{Depts. of Statistics \& Forest and Wildlife Ecology, University of Wisconsin--Madison}  and Claudia Sol\'{i}s-Lemus\thanks{Wisconsin Institute for Discovery \& Dept. of Plan Pathology, University of Wisconsin--Madison, Correspondence to: solislemus@wisc.edu}}

\begin{document}
\def\bY{\bm{Y}}
\def\by{\bm{y}} 

\def\bz{\bm{z}}
\def\bX{\bm{X}}
\def\bx{\bm{x}} 

\def\R{\mathbb{R}}
\def\N{\mathcal{N}}
\def\P{\mathbb{P}}
\def\E{\mathbb{E}}

\def\Xcal{\mathcal{X}}

\maketitle

\begin{abstract}

Microbiome data require statistical models that can simultaneously decode microbes' reaction to the environment and interactions among microbes.
While a multiresponse linear regression model seems  like  a  straight-forward  solution,  we  argue  that  treating  it  as  a  graphical model  is  flawed  given that the regression coefficient matrix does not encode the conditional dependence structure between response and predictor nodes as it does not represent the adjacency matrix.
This observation is especially important in biological settings when we have prior knowledge on the edges from specific experimental interventions that can only be properly encoded under a conditional dependence model.
Here, we propose a chain graph model with two sets of nodes (predictors and responses) whose solution yields a graph with edges that indeed represent conditional dependence and thus, agrees with the experimenter's intuition on the average behavior of nodes under treatment.
The solution to our model is sparse via Bayesian LASSO. 
In addition, we propose an adaptive extension so that different shrinkage can be applied to different edges to incorporate edge-specific prior knowledge. Our model is computationally inexpensive through an efficient Gibbs sampling algorithm and can account for binary, counting and compositional responses via appropriate hierarchical structure. 
We apply our model to a human gut and a
soil microbial compositional datasets and we highlight that \ysrevision{CG}-LASSO can estimate biologically meaningful network structures in the data. The \ysrevision{CG}-LASSO software is available
as an R package at \url{https://github.com/YunyiShen/CAR-LASSO}.
\end{abstract}


\section{Introduction}
\label{sec:intro}
Recent years have seen an explosion of microbiome research studies given that microbial communities are among the main driving forces of all biogeochemical processes on Earth. On one side, many critical soil processes such as mineral weathering, and soil cycling of mineral-sorbed organic matter are governed by mineral-associated microbes \citep{Fierer2012, Whitman2018, Cates2019, Kranz2019, Whitman2019}. On another side, plant and soil microbiome drive phenotype variation related to plant health and crop production \citep{Allsup2019, Rioux2019, Lankau2020, Lankau2020b}.
In addition, as evidenced by The Human Microbiome Project \citep{turnbaugh2007human}, the microbes that live on the human body are key determinants of human health and disease \citep{Dave2012}.

Understanding the composition of microbial communities and what environmental or experimental factors play a role in shaping this composition is crucial to comprehend biological processes in humans, soil and plants alike, and to predict microbial responses to environmental changes.
However, the inter-connectivity of microbes-environment is still not fully understood. One of the reasons for this gap in knowledge is the lack of statistical tools \ysrevision{to simultaneously infer connections among microbes and their \textit{direct} reactions to different environmental factors in an unified framework.}

\ysrevision{Graphical models conveniently represent dependency structures among several variables. In a nutshell, each variable is represented by a node, edges represent \textit{conditional} dependence between nodes, and absence of such edges represents conditional independence. Gaussian graphical models (GGMs) are widely used in microbiome studies for its interpretability and scalability \citep{layeghifard2017disentangling}, and a sparse solution can be obtained via LASSO \citep{friedman2008sparse,glasso}. 
However, GGMs only have one type of node with undirected edges between the nodes, and thus, GGMs are ill-equipped to estimate graphs in the context of microbe community where there are often two types of nodes (microbes and environmental variables) with edges representing direct effects on the microbes.} 

\begin{figure}[h]
	\centering
	\includegraphics[width=.8\linewidth]{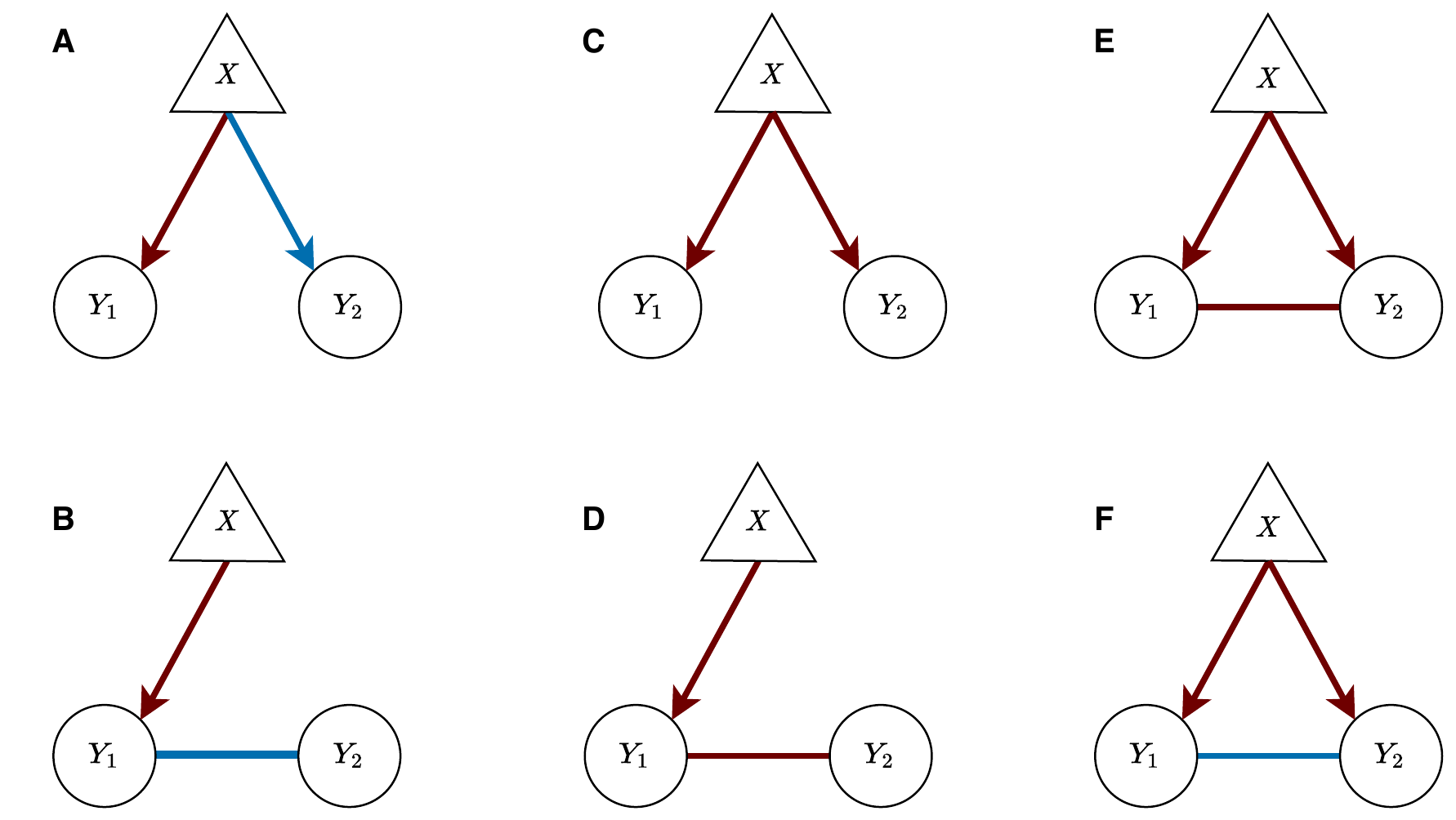}
	\caption{\textbf{Simple networks with predictors.} We use a triangle to represent a predictor $X$ and circles to represent the responses $Y_1$ and $Y_2$. Red edges correspond to positive links between nodes while blue edges correspond to a negative links. Networks A and B (likewise networks C and D) can produce a similar marginal correlation structure between any two nodes. Distinguishing edges in E can be difficult since all edges have the same direction. Finally, in network F, $X$ and $Y_2$ are conditionally correlated, yet they might not have a marginal correlation. For example, if $Y_1,Y_2$ has marginal variance of $1$, covariance $\rho=-0.5$, the conditional regression coefficient between $Y_1$ and $X$ conditioned on $Y_2$ is $\beta_1=2>0$ and conditional regression coefficient between $Y_2$ and $X$ conditioned on $Y_1$ is $\beta_2=1>0$, we can show that the marginal regression coefficient between $Y_2$ and $X$ when integrating out $Y_1$ is $\rho\beta_1+\beta_2=0$ (more in Section \ref{sec:cond_indp}).}
	\label{fig:egnetwork}
\end{figure}

\ysrevision{Intuitively, a multiresponse linear regression (sometimes referred as a covariate-adjusted GGM) with a LASSO prior on the regression coefficients in combination with a graphical LASSO prior on the precision matrix can provide sparse regression coefficients among responses and predictors. In addition, sparse graphical models can be used to estimate the sparse graphical structure among responses.}  
\ysrevision{In this formulation, however, the regression coefficients between responses and predictors represent \textit{marginal} effects rather than \textit{conditional} effects. Here, our goal is to estimate the graphical structure between the set of predictors (environment) and responses (microbes), as well as the graphical structure among responses while keeping the \textit{conditional} interpretation of both parameters.} 
The distinction between marginal effect and conditional effect is crucial when we would like to biologically interpret the result or to include biological prior knowledge to the model (e.g. as in \citet{lo2017mplasso}). For instance, penicillin has no biological effect on Gram-negative bacteria (a specific class of microbe), yet it might still promote the abundance of such bacteria by inhibiting their Gram-positive competitors. In this example, penicillin has no \textit{conditional} effect on Gram-negative nodes (conditioned on all other microbes), but it may have a \textit{marginal} effect on them when marginalizing over all other microbes (Figure \ref{fig:egnetwork} B). The inverse is also possible. A response can be conditionally dependent on a predictor, but marginally independent when another response has a similar dependence with that predictor (Figure \ref{fig:egnetwork} F). In this case, the link between the responses could marginally cancel out the effect of the predictor.

Here, we introduce a model framework to infer a sparse network structure that represents interactions among responses and effects of a set of predictors. 
\ysrevision{Specifically, our model simultaneously estimates the conditional effect of a set of predictors (e.g. diet, weather, experimental treatments) that influence the responses (e.g. abundances of microbes) and connections among responses.} 
Our model is represented by a chain graph \citep{lauritzen1989graphical,lauritzen2002chain} with two sets of nodes: predictors and responses (Figure \ref{fig:egnetwork}). Directed edges between a predictor and a response represent conditional links, and undirected edges among responses represent correlations. 
While chain graph models are not new, we can argue that they are under-utilized in microbiome studies and our work will serve to illustrate its potential to elucidate ties between microbial interactions and experimental or environmental predictors.

In addition to providing a more sensible representation compared to standard multiresponse linear regression, our model guarantees a sparse solution via Bayesian LASSO, \ysrevision{and with fixed penalty, the posterior is log-concave.} 
Furthermore, we propose an adaptive extension that allows different shrinkage to different edges to incorporate edge-specific knowledge into the model, and we use the Normal model as a core to build hierarchical structures upon it to account for binary, counting and compositional responses. Finally, our model is able to equally handle small and big data and is computationally inexpensive through an efficient Gibbs sampling algorithm.

\section{Methods}
\label{sec:proposed_method}
\subsection{Model specification and interpretation}
\label{sec:cond_indp}
Let $\mathbf{Y}_i \in \mathbb{R}^k$ be a multivariate response with $k$ entries for $i=1,\dots,n$ observations. Let $\mathbf{X}_i \in \mathbb{R}^{1\times p}$ be the row vector of predictors for $i=1,\dots,n$ (i.e. the $i^{th}$ row of the design matrix $\mathbf X\in \mathbb R^{n\times p}$). We assume that the design matrix is standardized so that each column has a mean of 0 and same standard deviation (set to be 1 in the simulations) so that same shrinkage parameter will not have different effect on different predictors. 

Let $\mathbf{Y}_i$ follow a Normal distribution with mean vector $\mathbf{\Omega}^{-1}(\mathbf{B}^T\mathbf{X}^T_i+\mu)$ and precision matrix $\mathbf{\Omega} \in \mathbb{R}^{k \times k}$ (positive definite) where $\mathbf B \in \mathbb{R}^{p \times k}$ corresponds to the regression coefficients connecting the responses ($\mathbf{Y}_i\in \mathbb{R}^k$) and the predictors ($\mathbf{X}_i\in \mathbb{R}^{1\times p}$) and $\mu \in \mathbb{R}^k$ corresponds to the intercept. We use the transpose $\mathbf{B}^T\mathbf{X}^T_i\in \mathbb R^{k\times 1}$ because samples are encoded as row vectors in the design matrix while by convention multivariate Normal samples are column vectors.

The likelihood function of the model is:
\begin{equation}
    p(\mathbf{Y}_i|\mathbf{X}_i,\mu, \mathbf{B}, \mathbf{\Omega}) \propto \exp[(\mathbf{B}^T\mathbf{X}^T_i+\mu)^T\mathbf{Y}_i-\frac{1}{2}\mathbf{Y}_i^T\mathbf{\Omega}\mathbf{Y}_i].
\end{equation}

Note that in this parametrization, $\mathbf{B}$ encodes \textit{conditional dependence} between $\mathbf{Y}$ and $\mathbf{X}$ because in the kernel of the density, $B_{jq}$ is the coefficient of product between $X_j$ and $Y_q$. Thus, if $B_{jq}=0$, then $X_j$ and $Y_q$ are \textit{conditionally independent}. 
This is analogous to the case of $\mathbf{\Omega}$ whose off-diagonal entries encode the conditional dependence between responses $Y_q$ and $Y_{q'}$. \ysrevision{To see the analogy, we provide an interpretation of the parameters in the same manner as in univariate linear regression. Let $\mathbf{Y}_{-q}$ be the vector of responses without the $q$th component, let $\mathbf{Y}_{-(q',q)}$ be the vector of responses without the $q$th and $q'$th components, and let $\mathbf{X}_{-j}$ be the vector of predictors without the $j$th component.}
\begin{equation}
\begin{aligned}
    \ysrevision{\E[Y_{q} \vert X_{j} = x_{j} + 1, \mathbf Y_{-q}, \mathbf X_{-j}, \mathbf{B}, \mathbf\Omega]} &\ysrevision{- \E[Y_{q} \vert X_{j} = x_{j}, \mathbf Y_{-q}, \mathbf X_{-j}, \mathbf{B}, \mathbf\Omega] = B_{jq}/(-\omega_{qq})}\\
    \ysrevision{\E[Y_{q} \vert y_{q'} = y_{q'}+1, \mathbf Y_{-(q',q)}, \mathbf X, \mathbf{B}, \mathbf\Omega]} &\ysrevision{- \E[Y_{q} \vert \vert Y_{q'} = y_{q'}, \mathbf Y_{-(q,q')}, \mathbf X, \mathbf{B}, \mathbf\Omega] = \omega_{qq'}/(-\omega_{qq})}\\
\end{aligned}
\end{equation}

\ysrevision{That is, fixing the values of all but one predictor and all of the other responses, an increase of one unit in $X_{j}$ is associated with $B_{jq}/(-\omega_{qq})$ unit increase in the expectation of $Y_q$, hence conditioned on the values of all other responses $Y_{q'}.$} 

\ysrevision{On the other hand, the regression coefficients in multiresponse linear regression, denoted as $\mathbf{\tilde{B}}=\mathbf{B\Omega}$, are marginal effects. That is,}
\begin{equation}
\begin{aligned}
    \ysrevision{\E[Y_{q} \vert X_{j} = x_{j} + 1, \mathbf X_{-j}, \tilde{\mathbf{B}}, \mathbf\Omega]} &\ysrevision{- \E[Y_{q} \vert X_{j} = x_{j}, \mathbf X_{-j}, \tilde{\mathbf{B}}, \mathbf\Omega] = \tilde{B}_{jq}}\\
\end{aligned}
\end{equation}

\ysrevision{Note that the crucial difference here is that we do not condition on $\mathbf Y_{-q}$. In microbiome specifically, the presence of $\mathbf \Omega^{-1}$ in the mean represents the ecological knowledge that the responses of a species (e.g. relative abundances of microbes) depend on both its reaction to the environment ($\mathbf{B}$) and interactions with other species ($\mathbf{\Omega}$). } \ysrevision{
In our model, the regression coefficients matrix $\mathbf B$ encode conditional dependence among the responses (scaled by variance) and the predictors that, arguably has a more mechanistic interpretation \citep{andersson2001alternative,lauritzen1989graphical,frydenberg1990chain}.
}

\ysrevision{In general, it is not possible to find sparse marginal prediction of single node and sparse graph simultaneously.
That is, the marginal effect $\tilde{\mathbf B}=\mathbf B \mathbf \Omega^{-1}$ typically has different support to the direct effect $\mathbf{B}$. 
It is possible for both parameters $\mathbf{B}$ and $\mathbf{\tilde B}$ to be sparse when $\mathbf \Omega^{-1}$ is diagonal. In this case, the responses are independent, and thus, $\mathbf B_{jq}=0$ implies $\tilde{\mathbf B}_{jq}=0$ for any $\mathbf B$. 
}

\ysrevision{While our model focuses on finding the sparse graph, we also implement a model for sparse marginal regression coefficient and sparse precision matrix by combining the Gibbs sampling for the precision matrix in \citet{glasso} with the Gibbs sampler in \citet{bayeslasso}. We denote this model Simultaneous Regression and Graphical LASSO (SRG-LASSO) and we use it to compare to the CG-LASSO in the simulation study (Section \ref{sec-sims}). }


\subsection{Prior specification}
We assume a Laplace prior on the entries of $\mathbf{B}$ and graphical LASSO prior on $\mathbf{\Omega}$ \citep{bayeslasso, glasso}. 
Using the Normal scale mixture representation of Laplace distribution \citep{bayeslasso,glasso,andrews1974scale}, let $\eta_{ml}$ be the latent scale parameters for $\mathbf{\Omega}$ for $1\le q<q'\le k$ since $\mathbf\Omega$ is symmetric  and let $\tau_{jq}$ ($1\le j\le p,1\le q\le k$) be the latent scale parameters for $\mathbf B$. 

The full model specification is then:
\begin{equation}
    \begin{aligned}
        \mathbf{Y}_i|\mathbf{X}_i,\mu, \mathbf{B},\mathbf{\Omega}& \sim N(\mathbf{\Omega}^{-1}(\mathbf{B}^T\mathbf{X}^T_i+\mu),\mathbf{\Omega}^{-1})\\
        B_{jq}|\tau_{jq}, & \sim N(0,\tau_{jq}^2)\\ 
        \tau_{jq}&\sim \frac{\lambda_{\beta}^2}{2}e^{-\lambda_{\beta}^2\tau_{jq}}\\
        p(\mathbf \Omega|\boldsymbol \eta,\lambda_{\Omega})&=C_{\eta}^{-1}\prod_{q<q'}\left[\frac{1}{\sqrt{2\pi\eta_{qq'}}}\exp\left(-\frac{\omega_{qq'}^2}{2\eta_{qq'}}\right) \right]\prod_{q=1}^{q}\left[\frac{\lambda_{\Omega}}{2}\exp\left(-\frac{\lambda_{\Omega}\omega_{qq}}{2}\right)\right]I_{\Omega\in M^+}\\
        p(\boldsymbol \eta|\lambda_{\Omega})&\propto C_{\eta}\prod_{q<q'}\frac{\lambda_{\Omega}^2}{2}\exp\left(-\frac{\lambda^2_{\Omega}\eta_{qq'}}{2}\right)\\
    \end{aligned}
    \label{eqn-prior}
\end{equation}
where $I_{\Omega\in M^+}$ means that $\mathbf{\Omega}$ must be positive definite.





\subsection{Implementation}

\subsubsection{Sampling scheme}

We derive an efficient Gibbs sampler for all parameters in this model
due to the scale mixture representation of the graphical LASSO prior \citep{glasso}.
Details on derivation of the sampling scheme can be found in the Appendix and are summarized in Algorithm \ref{algo:gibbs} with all the extensions (Section \ref{extensions}).

\begin{algorithm}
\caption{Implementation of Gibbs sampling}
\label{algo:gibbs}

\SetAlgoLined
\KwResult{MCMC samples of the posterior distribution of parameters of interest}
Initialization\;
\While{not enough samples}{
  \If{responses not Normal}{
   Update the latent Normal variables using Adaptive Rejection Sampling (ARS) for counting and compositional data and truncated Normal for binary data (Section \ref{app:nongaussian})\;
   }
   \For{q$^{th}$ diagonal entries in $\mathbf{\Omega}$}{
   \tcc{blockwise update for $\mathbf{\Omega}$}
    Sample the determinant of $\mathbf{\Omega}$: $\gamma | \mathbf \Omega_{[q]}, \eta, \lambda_{\Omega} \sim$ Generalize Inverse Gaussian (GIG) distribution (Equation \ref{gamma}) where $\mathbf \Omega_{[q]}$ corresponds to some partition of $\mathbf \Omega$ based on the diagonal entry $\omega_{qq}$\;
    Update the off diagonal entries in the q$^{th}$ row (column): $\boldsymbol\omega_{-qq} | \gamma, \mathbf \Omega_{[q]}, \eta, \lambda_{\Omega} \sim$ Normal distribution (Equation \ref{omega12})\;
    Compute the updated q$^{th}$ diagonal entry ($\omega_{qq}$) with Equation \ref{omega22} that depends on the determinant ($\gamma$), the off diagonal entries ($\boldsymbol\omega_{-qq}$) and the partition of $\mathbf \Omega$ ($\mathbf \Omega_{[q]}$)\;
   }
   Update $\mathbf{B} | \boldsymbol \tau^2,\mathbf{\Omega},\boldsymbol \mu,\mathbf X,\mathbf Y \sim$ Normal distribution (Equation \ref{distB})\;
   Update $\boldsymbol{\mu} \sim$ Normal$((\mathbf{ Y\Omega} - \mathbf{XB})^T, \mathbf \Omega /n)$\;
   Update latent variables in the scale mixture representation of the two LASSO priors: $\boldsymbol \eta = \{\eta_{jq}\}, \boldsymbol \tau = \{\tau_{jq}\}$ following an Inverse Gaussian distribution (Equation \ref{zij})\;
   \eIf{adaptive shrinkage}{
    Update the shrinkage parameters on $\mathbf{\Omega}$ ($\lambda_{jq,\Omega}$) and on $\mathbf{B}$ ($\lambda^2_{jq,\beta}$) for individual entries (Section \ref{adaptive})\;
   }{
    Update the shrinkage parameters on $\mathbf{\Omega}$ ($\lambda_{\Omega}$) and on $\mathbf{B}$ ($\lambda^2_{\beta}$) uniformly for all entries following a Gamma distribution (Section \ref{hyperparam})\;
   }
 }

\end{algorithm}

\subsubsection{Choice of hyperparameters}
\label{hyperparam}
The shrinkage parameters $\lambda_{\Omega}$ and $\lambda_{\beta}$ (Equation \ref{eqn-prior}) are hyperparameters to be determined. 
As in \citet{bayeslasso,glasso}, we assume these shrinkage parameters have a hyperprior Gamma distribution with shape parameter $r$ and rate parameter $\delta$ which can be set to produce a relatively flat density for a non-informative prior scenario. Note that since the prior on $\mathbf\Omega$ is not a Laplace but a graphical LASSO prior \citep{glasso}, the Gamma prior is on $\lambda$, not on $\lambda^2$ as it would be under a Laplace prior.
\[
\begin{aligned}
    \lambda^2_{\beta}&\sim \GA(r_{\beta},\delta_{\beta})\\
    \lambda_{\Omega} &\sim \GA(r_{\Omega},\delta_{\Omega})
\end{aligned}
\]

The shrinkage parameters $\lambda_{\Omega}$ and $\lambda_{\beta}$ are included in the Gibbs sampler with full conditional distribution still Gamma with shape parameters $r_{\beta}+kp, \delta_{\beta}+\sum \tau_i/2$ and rate parameters $r_{\Omega}+k(k+1)/2,\delta_{\Omega}+||\mathbf \Omega||_1/2$ respectively.


\subsubsection{Learning the graphical structure}
\label{sec:graph_learning}
Our model has a zero posterior probability for a parameter to be zero given the continuous priors. Yet, we still need to determine the cases when the edges of the graph will be considered non-existent.
Here, we infer the graph structure using the horseshoe method in \citet{carvalho2010horseshoe,glasso} which compares the LASSO estimate for the regression coefficient with the posterior mean of a standard conjugate (non-shrinkage) prior \citep{jones2005experiments}. 

Let $\pi=\frac{\tilde \theta}{E_{g}(\theta|\mathbf Y)}$ where $\tilde \theta$ represents the estimate of the parameter under the LASSO prior and $E_{g}(\theta|\mathbf Y)$ is the posterior mean of that parameter under non-shrinkage prior (e.g. Normal for $\mathbf{B}$ and Wishart for $\mathbf \Omega$). The statistics $1-\pi$ characterizes the amount of shrinkage due to the LASSO prior. We use $\pi >0.5$ as the threshold to decide that $\theta\ne 0$ as in \citet{glasso}.


\subsection{Extensions}
\label{extensions}
\subsubsection{Adaptive LASSO}
\label{adaptive}
One simple extension to LASSO is Adaptive LASSO, in which the shrinkage parameter $\lambda$ can be different for all elements in $\mathbf B$ and $\mathbf \Omega$ \citep{Leng2014adaptivelasso,glasso}. This extension is particularly useful when we have prior knowledge of independence among certain nodes.  For example, 
larger shrinkage parameters ($\lambda$) on specific entries can be used to indicate prior knowledge of independence.

As suggested in \citet{Leng2014adaptivelasso,glasso}, we set the hyperpriors on $\lambda^2_{jq,\beta}$ as Gamma distributions with shape parameters $r_{jq,\beta}$ and rate parameter $\delta_{jq,\beta}$. We also set the prior suggested in \citet{glasso} for $\lambda_{qq',\Omega} $ (with $q\ne q'$). While in \citet{glasso} shrinkage on diagonal $\lambda_{qq,\Omega}$ is a hyperparameter, we set it here to 0. That is, we are not shrinking the diagonal entries of $\mathbf\Omega$. But such shrinkage can be included by multiplying the prior of $\mathbf \Omega$ by $\prod_{q=1}^k \frac{\lambda_{qq}}{2}exp(-\lambda_{qq,\Omega}\omega_{qq})$.

The prior for $\mathbf \Omega$ is

\[
\begin{aligned}
p(\mathbf \Omega|\{\lambda_{qq',\Omega}\}_{q<q'})&=C^{-1}_{\{\lambda_{qq',\Omega}\}_{q<q'}}\prod_{q<q'}\frac{\lambda_{qq',\Omega}}{2}\exp(-\lambda_{qq',\Omega}|\omega_{qq'}|) I_{\mathbf{\Omega}\in M_+} \\
p(\{\lambda_{qq',\Omega}\}_{q<q'})&\propto 
C_{\{\lambda_{qq',\Omega}\}_{q<q'}}\prod_{q<q'}\frac{1}{\Gamma(r_{qq',\Omega})}\lambda_{qq',\Omega}^{r_{qq',\Omega}-1}\exp(-\delta_{qq',\Omega}\lambda_{qq',\Omega}).
\end{aligned}
\]

The full conditional distribution of the shrinkage parameters is then Gamma (shape and rate parametrization):

\[
\begin{aligned}
\lambda_{qq',\Omega}| \mathbf{\Omega} &\sim \GA(r_{qq',\Omega}+1,\delta_{qq',\Omega}+|\omega_{qq'}|), q\ne q'\\
\lambda^2_{qq',\beta}|\tau &\sim \GA(r_{qq',\beta}+1,\delta_{qq',\beta}+\tau_{qq'}/2).
\end{aligned}
\]

We set the hyperparameters as $r=10^{-2}$ and $\delta=10^{-6}$ for both $\mathbf \Omega$ and $\mathbf B$ \citep{glasso,Leng2014adaptivelasso} with a small value of $\delta$ selected to take advantage of the adaptiveness of the shrinkage.

The model has been defined for continuous responses, yet there are different extensions for the case of binary data,
counts and compositional data that we describe in Appendix \ref{app:nongaussian}.


\section{Simulation studies}
\label{sec-sims}


\subsection{Simulation design}
\begin{figure}[htp]
    \centering
\begin{subfigure}[b]{0.15\textwidth}
\centering
\includegraphics[width = \textwidth]{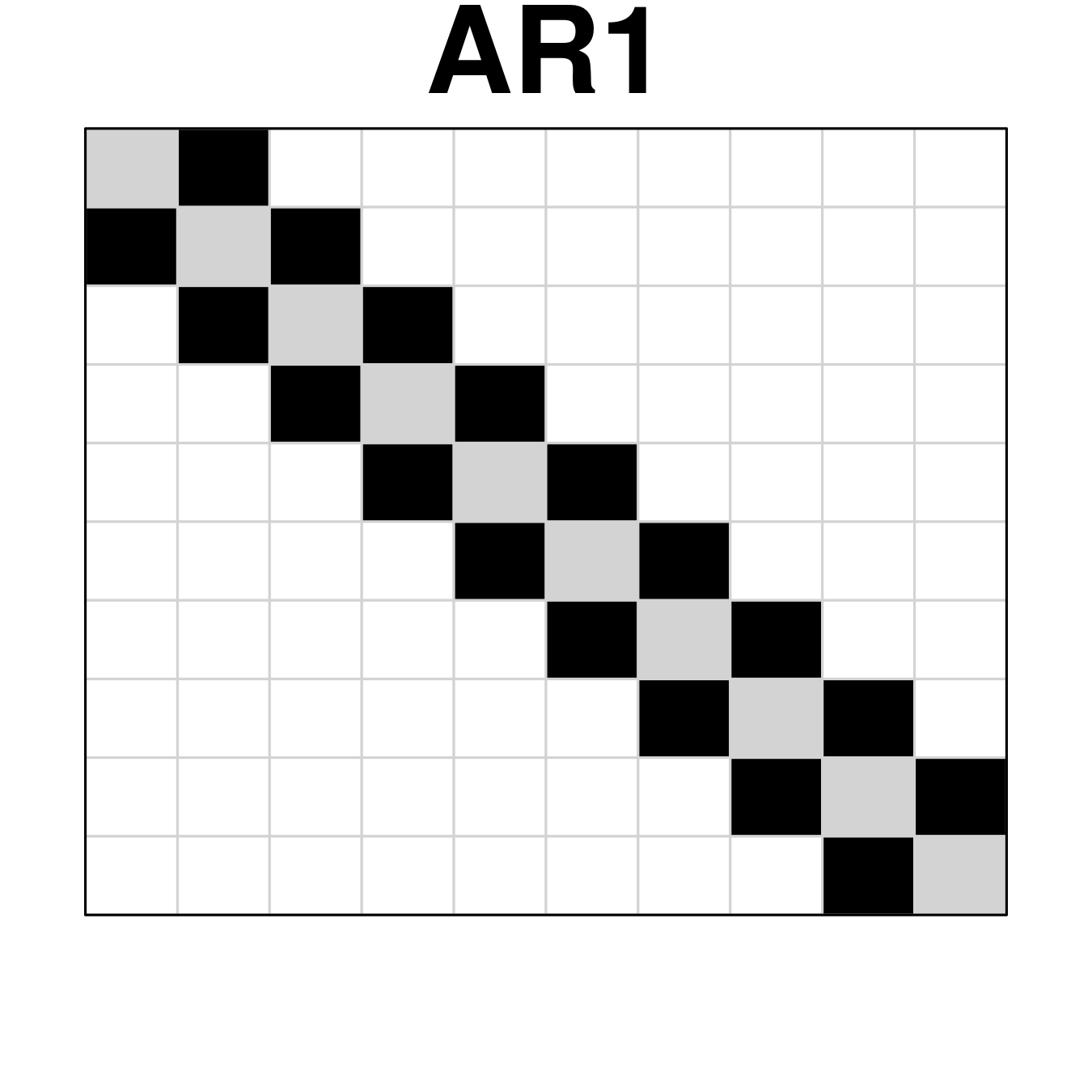}
\end{subfigure}
\begin{subfigure}[b]{0.15\textwidth}
\centering
\includegraphics[width = \textwidth]{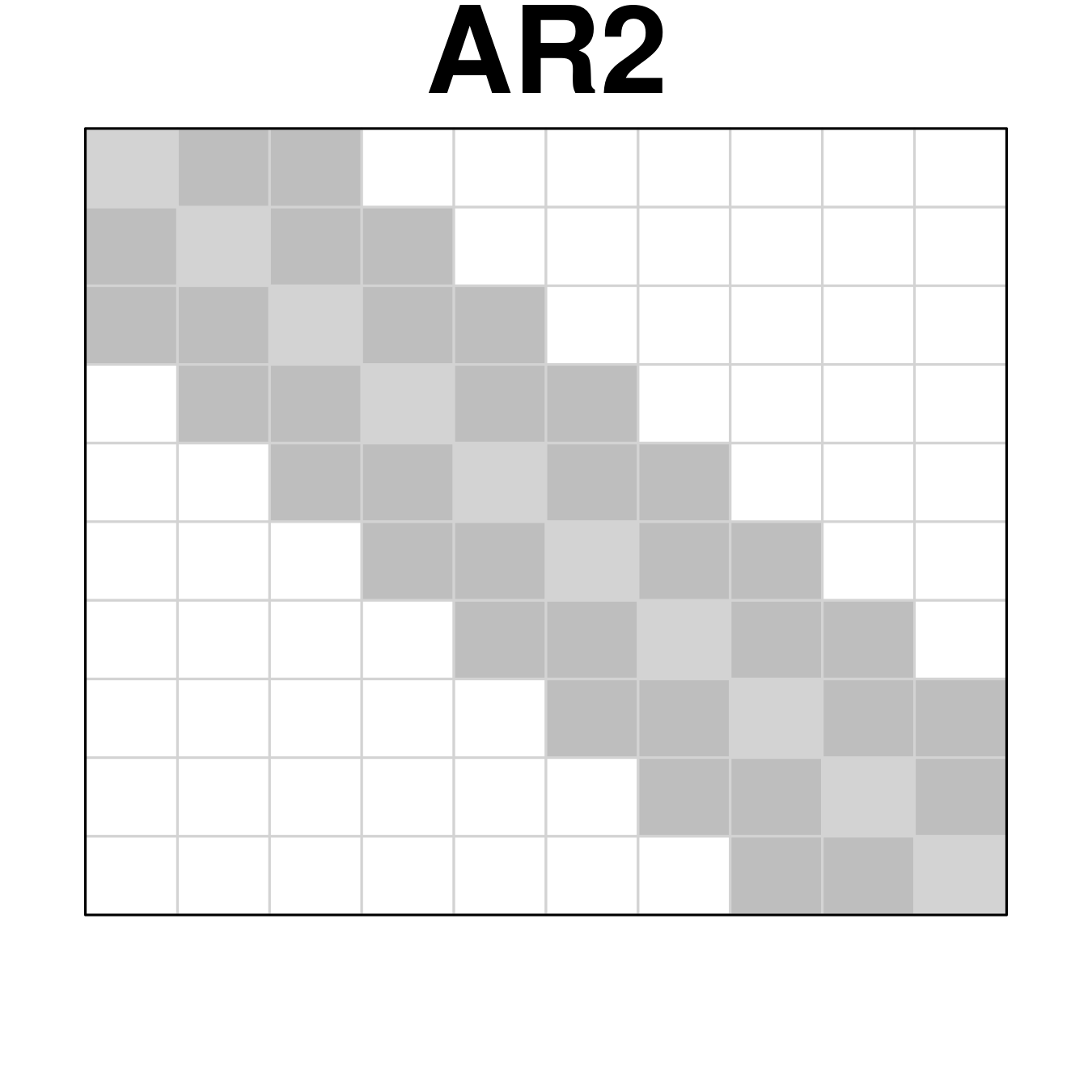}
\end{subfigure}
\begin{subfigure}[b]{0.15\textwidth}
\centering
\includegraphics[width = \textwidth]{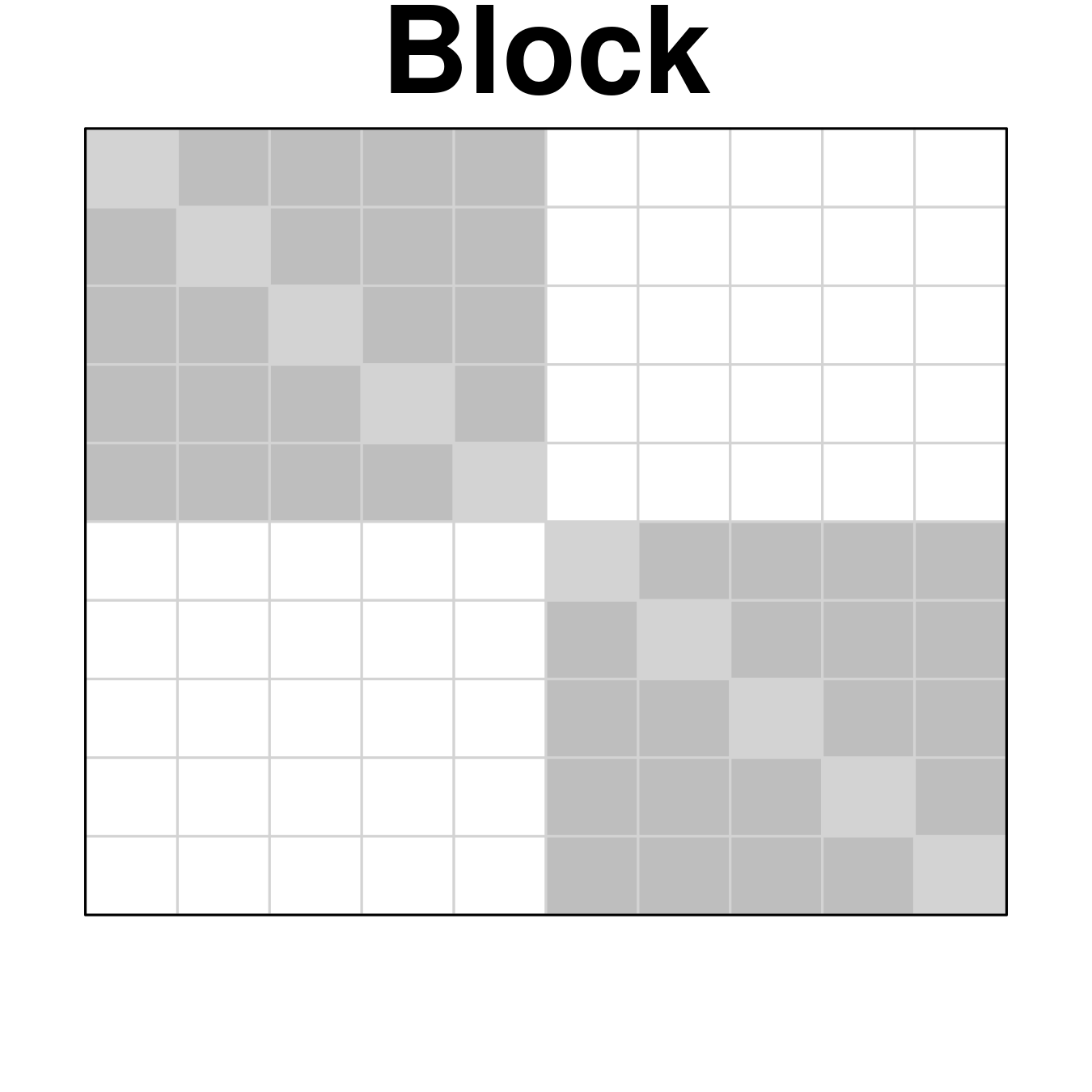}
\end{subfigure}
\begin{subfigure}[b]{0.15\textwidth}
\centering
\includegraphics[width = \textwidth]{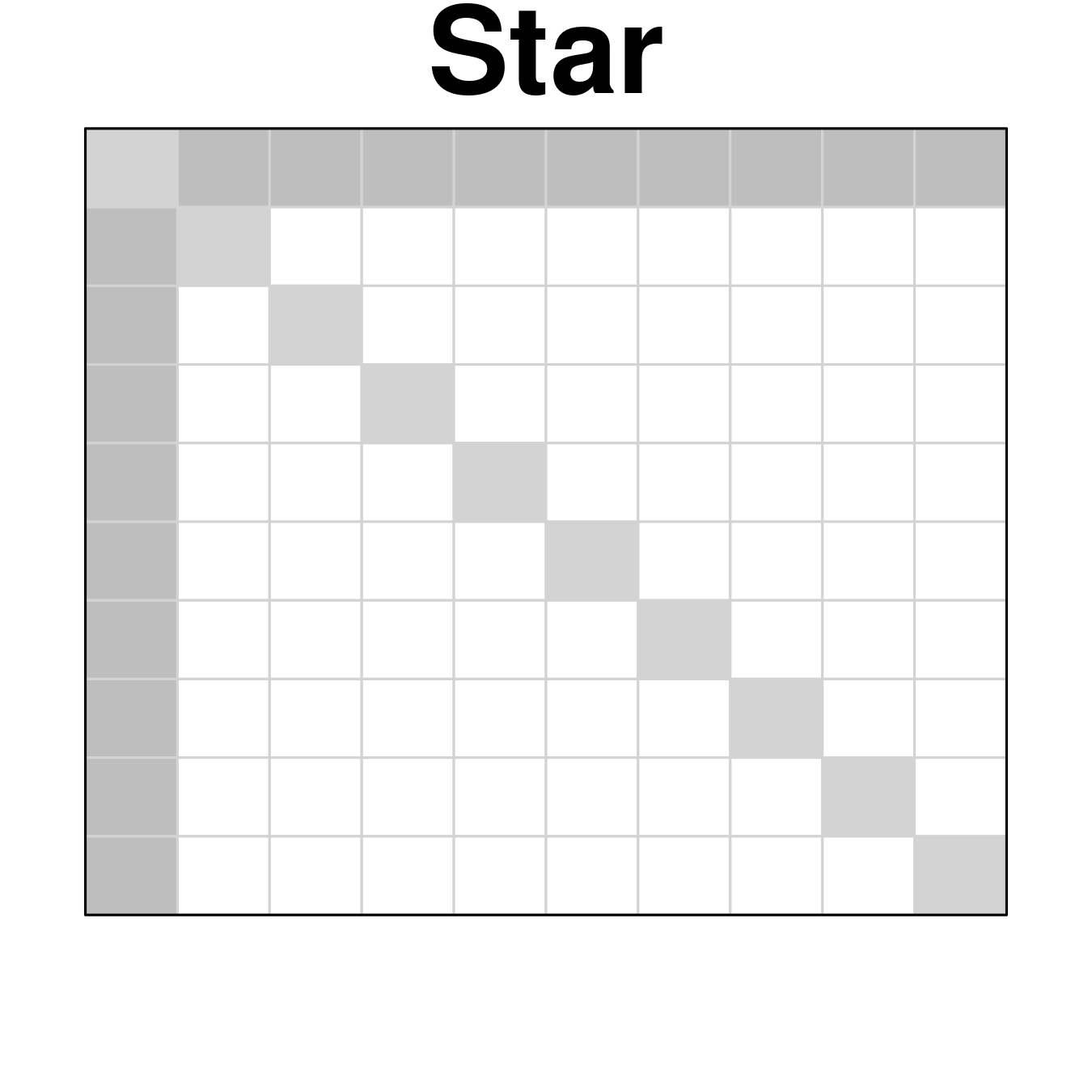}
\end{subfigure}
\begin{subfigure}[b]{0.15\textwidth}
\centering
\includegraphics[width = \textwidth]{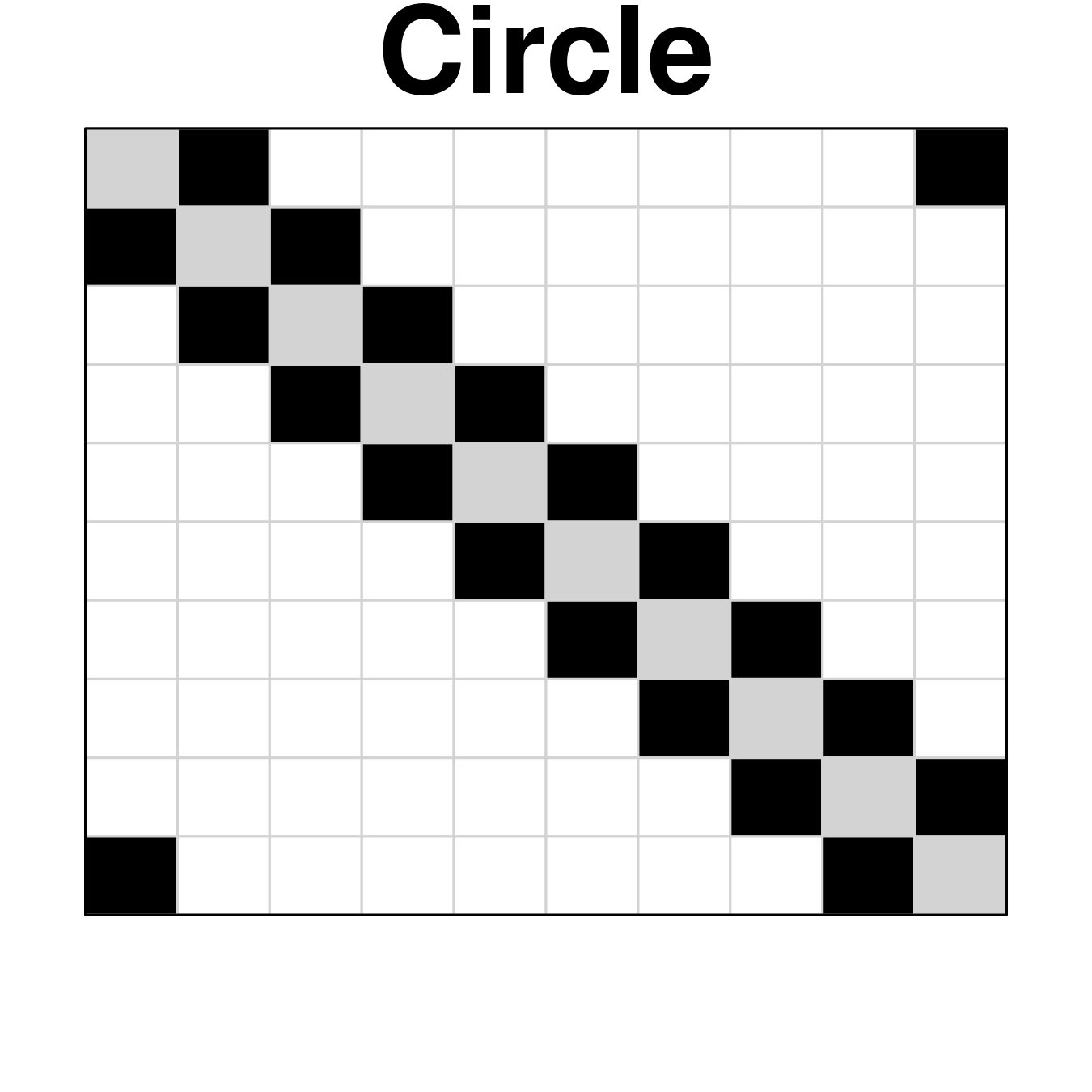}
\end{subfigure}
\begin{subfigure}[b]{0.15\textwidth}
\centering
\includegraphics[width = \textwidth]{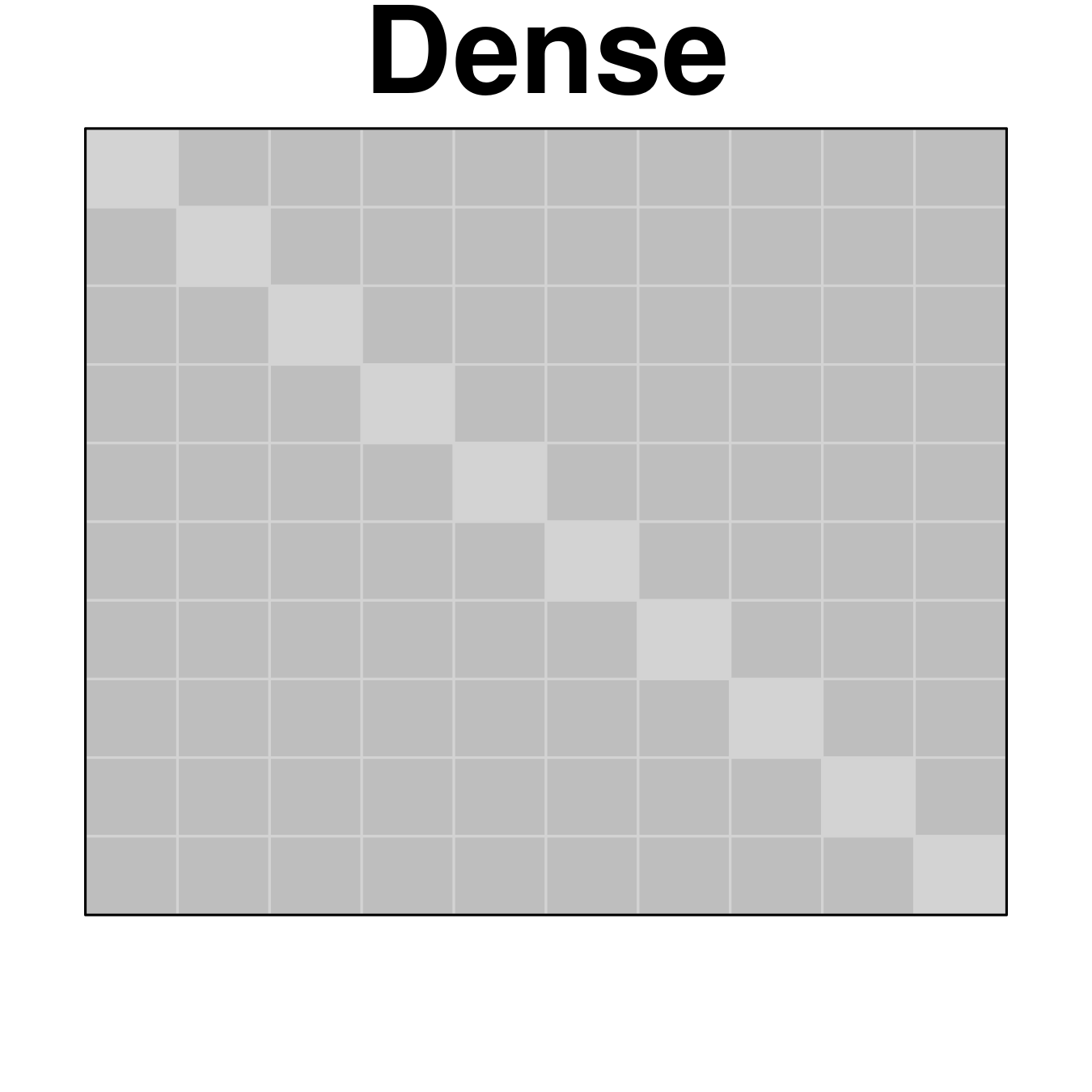}
\end{subfigure}	

\begin{subfigure}[b]{0.15\textwidth}
\centering
\includegraphics[width = \textwidth]{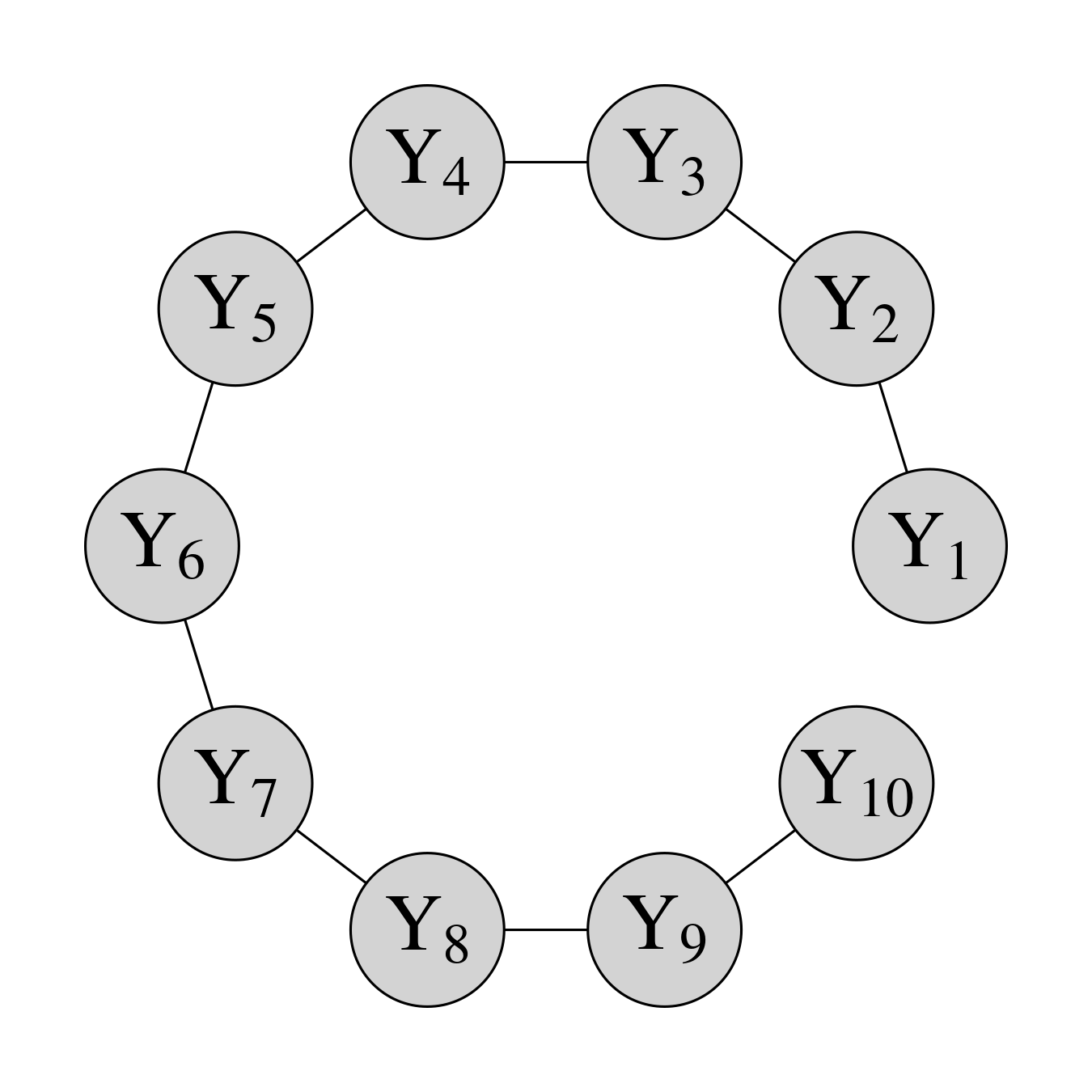}
\end{subfigure}
\begin{subfigure}[b]{0.15\textwidth}
\centering
\includegraphics[width = \textwidth]{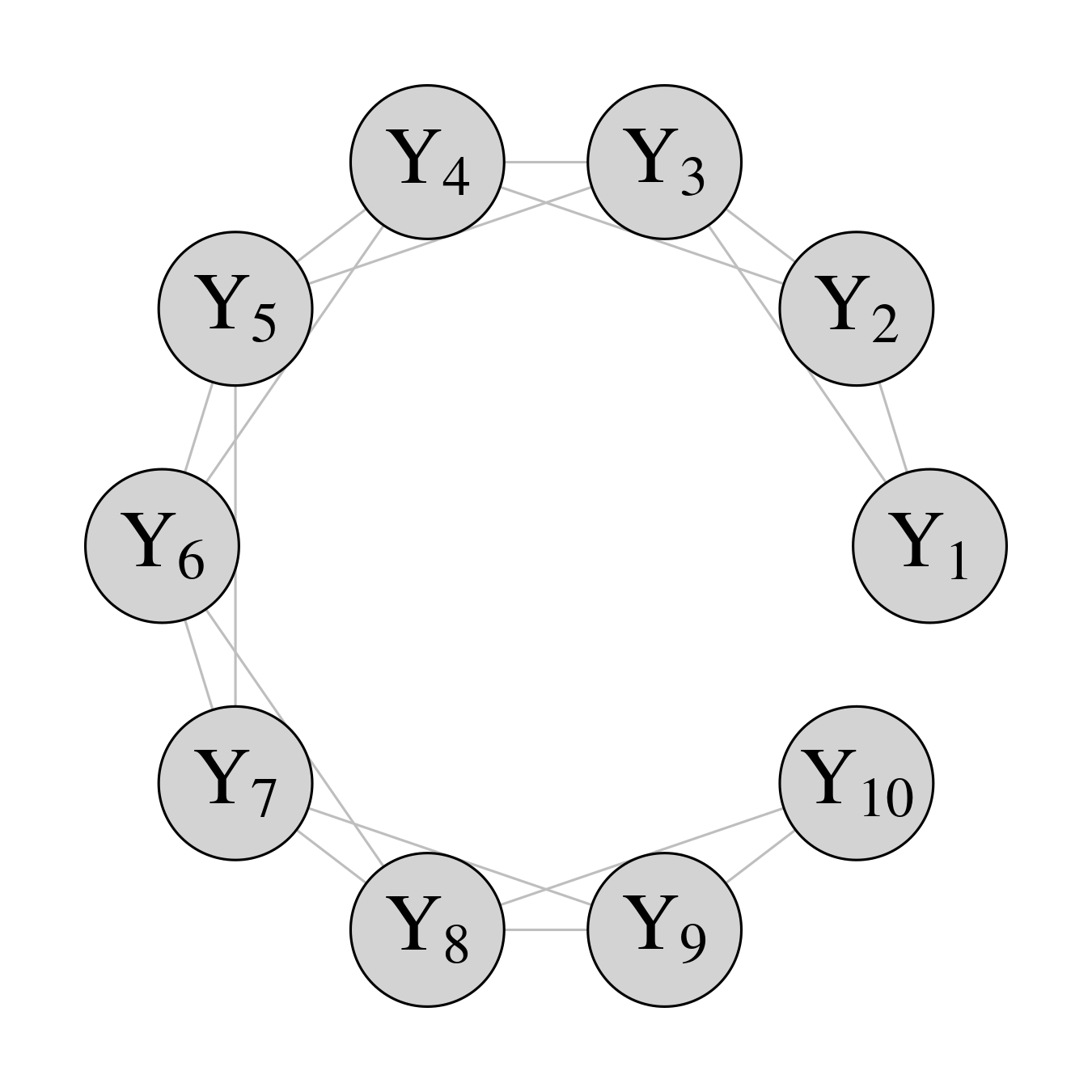}
\end{subfigure}
\begin{subfigure}[b]{0.15\textwidth}
\centering
\includegraphics[width = \textwidth]{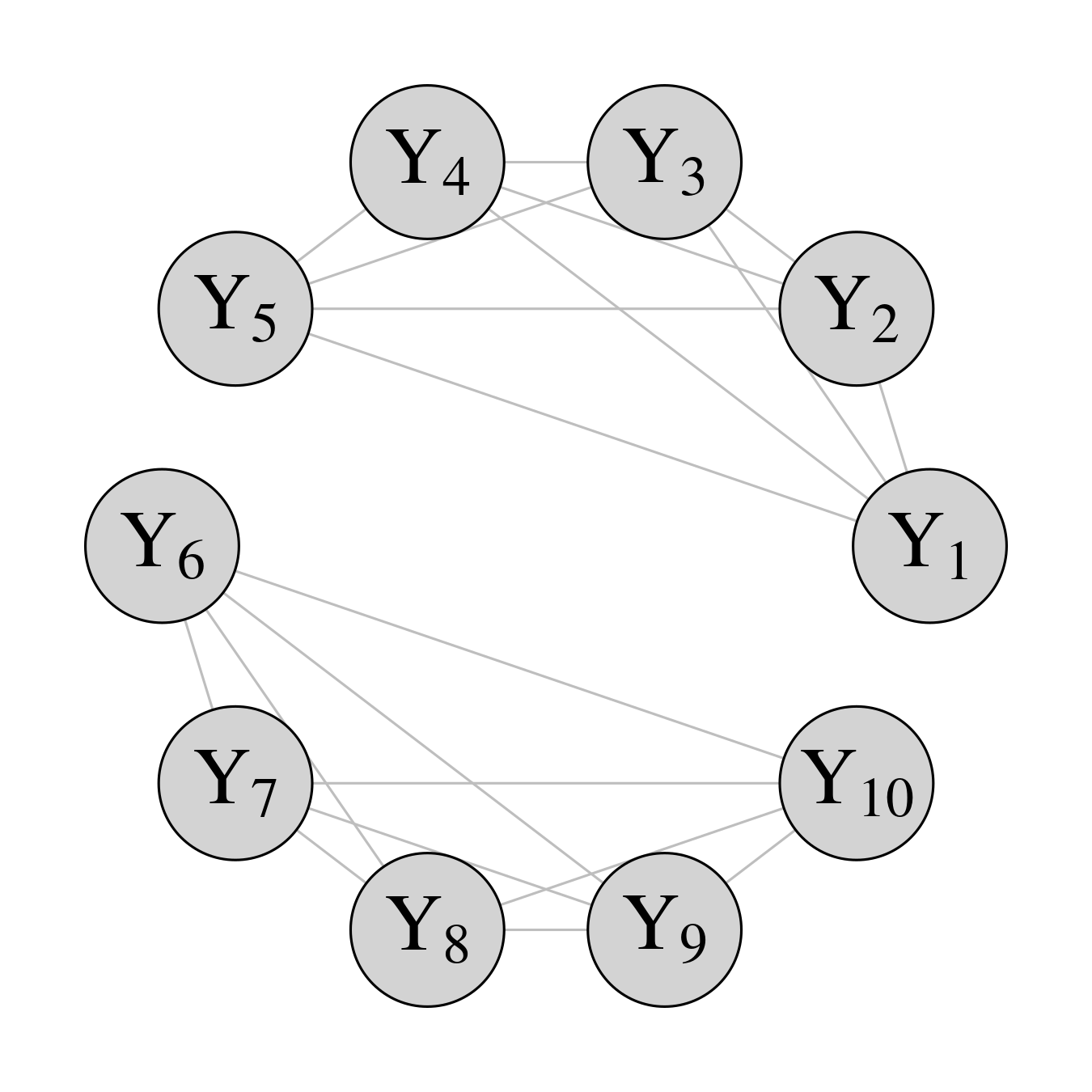}
\end{subfigure}
\begin{subfigure}[b]{0.15\textwidth}
\centering
\includegraphics[width = \textwidth]{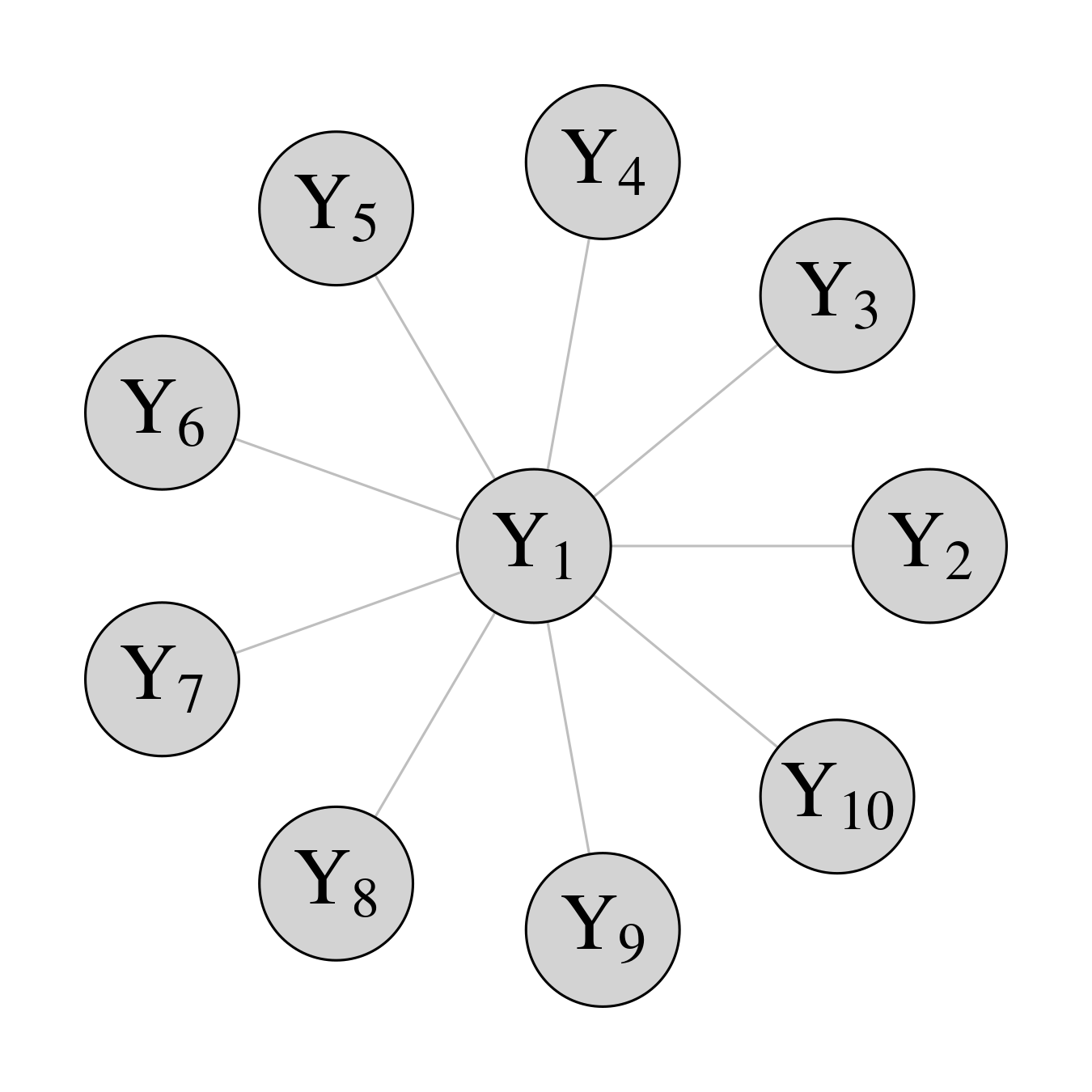}
\end{subfigure}
\begin{subfigure}[b]{0.15\textwidth}
\centering
\includegraphics[width = \textwidth]{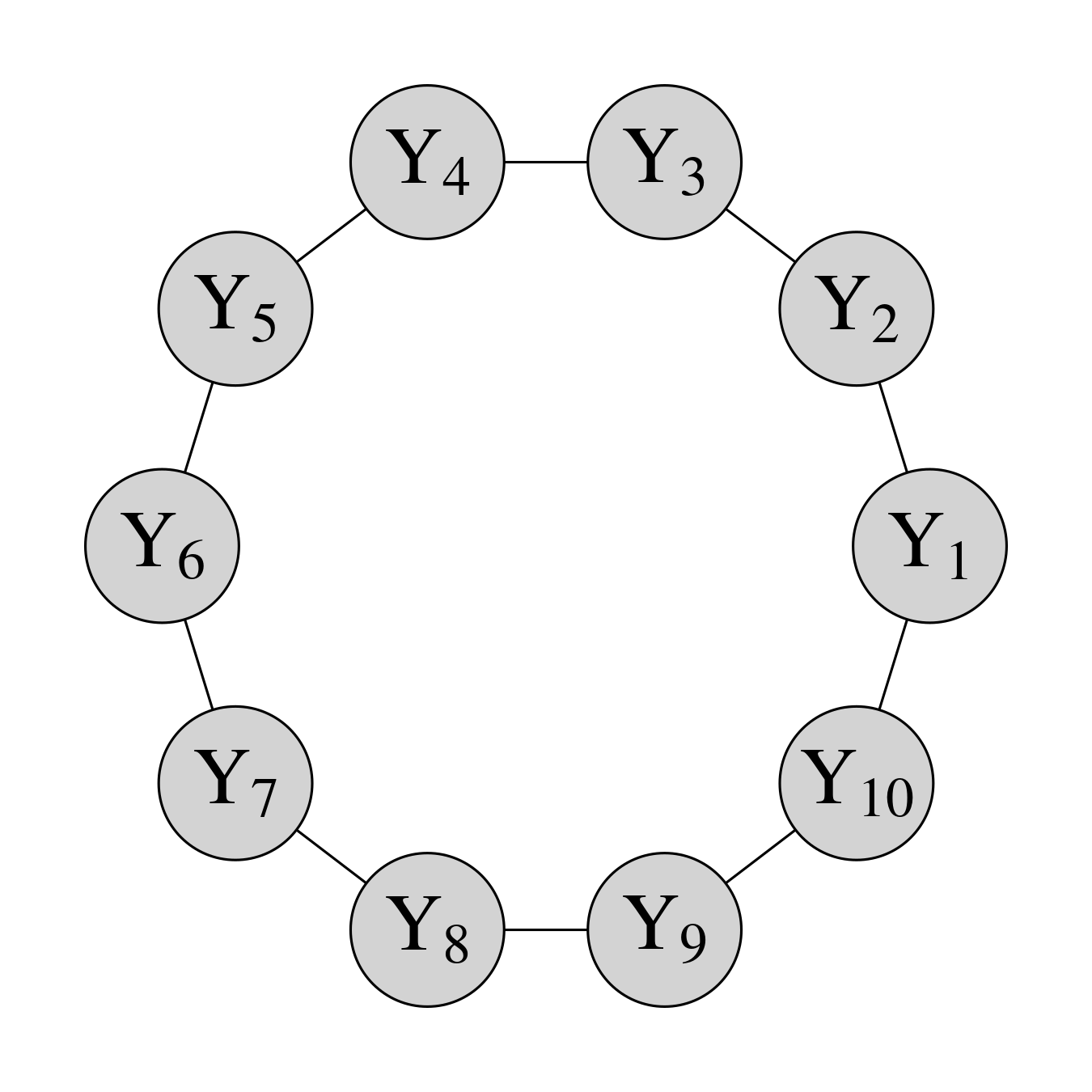}
\end{subfigure}
\begin{subfigure}[b]{0.15\textwidth}
\centering
\includegraphics[width = \textwidth]{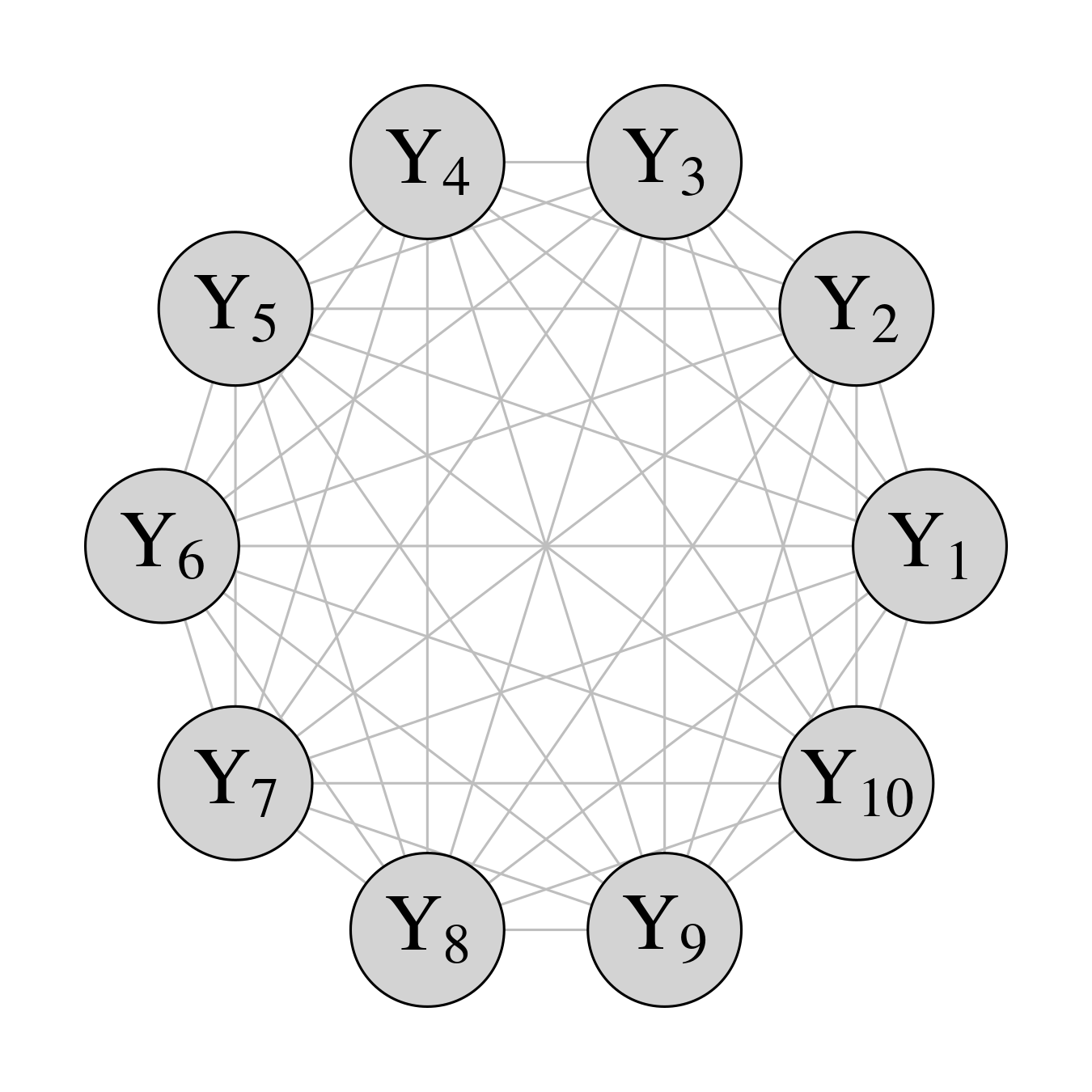}
\end{subfigure}	
    \caption{Precision matrices $\mathbf \Omega$ and corresponding graphical structure used in the simulations for a multivariate response of dimension $k=10$. These six models correspond to the six graphical structures in \citet{glasso}. \ysrevision{Positive entries represented in gray and negative entries in black.}}
    \label{fig:models}
\end{figure}

\ysrevision{One of the crucial differences between chain graphs and multiresponse regression models is the interplay between $\mathbf \Omega$ and $\mathbf B$, thus,} we simulate data under the six graphical structures in \citet{glasso}, \ysrevision{i.e. 1) an AR(1) model so that $\mathbf \Omega$ is tridiagonal; 2) an AR(2) model such that $\omega_{k,k'}=0$ whenever $|k-k'|>2$; 3) a block model so that there are two dense blocks along the diagonal; 4) a star model with every node connected to the first node; 5) a circle model so that the graph forms a circle, and 6) a full dense model. See Figure \ref{fig:models} for a visualization of the six graphical structures and Appendix \ref{sec:graphical_structure} for more details. We vary the sparsity of $\mathbf B$ with 80\% or 50\% entries equal to zero (denoted beta sparsity of 0.8 and 0.5 in the figures) with non-zero entries sampled from a standard Normal distribution. The multivariate response is sampled with dimension $k=10,30$, $p=5,10$ predictors and $n=50$ samples with zero mean. Design matrices are sampled from standard Normal distributions. Each simulation setting was repeated 50 times.}

\noindent On the simulated data, we compare the performance of 12 methods \ysrevision{that generally fall into two categories: 1) methods that estimate both $\mathbf{B}$ and $\mathbf{\Omega}$ (or that can get the parameters by transformation) and 2) methods that only estimate $\mathbf{\Omega}$.} 

The methods that estimate both $\mathbf{B}$ and $\mathbf{\Omega}$ are
\begin{itemize}
    \item[1.] \ysrevision{CG}-LASSO: our proposed model, \ysrevision{denoted as \texttt{CG-LASSO} or \texttt{CG} in the figures};
    \item[2.] Adaptive \ysrevision{CG}-LASSO: our proposed model with different shrinkage parameters for different entries in $\mathbf{B}$ and $\mathbf{\Omega}$, \ysrevision{denoted \texttt{CG-ALASSO} and \texttt{CG-A} in the figures};
    \item[3.] SRG-LASSO: \ysrevision{Bayesian LASSO with standard mean-covariance parametrization. That is, sparsity is on $\mathbf{\tilde{B}}=\mathbf{B\Omega}^{-1}$ and $\mathbf\Omega$ (see Section \ref{sec:cond_indp}), denoted as \texttt{SRG-LASSO} and \texttt{SRG} in the figures;}
    \item[4.] Bayesian multiresponse regression with conjugate priors, 
    \ysrevision{denoted as \texttt{multireg} in the figures};
    \item[5.] Bayesian multiresponse regression with conjugate priors that assume the marginal mean is 0 (similar to Graphical LASSO), \ysrevision{denoted as \texttt{multireg\_mu0} in the figures}.
\end{itemize}

\ysrevision{Note that methods 3-5 are based on the idea of fitting the marginal model (similar to covariate-adjusted GGMs, e.g.  \citet{chen2016asymptotically,zhang2022high}) and getting the chain graph parameter by transformation $\mathbf{B}=\tilde{\mathbf{B}}\mathbf\Omega$. While methods 4-5 put no sparsity assumptions, method 3 indeed imposes sparsity on $\tilde{\mathbf B}$ rather than on $\mathbf B$. All of our comparisons in the simulations are based on the conditional parameter $\mathbf B$, but see the results on real data (Section \ref{sec:real_data_experiments}) for comparisons among the marginal, the conditional and the conditional obtained from the marginal parameters.}

The methods that only estimate $\mathbf{\Omega}$ are

\begin{itemize}
    \item[6.] Graphical LASSO in \citet{glasso}, \ysrevision{denoted as \texttt{GLASSO} in the figures};
    \item[7.] Adaptive Graphical LASSO: adaptive version in \citet{glasso}, \ysrevision{denoted as \texttt{GALASSO} in the figures};
    \item[8.] Augmented Graphical LASSO: Graphical LASSO including responses and predictors (assumed Normally distributed), \ysrevision{denoted as \texttt{GLASSO-aug} in the figures};
    \item[9.] Adaptive version of Augmented Graphical LASSO, \ysrevision{denoted as \texttt{GALASSO-aug} in the figures};
    \item[10.] Bayesian multiresponse regression with conjugate priors \ysrevision{(Wishart prior on the precision matrix and a Normal prior on the mean)} that assume the marginal mean is 0 (similar to Graphical LASSO), but using all the responses and predictors as ``responses" in the model \ysrevision{and no predictors, denoted as \texttt{multireg\_mu0-aug} in the figures};
    \item[11.] Calculate the inverse of the empirical covariance matrix, denoted \texttt{ad-hoc} in the figures;
    \item[12.] Calculate the inverse of the empirical covariance matrix with both responses and predictors, \ysrevision{denoted as \texttt{ad-hoc-aug} in the figures}.
\end{itemize}

As in \citet{glasso} and \citet{Leng2014adaptivelasso}, we set the hyperparameters of the Gamma hyperprior for the shrinkage parameters of both $\mathbf B$ and $\mathbf \Omega$ as $r=0.01,\delta=10^{-6}$ for the adaptive versions, and $r=1,\delta=0.01$ for the non-adaptive versions. In the multiresponse regression models (methods 8-10), we consider any edge with weight $<1\times 10^{-3}$ to be 0.

To evaluate the performance of the methods, we compute the Frobenius loss of the estimate of $\mathbf B$ and the Stein's loss of the estimate of $\mathbf \Omega$. 
We use Stein's loss for $\mathbf \Omega$ since it is the KL-divergence when the mean vector is $0$.
In addition, we evaluate the reconstruction of the graphical structures based on the Matthews Correlation Coefficient (MCC) \citep{fan2009network} which range from $-1$ to $1$ with $1$ representing a perfect prediction. 

Finally, we calculate the proportion of true positive edges and false positive edges in the reconstructed graphs \ysrevision{(both $\mathbf B$ and $\mathbf \Omega$)}. The true positive rate is calculated as the proportion of times a true edge is reconstructed and the false positive rate is calculated as the proportion of times an edge appears in the estimated graph that is not present in the true graph. The false positive rate is presented as a negative quantity \ysrevision{in the figures} aligned with standard network reconstruction practice \citep{Xie2021}. 

\subsection{Simulation results}

\subsubsection{\ysrevision{Performance on the inference of $\mathbf B$}}
\begin{figure}[h]
	\centering
	\includegraphics[scale=0.5]{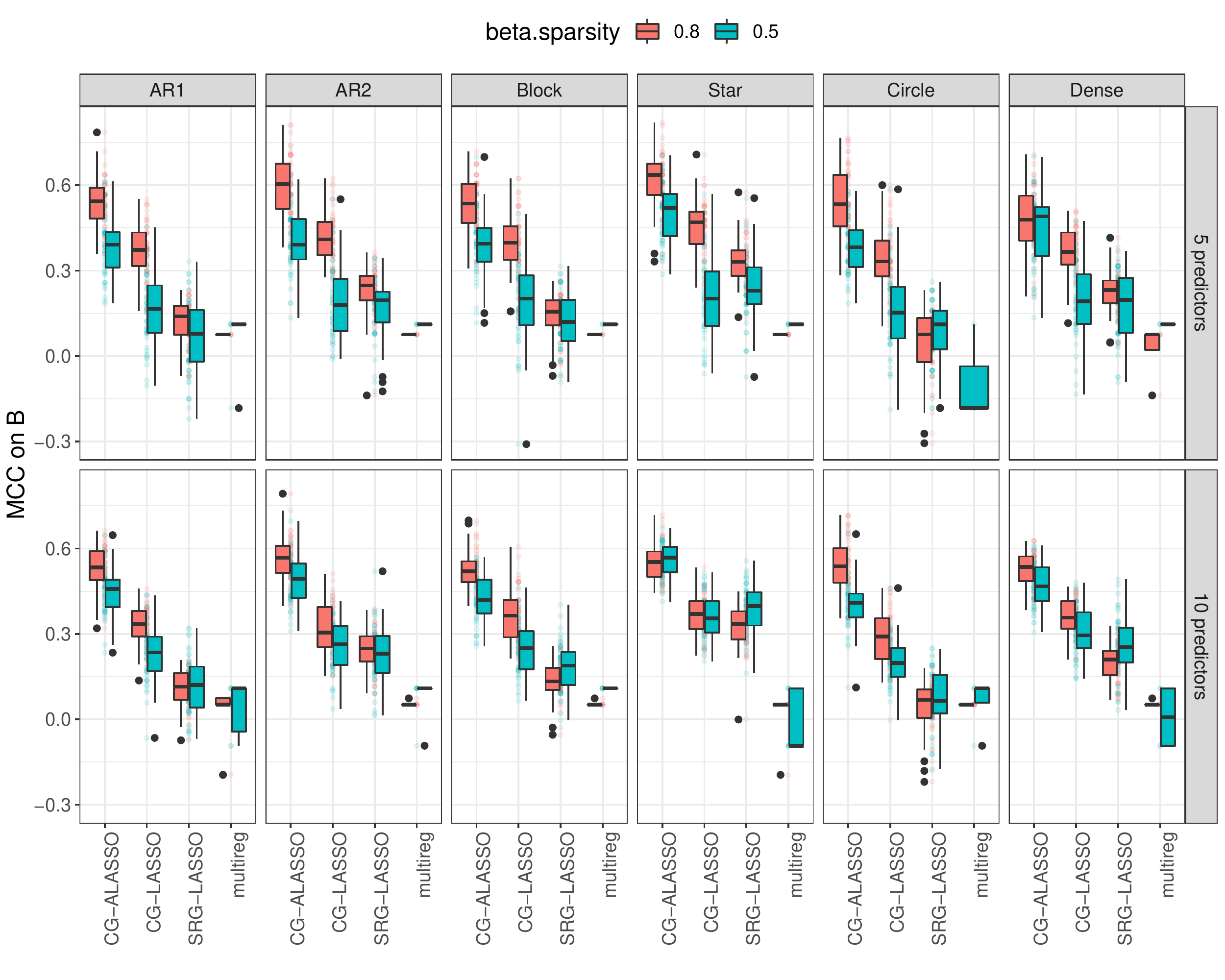}
	\caption{\textbf{Matthews Correlation Coefficients for} $\mathbf{B}$ for simulated datasets with 10 nodes and 50 samples under two levels of beta sparsity (red 0.8 and blue 0.5), two different number of predictors (10 in bottom row and 5 in top row) and six covariance models (columns, fully connected covariance model was omitted from $\mathbf \Omega$ result since MCC was not defined). X-axis corresponds to the models compared. MCC=1 means a perfect reconstruction. Our model Adaptive \ysrevision{CG}-LASSO gets the highest MCC in most cases. \ysrevision{We omit the \texttt{multireg\_mu0} model because it performs poorly across all cases (MCC close to 0).}}
	\label{fig:MCC_beta10}
\end{figure}

Our proposed models (\ysrevision{CG}-LASSO and adaptive \ysrevision{CG}-LASSO) outperform the other models \ysrevision{to reconstruct $\mathbf{B}$ as evaluated by MCC in Figure \ref{fig:MCC_beta10} for $k=10$. For $k=30$, both CG-LASSO and SRG-LASSO have comparable performance (Figure \ref{fig:MCC_beta}), but adaptive CG-LASSO continues to outperform all other methods} in almost every \ysrevision{combination of sparsity settings on $\mathbf{B}$ and structure of $\mathbf{\Omega}$ (Figure \ref{fig:models}).} 

\ysrevision{Regarding the Frobenius loss of the estimate of $\mathbf B$ (Figure \ref{fig:beta_10} for $k=10$ and Figure \ref{fig:beta_30} for $k=30$), both adaptive and non-adaptive CG-LASSO outperform all other methods. This is particularly true for the circle model for which the Frobenius loss of SRG-LASSO is much higher than any other method.} 

\begin{figure}[htp]
    \centering
    \includegraphics[width=\linewidth]{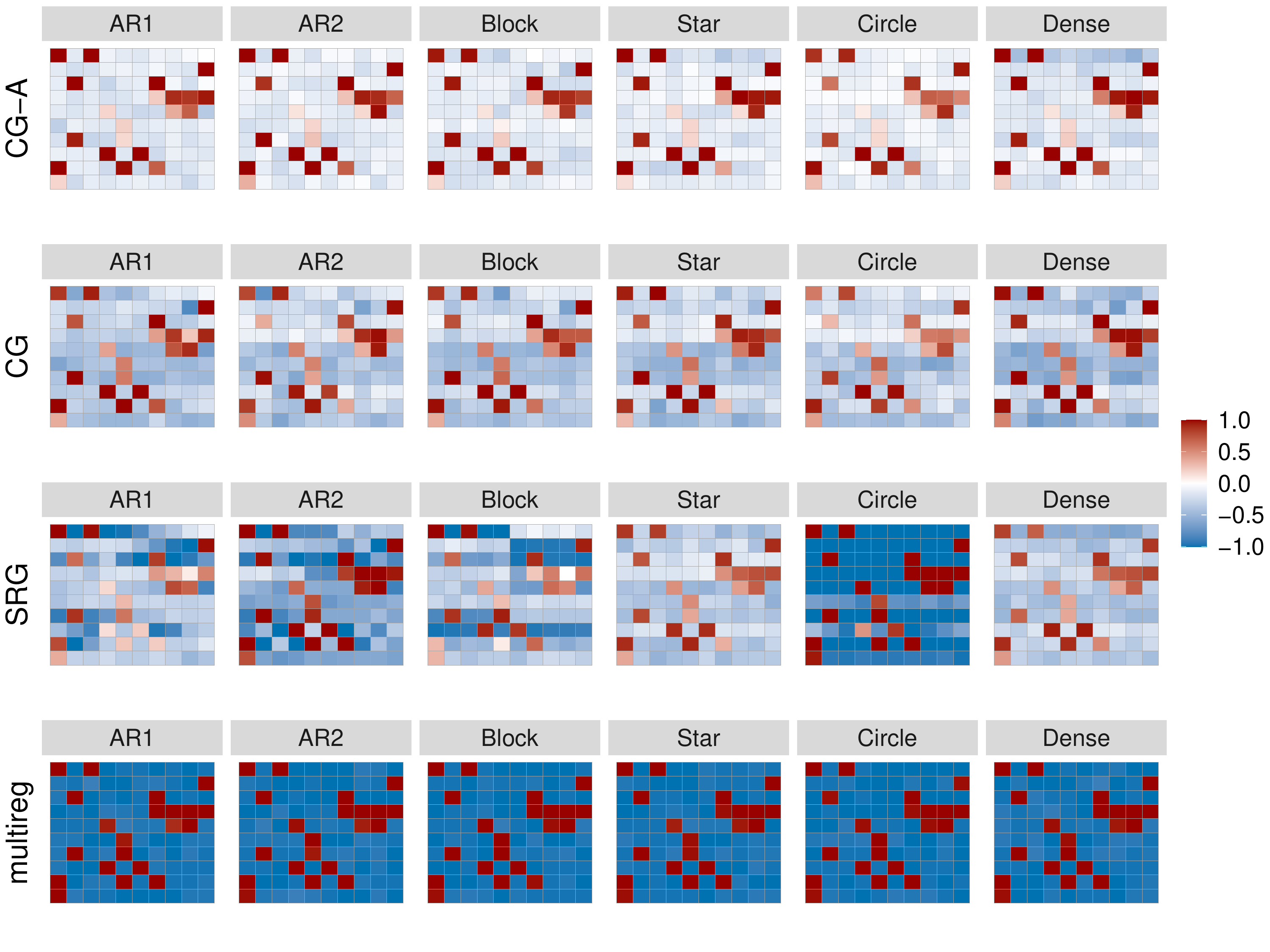}
    \caption{\textbf{Reconstruction accuracy of the graph between responses and predictors ($\mathbf B$)} for $k=10$ nodes, $p=10$ predictors and sparsity of $0.8$. Red entries correspond to true positive edges and blue entries correspond to false positive edges. Darker color means higher frequency of being estimated in 50 reconstructions. Our proposed method Adaptive \ysrevision{CG}-LASSO (\ysrevision{CG}-A) outperforms the other methods by displaying the lowest false positive rate (less blue).}
    \label{fig:vis_learnings}
\end{figure}

\ysrevision{Next, we select the four best performing methods based on MCC plus Bayesian multiresponse linear regression as a reference (method 8) to plot the true/false positive rates in the estimation of edges. Figure \ref{fig:vis_learnings} shows the structure reconstruction for $\mathbf B$ with $k=10$ nodes, $p=10$ predictors and sparsity level of $0.8$ (see Figures \ref{fig:visbeta_learnings.2_5}, \ref{fig:visbeta_learnings.5_5}, \ref{fig:visbeta_learnings.5_10} for other number of nodes, predictors and sparsity). Adaptive CG-LASSO (CG-A) produces the lowest false positive rate (represented as blue entries) compared to the other methods with Bayesian multiresponse linear regression having the highest false positive rates. Similarly to our conclusion before, the circle model is particularly difficult for SRG-LASSO as shown here (Figure \ref{fig:vis_learnings}) with high false positive rates.}

\ysrevision{Note that SRG-LASSO and Bayesian multiresponse linear regression use the formulation of covariate-adjusted GGMs to fit the marginal model and obtain the conditional regression coefficients by transformation. The simulation results show that SRG-LASSO and Bayesian multiresponse linear regression models have lower MCC (Figure \ref{fig:MCC_beta10}) and high false positive rates (Figure \ref{fig:vis_learnings}), and thus, these models are not very effective to infer $\mathbf B$ when $\mathbf{B}$ is indeed sparse which justifies the use of the chain graph model (CG-LASSO). After all, the chain graph model puts a different sparsity assumption and has a different interpretation than the marginal model and the covariate-adjusted GGM.} 

\subsubsection{\ysrevision{Performance on the inference of $\mathbf \Omega$}}

\begin{figure}[htp]
	\centering
	\includegraphics[scale=0.5]{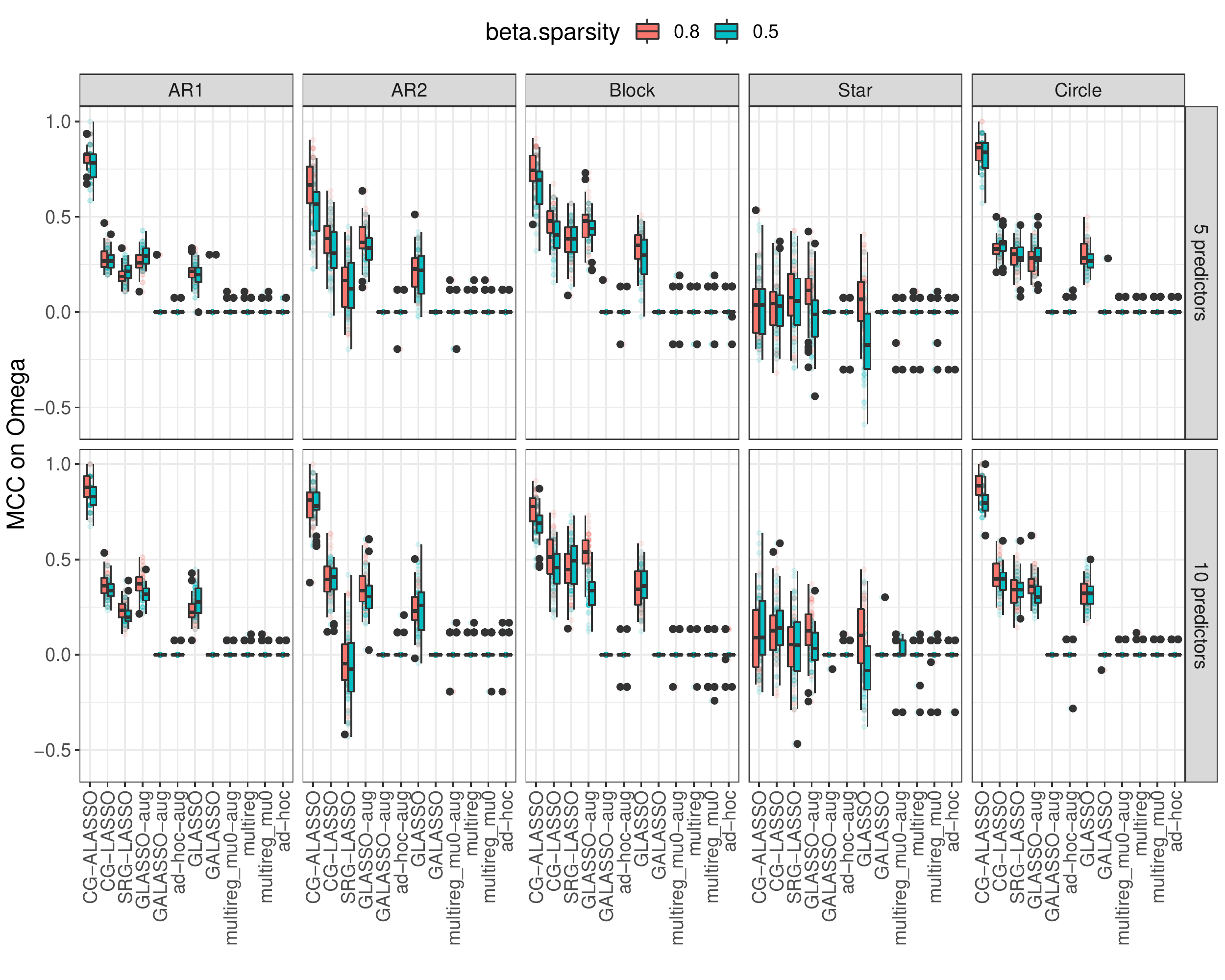}
	\caption{\textbf{Matthews Correlation Coefficients for} $\mathbf{\Omega}$ for simulated datasets with 10 nodes and 50 samples under two levels of beta sparsity (red 0.8 and blue 0.5), two different number of predictors (10 in bottom row and 5 in top row) and six covariance models (columns, fully connected covariance model was omitted from $\mathbf \Omega$ result since MCC was not defined). X-axis corresponds to the models compared. MCC=1 means a perfect reconstruction. Our model Adaptive \ysrevision{CG}-LASSO gets the highest MCC in most cases. We omit the dense model because MCC is not defined. }
	\label{fig:MCC_Omega10}
\end{figure}

\ysrevision{Our simulations show that adaptive CG-LASSO outperforms all other alternatives to accurately reconstruct the structure of $\mathbf\Omega$ in most settings (evaluated by MCC in Figure \ref{fig:MCC_Omega10} for $k=10$, and in Figure \ref{fig:MCC_Omega} for $k=30$). Non-adaptive CG-LASSO and SRG-LASSO show comparable performance, as well as augmented GLASSO. The star model proves to be equally difficult for all methods.} In the star model, the off diagonal signal is weak compared to other models (off diagonal entries are close to 0.1 while diagonal entries are close to 1). It is also the sparsest setting with 80\% of off diagonal entries to be zero when $k=10$ and 93\% when $k=30$. This setup might cause penalized methods to over penalize the off diagonal entries. \ysrevision{The AR2 model for $k=30$ nodes (Figure \ref{fig:MCC_Omega}) also proves to be difficult for all methods, except for the adaptive CG-LASSO.}

It is worth noting that (augmented) Graphical LASSO show good performance in most cases (Figure \ref{fig:MCC_Omega10}). However, this method assumes that the joint distribution of responses and predictors is Normal. This is the case in our simulation setup, and hence, the good performance of the method. However, our \ysrevision{CG}-LASSO model is more flexible because does not make any assumptions on the design matrix. 

\ysrevision{In terms of the Stein's loss of the estimate of $\mathbf \Omega$, (adaptive) CG-LASSO, SRG-LASSO and augmented GLASSO all have comparable performance with loss close to zero under most settings (Figure \ref{fig:stein_10} for $k=10$ and Figure \ref{fig:stein_30} for $k=30$)}

\begin{figure}[H]
    \centering
    \includegraphics[width=\linewidth]{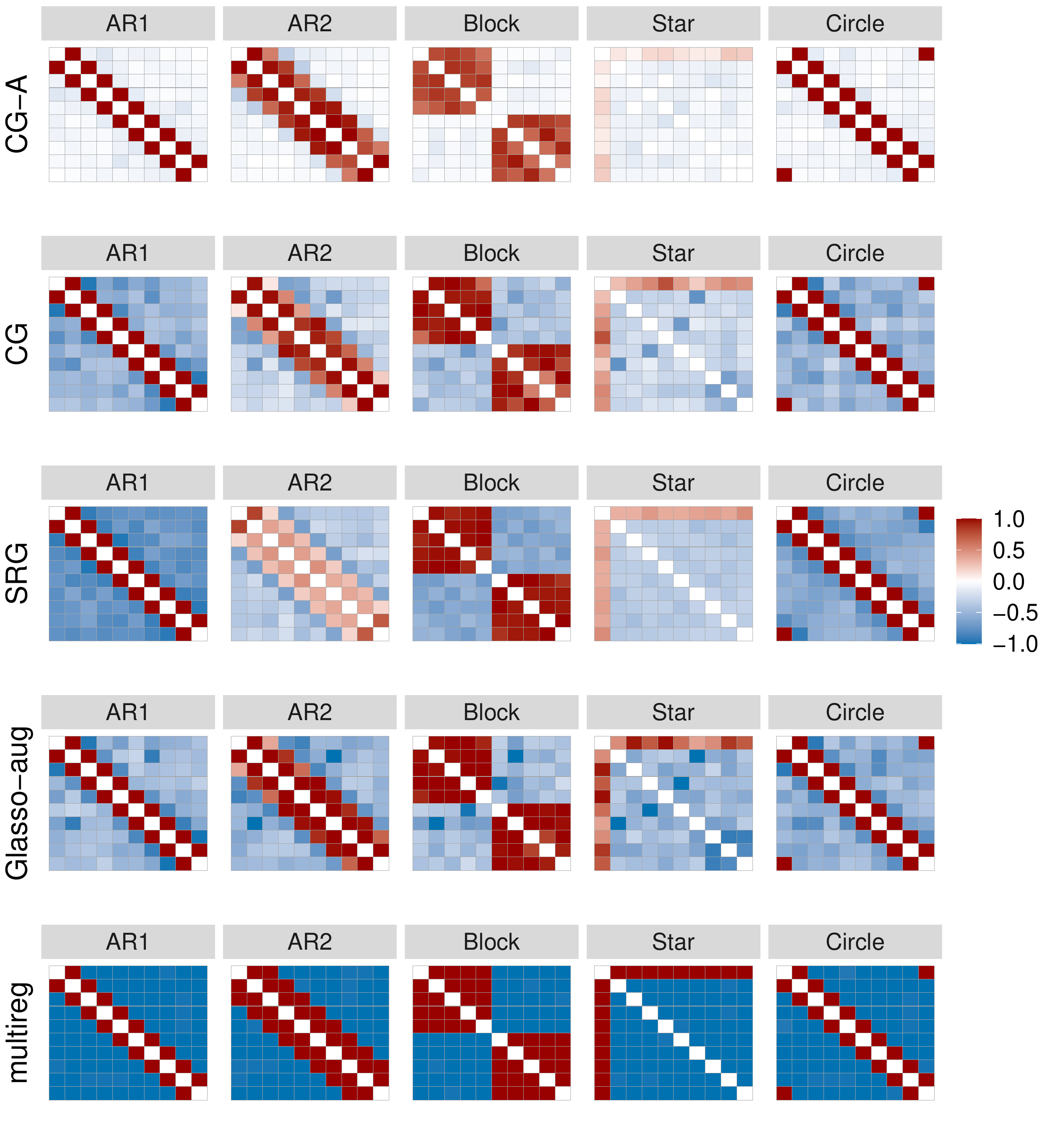}
    \caption{\textbf{Reconstruction accuracy of the graph among responses ($\mathbf \Omega$)} for $k=10$ nodes, $p=10$ predictors and sparsity of $0.8$. Red entries correspond to true positive edges and blue entries correspond to false positive edges. Darker color means higher frequency of being estimated in 50 reconstructions. Our proposed method Adaptive \ysrevision{CG}-LASSO (\ysrevision{CG}-A) outperforms the other methods by displaying the lowest false positive rate (less blue). We omit the dense model because it has no false positive or true negatives. }
    \label{fig:vis_learnings.2_10}
\end{figure}

\ysrevision{Last, in terms of false positive rates in the reconstruction of $\mathbf \Omega$, we select the four best methods in terms of MCC: adaptive CG-LASSO, CG-LASSO, SRG-LASSO, augmented GLASSO, plus Bayesian multiresponse linear regression as a reference (method 8). Figure \ref{fig:vis_learnings.2_10} (for $k=10$ nodes, $p=10$ predictors and sparsity level of $0.8$) shows that adaptive CG-LASSO outperforms all other methods in terms of controlled false positive edges with Bayesian multiresponse linear regression performing the worst. See Figures \ref{fig:vis_learnings.2_5}, \ref{fig:vis_learnings.5_5} and \ref{fig:vis_learnings.5_10} for other numbers of nodes, predictors and sparsity levels.}

\subsection{Computational speed and scaling}
We test the scalability of our estimation procedure by simulating 500 and 1000 samples with 5, 10, 25, 50, 100 nodes. We sample 1000 generations with 100 burn-in on a machine with Core-i7 4790 CPU and Windows 7 operating system. We recorded CPU seconds in R. 

While our models are slower than Graphical LASSO or multiresponse regression, running time is not severely impacted by sample size (Figure \ref{fig:scaling}). Instead, speed is mostly influenced by the number of nodes and the number of predictors. However, even the case of 100 nodes and 10 predictors is successfully completed in less than 10 minutes.

\section{Application to real microbiota data}
\label{sec:real_data_experiments}

\subsection{Human gut microbiota compositional data}
\label{human}

The microbiota of older people displays greater inter-individual variation than that of younger adults. The study in \citet{claesson2012gut} collected faecal microbiota composition from 178 elderly subjects, together with subjects' residence type (in the community, day-hospital, rehabilitation or in long-term residential care) and diet (data at \citet{gut_dataset}). Researchers studied the correlation between microbes and other measurements. They found that individual microbiota of people in long-stay care was significantly less diverse and loss of diversity might associate with increased frailty. They clustered microbes based on co-abundances and performed dimension reduction techniques to infer relationships between composition and health. However, co-abundances might not appropriately infer interactions because of the existence of other microbes and environment \citep{Blanchet2020_coexist}. Partial correlations between microbes and environment or other microbes are more ecologically meaningful. Here, we infer the partial correlation between environments and among microbes in those elderly subjects by reconstructing a sparse network via the adaptive \ysrevision{CG}-LASSO model.

We use the MG-RAST server \citep{meyer2008metagenomics} for profiling with an e-value of 5, 60\% identity, alignment length of 15 bp, and minimal abundance of 10 reads. Unclassified hits are not included in the analysis. Genus with more than 0.5\% (human) or 1\% (soil) relative abundance in more than 50 samples is selected as the focal genus and all other genus serve as the reference group. 

We reconstruct the weighted graph using the conditional regression coefficient between any two nodes. The $\alpha-$centrality \citep{bonacich2001eigenvector} is used to identify the importance of nodes. Weighted adjacency matrix is constructed with the posterior mean of the conditional regression coefficients of those that showed significance with the horseshoe method described in Section \ref{sec:graph_learning}.

\begin{figure}[htp]
	\centering
	\includegraphics[scale = 0.6]{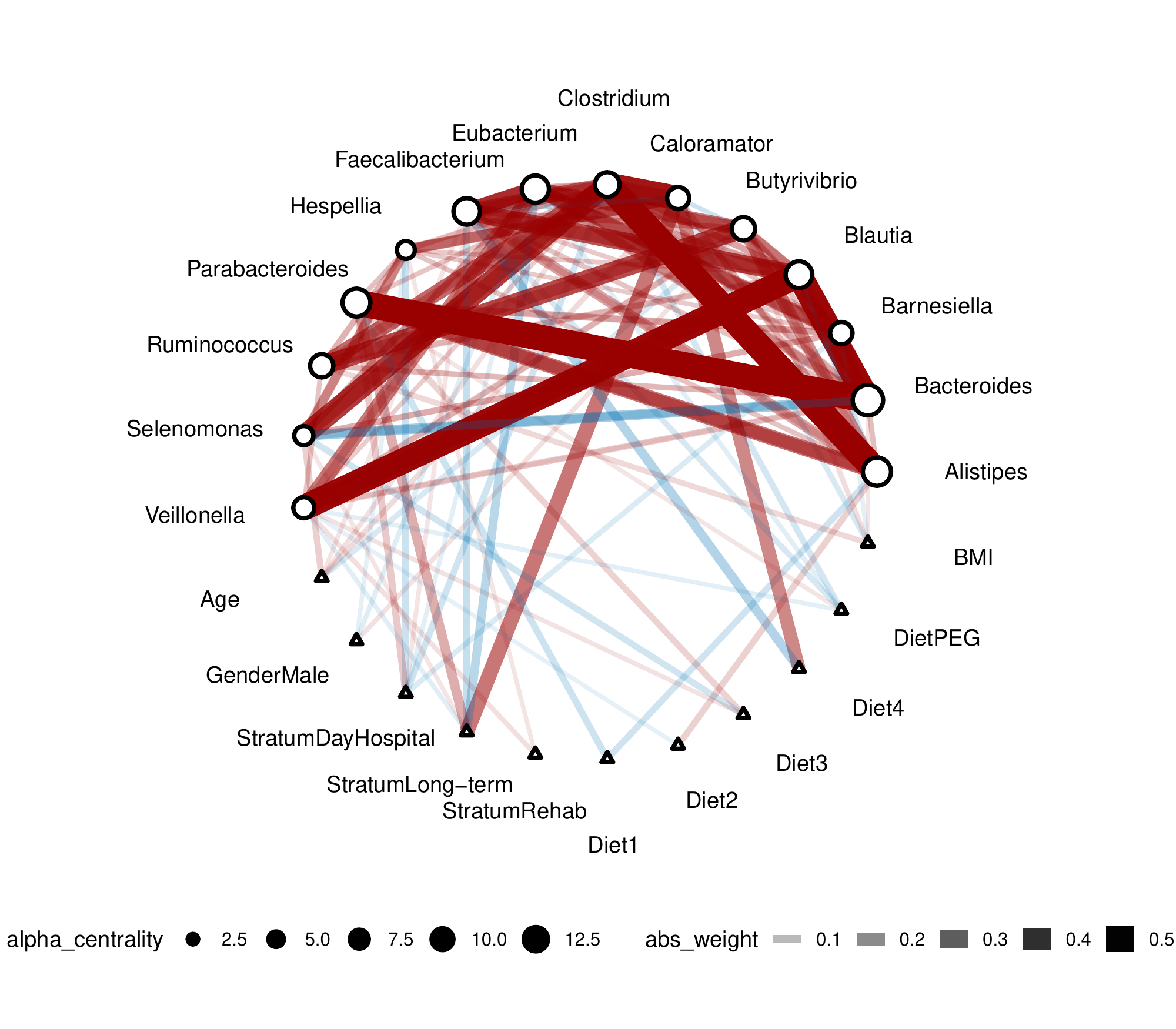}
	\caption{\textbf{Reconstructed genus conditional network for human gut microbiota using the adaptive \ysrevision{CG}-LASSO model.} Triangle nodes correspond to predictors and circle nodes correspond to relative abundances of genus. The node size on the circle nodes correspond to the $\alpha-$centrality values \citep{bonacich2001eigenvector}. The width of the edges correspond to the absolute weight, and the color to the type of interaction (red positive, blue negative).}
    \label{fig:real_data_a}
\end{figure}

Figure \ref{fig:real_data_a} shows the estimated human gut microbiota network under the adaptive \ysrevision{CG}-LASSO model where the edges with the most weight correspond to connections between genus nodes, not so much with predictors. The most important predictor is whether the patient's residence was a long-term residential care which positively affected genus \textit{Caloramator}. This result agrees with the original analysis that also separated elderly subjects based upon where they live in the community. Another important predictor is Diet Group 4 which corresponds to the high fat/low fiber group. This diet positively affected genus \textit{Caloramator} as well.

\subsubsection{Comparison with a marginal network}
\begin{figure}[htp]
	\centering
	\includegraphics[scale = 0.6]{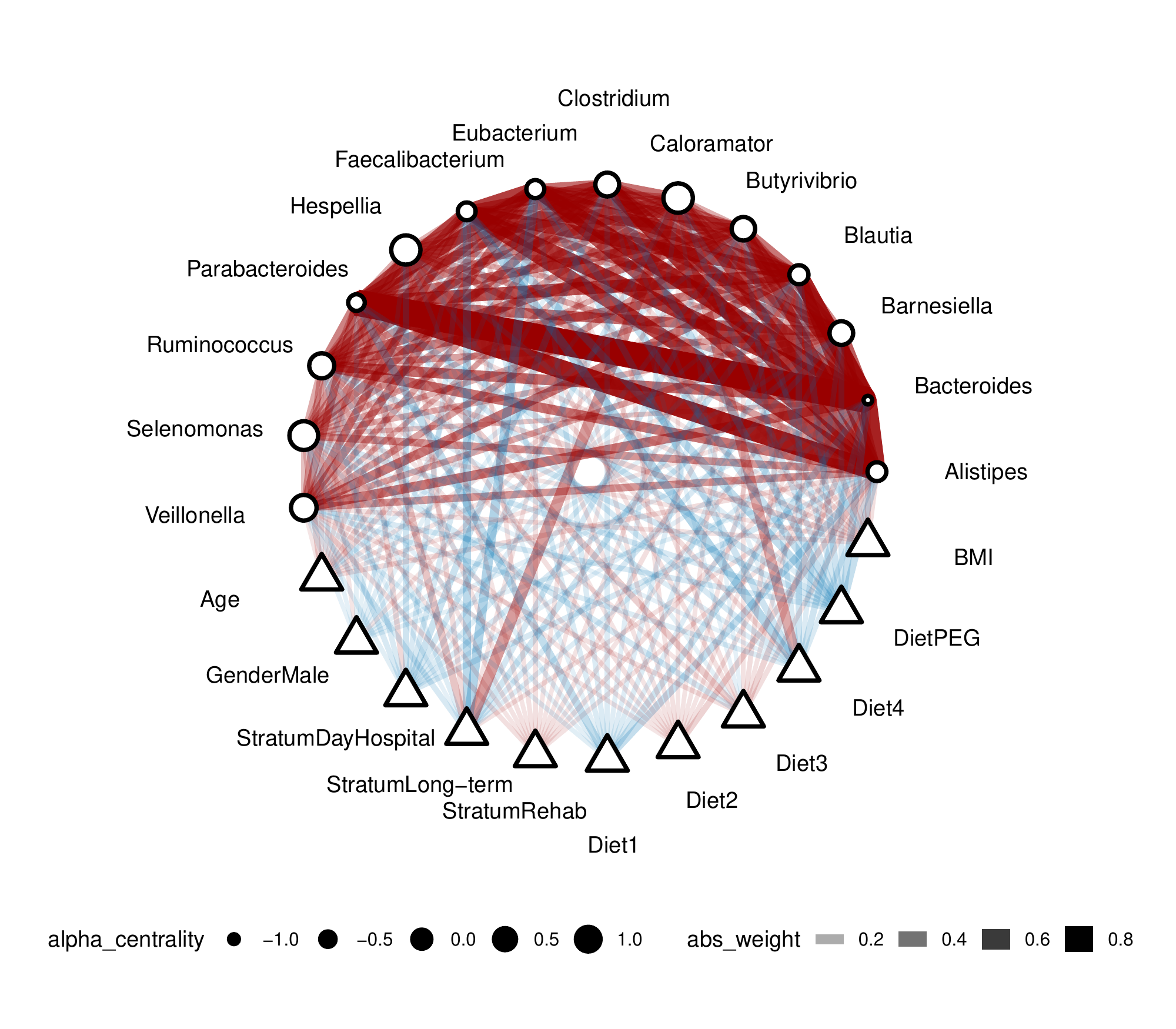}
	\caption[Gut marginal network]{\textbf{Reconstructed genus marginal network for human gut microbiota using multiresponse regression.} Triangle nodes correspond to predictors and circle nodes correspond to relative abundances of genus. The node size on the circle nodes correspond to the $\alpha-$centrality values \citep{bonacich2001eigenvector}. The width of the edges correspond to the absolute weight, and the color to the type of interaction (red positive, blue negative). Edges are marginal rather than conditional. Note that the larger size of the triangles compared to the triangles in the conditional network (Figure \ref{fig:real_data_a}) are an effect of the shrinkage of the circle nodes in this network.}
    \label{fig:gut_margin}
\end{figure}

As comparison, we estimate the marginal network (Figure \ref{fig:gut_margin}) to observe the differences with the conditional network in Figure \ref{fig:real_data_a}. To obtain the marginal network, we start with the conditional network and use the equivalence $\mathbf{\tilde{B}} = \mathbf{B \Omega^{-1}}$ in Section \ref{sec:cond_indp}. Edges between predictors and responses are marginal regression coefficients while edges between responses are covariances.
Edges between responses and predictors generally agree within two cases given the partial correlation between responses are mostly positive (Figure \ref{fig:gut_margin}). However, marginally the connection between responses and predictors is very dense. This is because all responses (genus) are connected by their partial correlations so that as long as the predictor can influence one of the responses conditionally, it should be able to affect all the responses marginally.
We observe that some edges flip color when comparing the conditional and the marginal network. For example, the link between Diet Group 2 corresponding to both complex (wholegrain breakfast cereals and breads, boiled potatoes) and simple
carbohydrates (white bread) and \textit{Veillonella} is blue (negative) in the conditional network (Figure \ref{fig:real_data_a}) and red (positive) in the marginal network (Figure \ref{fig:gut_margin}). \textit{Veillonella} is well known for its lactate fermenting abilities, so a negative link with a carbohydrate diet is reasonable.
Edges flipping color could be explained by interactions with other genus. 
Another example is that marginally, all links are blue (negative) from Diet Group PEG (percutaneous endoscopic gastrostomy (PEG)-fed subjects) to the genus whereas conditionally there is a positive (red) link with \textit{Parabacteroides}. These observations further reiterate that we should distinguish marginal and conditional effects of predictors in these research scenarios, especially when sparsity is assumed since it is usually impossible to have both marginal and conditional effects sparse. 

\ysrevision{Last, we also compare our conditional network (Figure \ref{fig:real_data_a}) with the conditional network obtained by fitting a multiresponse regression and then transform the parameters into the conditional chain graph representation using the equivalence $\tilde{\mathbf B} = \mathbf{B \Omega}$ in Section \ref{sec:cond_indp} (Figure \ref{fig:human_cond_from_marg}). This last network (transformed conditional network) is extremely dense, even though biologically, we expect the network to be sparse (hence the sparsity imposed by the chain graph model). This result shows the importance of having a proper sparsity prior.}

\subsection{Soil microbiota compositional data}

The objective of this study \citep{soil_dataset,bach2018greatest} is to examine soil microbial community composition and structure of both bacteria and fungi at a microbially-relevant scale. The researchers isolated soil aggregates from three land management systems in central Iowa to test if the aggregate-level microbial responses are related to plant community and management practices. The clean dataset has 120 samples with 17 genus under consideration. We focus on the bacteria to further evaluate the partial association among them and the environmental factors. 
    
We use the MG-RAST server \citep{meyer2008metagenomics} with the same settings as in the human gut microbiome data described in Section \ref{human}. In addition, weighted adjacency matrix is also constructed with the posterior mean of the conditional regression coefficients of those that showed significance with the horseshoe method described in Section \ref{sec:graph_learning}.

\begin{figure}[htp]
	\centering
	\includegraphics[scale=0.6]{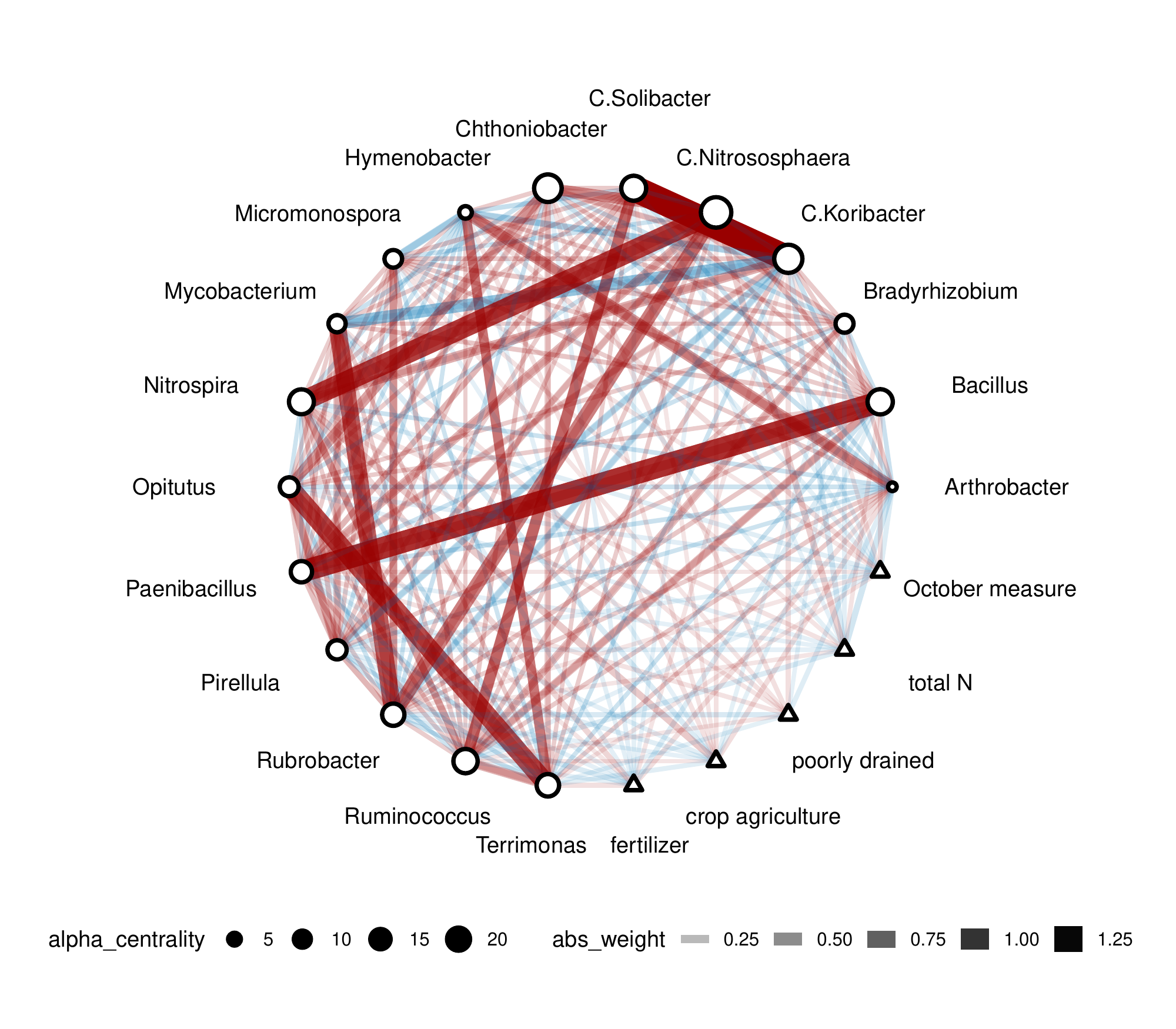}
	\caption{\textbf{Reconstructed genus conditional network for soil microbiota using adaptive \ysrevision{CG}-LASSO.} Triangle nodes correspond to predictors and circle nodes correspond to relative abundances of genus. The node size on the circle nodes correspond to the $\alpha-$centrality values \citep{bonacich2001eigenvector}. The width of the edges correspond to the absolute weight, and the color to the type of interaction (red positive, blue negative). \ysrevision{Weak links with environment (triangle nodes) agree with the original research \citep{soil_dataset,bach2018greatest} showing stable microbial community to environmental perturbations.}}
	\label{fig:real_data_b}
\end{figure}

Figure \ref{fig:real_data_b} shows the soil microbiota estimated network using the adaptive \ysrevision{CG}-LASSO model. In this network, the most important link is between \textit{Candidatus Solibacter} and \textit{Candidatus Koribacter}. There are not important connections with predictors in this case which seem to suggest that the soil microbial community is robust to environmental perturbations.
These results agree with the original research \citep{soil_dataset,bach2018greatest} that indicated that core microbial communities within soil aggregates are likely driven by stable and long-term factors such as clay content rather than relative short time scaled land management as the ones considered as predictors in this study. 
We note that the original research concentrated on the diversity of the community while our analysis focuses on the structure and correlations within the community.

\subsubsection{Comparison with a marginal network}
\begin{figure}[htp]
	\centering
	\includegraphics[scale=0.6]{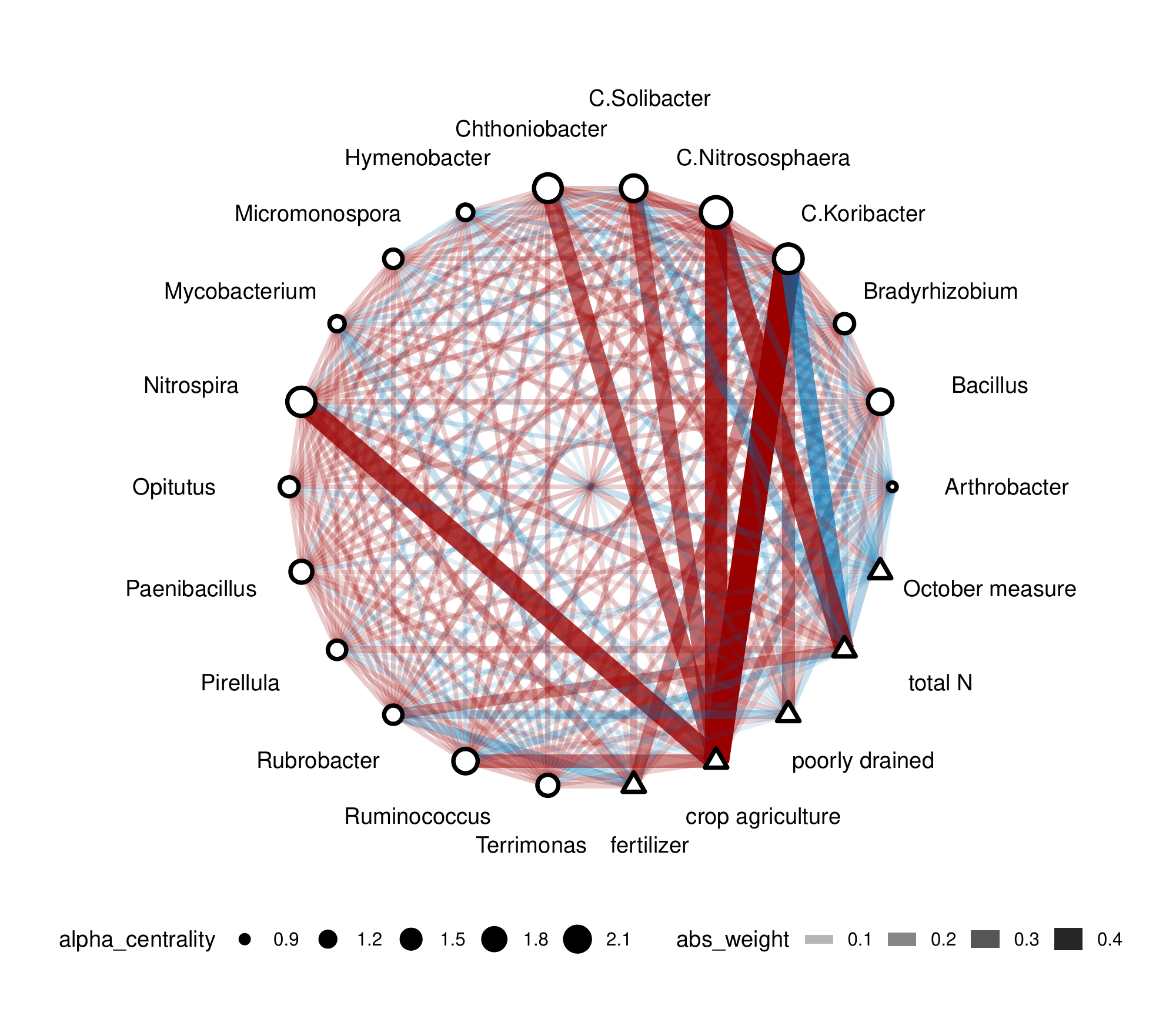}
	\caption[Soil marginal network]{\textbf{Reconstructed genus marginal network for soil microbiota using multiresponse regression.}  Triangle nodes correspond to predictors and circle nodes correspond to relative abundances of genus. The node size on the circle nodes correspond to the $\alpha-$centrality values \citep{bonacich2001eigenvector}. The width of the edges correspond to the absolute weight, and the color to the type of interaction (red positive, blue negative). Edges are marginal rather than conditional. \ysrevision{Strong links with environment (triangle nodes) disagree with the original research \citep{soil_dataset,bach2018greatest} that showed the microbial community should be stable to environmental perturbations.}}
	\label{fig:soil_margin}
\end{figure}

Again, we estimate the marginal network (Figure \ref{fig:soil_margin}) starting with the conditional network and using the equivalence $\mathbf{\tilde{B}} = \mathbf{B \Omega^{-1}}$ in Section \ref{sec:cond_indp}. Edges between predictors and responses are marginal regression coefficients while edges between responses are covariances. In the estimated network (Figure \ref{fig:soil_margin}),
we see a much stronger marginal effect of environmental predictors than the one we had seen with conditional effects (Figure \ref{fig:real_data_b}). The LASSO penalty might contribute to this behavior, but it might also be because the observed strong dependence between microbes and environment is due to the strong partial correlation among microbes. In addition, 
marginally, crop agriculture and total nitrogen (total N) have strong correlations to multiple genus. However, this pattern is not obvious in the conditional network (nor on the original research). It might suggest that the influence of crop agriculture and total N might be enhanced by strong partial correlation among microbes. The largest partial regression coefficient for total N is to genus \textit{Nitrososphaera} which is indeed a N-fixing genus and it has the highest $\alpha-$centrality (Figure \ref{fig:centrality}).

\ysrevision{Last, we also compare our conditional network (Figure \ref{fig:real_data_b}) with the conditional network obtained by fitting a multiresponse regression and then transform the parameters into the conditional chain graph representation using the equivalence $\tilde{\mathbf B} = \mathbf{B \Omega}$ in Section \ref{sec:cond_indp} (Figure \ref{fig:soil_cond_from_marg}). This last network (transformed conditional network) is extremely dense, even though biologically, we expect the network to be sparse (hence the sparsity imposed by the chain graph model). Similarly to the human gut data, this result shows the importance of having a proper sparsity prior.}

To conclude, \ysrevision {both} the marginal network (Figure \ref{fig:soil_margin}) \ysrevision{and the conditional network transformed from marginal coefficients (Figure \ref{fig:soil_cond_from_marg})} disagree with the original research that core microbial communities within soil aggregates are likely driven by stable and long-term factors instead of the predictors measured in the data.
Unlike marginal effects, conditional effects of the environment (Figure \ref{fig:real_data_b}) can be more informative to biologists who would like to conduct research in understanding the environment's effect on certain microbes, for instance, the effect of environmental antibiotics. 

It is worth highlighting that our model can produce meaningful results from relative small sample sizes: 120 samples for the soil microbiota study and 178 samples for the human gut microbiota study.

\subsection{Comparison of $\alpha-$centrality and abundances in human and soil communities}

\begin{figure}[htp]
	\centering
	\includegraphics[width=\linewidth]{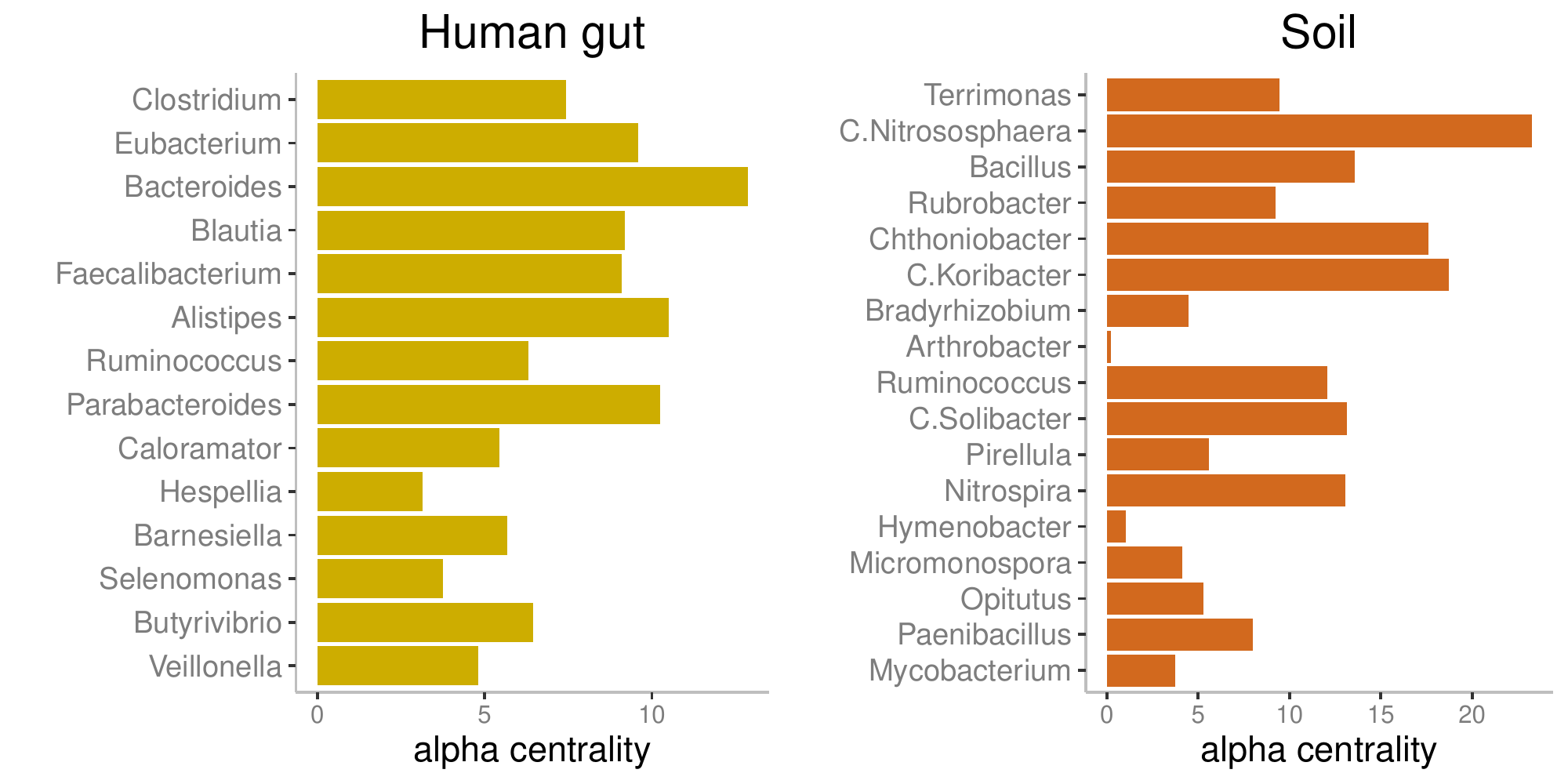}
	\caption{\textbf{$\alpha-$centrality of genus ranked by abundances.} We rank the genus based on estimated relative abundance (upper=higher abundance) and bars correspond to the estimated $\alpha-$centrality. We see a general trend of abundant genus having higher $\alpha-$centrality, but it is not definite.}
	\label{fig:centrality}
\end{figure}

We evaluate the $\alpha-$centrality based on the estimated network for both datasets (human gut and soil) to identify keystone genus. Figure \ref{fig:centrality} shows on the y-axis the ranking of genus based on the point estimation of the grand mean ($\mu$), i.e. the log relative abundances. That is, genus on the top correspond to the most abundant microbes (\textit{Clostridium} for human gut and \textit{Terrimonas} for soil). On the x-axis, we show the estimated $\alpha-$centrality for each genus. For the human gut data, the genus with the highest $\alpha-$centrality is \textit{Bacteroides} which is an abundant genus and known to have the ability to moderate the host's immune response \citep{blander2017regulation} and transfer antibiotic genes to other members of the community \citep{shoemaker2001evidence}. For the soil data, the genus with highest $\alpha-$centrality is \textit{Nitrososphaera}. Members of this genus have the ability to perform ammonia oxidizing which might play a major role in nitrification \citep{tourna2011nitrososphaera} which is crucial in the soil microbial community. We observe a general trend that abundant genus have higher $\alpha-$centrality, but this trend is not definite. For instance, neither \textit{Bacteroides} nor \textit{Nitrososphaera} have the highest $\alpha-$centrality estimate. The Spearman correlation coefficients between estimated relative abundance and $\alpha-$centrality in the human gut dataset is $0.68$ while it is $0.58$ in the soil dataset. Variation of $\alpha-$centrality measure is also larger in soil (standard deviation is $0.67$ times the mean in the soil data while the standard deviation is $0.38$ times the mean in the human gut data). The difference might due to the more variate environmental condition in soil making it difficult to have just one genus taking the central role.

\section{Discussion}
\label{sec:discussion}
\subsection{Importance of conditional dependence} It is crucial for any model dealing with predictors and multivariate responses to distinguish between marginal effects and conditional effects. A conditional construction coincides with the intuition that the marginal response of a node \textit{should} be influenced by both its and others' reaction to a common input. 
This distinction of marginal or conditional effects is particularly important when including biological prior knowledge. For example, species reactions to treatments can be measured under controlled experiments (e.g. \citet{lo2017mplasso}) and this knowledge would be properly encoded under a conditional dependence model. See more in the ``Agreement with experimenter's intuition on mean behavior" and ``Optimal model-based design of experiments" below.

\subsection{Flexibility of the Bayesian model}
Compared with the frequentist method, the Bayesian method allows an easier extension of the core Normal model to different types of responses via hierarchical structures. 
As long as one can sample from the full conditional distribution of the (latent) Normal variable, the posterior sampling is a straight-forward extension of the proposed Gibbs sampler. Though not shown here, other commonly encountered models in biology are also simple extensions of our model, e.g. zero-inflated Poisson and multinomial \citep{lambert1992zero}. By using the Normal distribution as the core model, we can automatically take into account the over-dispersion because the model considers the variance parameters explicitly. Note that one common complaint on the LASSO prior is that it does not put any mass on 0 for any edge. Though a spike-and-slab prior is possible, an efficient posterior sampling algorithm like the block Gibbs sampler presented in this work (also in \citet{glasso}) would be hard to derive due the intractable normalizing constant.

\subsection{Challenges in learning the graphical structure} Graphical selection can be difficult because of the confounding in its own structure. For example, recall Figure \ref{fig:egnetwork} A and B. These two graphs can produce a similar correlation between $Y_1$ and $Y_2$. One extreme example is when all links in A and B have no noise (e.g. $Y_1=X$, $Y_2=-Y_1$ versus $Y_1=X$, $Y_2=-X$). In this extreme example, it is impossible to distinguish graph A from B. Of particular difficulty are also cases like Figure \ref{fig:egnetwork} E where all partial correlations are positive (or negative). 
Additionally, when $\mathbf \Omega$ has bad condition numbers, then $\mathbf B$ might have large error in estimation since the marginal mean response and $\mathbf \Omega$ inform the estimation of $\mathbf B$, and a small change in the marginal mean response can have a large influence in $\mathbf B$. 
Future work could focus on how to use experiments to decouple the confounding of $\mathbf \Omega$ and $\mathbf B$ to address some of these challenges.

\subsection{Agreement with experimenter's intuition on mean behavior}
Intuitively, an experimenter should be able to make inferences about the interactions among responses from the behavior of the mean structures under treatment. For example, in Figure \ref{fig:egnetwork} D, an experimenter might knock out a gene as the treatment ($X=1$ for knock out and $X=0$ for not) and compare the gene expression levels of another gene ($Y_2$) via a t test. The result of this t test will provide information regarding the interaction between $Y_1$ and $Y_2$ because there are no other factors affecting $Y_1$ and $Y_2$ is conditionally independent with $X$. Thus, this experiment is specific to $Y_1$ and provides information on partial correlation between $Y_1$ and $Y_2$ by only affecting $Y_1$. That is, any change in $Y_2$ is due to the partial correlation with $Y_1$ rather than a reaction to $X$.
It is precisely the fact that the mean of $Y_2$ in this experiment depends on the correlation between $Y_1$ and $Y_2$ that allows experimenters to test differences in means of $Y_2$ under the effect of the treatment ($X$) through standard t tests. 
However, this intuition is violated under the standard linear regression setting. The vector $(Y_1,Y_2)$ is Normally distributed with mean $\mu=(X \beta_1, 0)$ and covariance $\mathbf{\Sigma}$ under the network in Figure \ref{fig:egnetwork} D, and thus, the mean of $Y_2$ is always 0 regardless of the value of $X$. In contrast, in the \ysrevision{CG} parametrization, the mean vector is $\mathbf\Sigma \mu$ whose second entry is given by $\rho\beta_1X$, i.e. the mean value of $Y_2$ depends on $\beta_1$ (the reaction of $Y_1$ to the treatment) as well as $\rho$ (the correlation between $Y_1$ and $Y_2$). 
Given that the experimenter's intuition on specificity is based on the notion of \textit{conditional (in)dependence} between $X$ and $Y_1$, $Y_2$, we conclude that it is desirable that the mean vector contains information on the correlation structure among responses and this is a characteristic of the \ysrevision{CG} model that we propose.

\subsection{Optimal model-based design of experiments}
An experimenter should be able to design experiments that decode the links among response nodes when specific experimental interventions towards one node are possible. In practice, when possible, experimenters will always prefer experiments with better specificity. 
However, this preference is not evident in the linear regression setting since the Fisher information matrix of the mean vector and the precision matrix is block-diagonal \citep{malago2015information}, and thus, any information that we have on $\mathbf B$ will not affect estimation of $\mathbf{\Sigma}$. In addition, the information of $\mathbf{\Sigma}$ is not a function of design ($\mathbf X$) no matter whether we have prior knowledge about effect of such experiment (prior on $\mathbf B$). The \ysrevision{CG} parametrization avoids this disagreement because the Fisher information matrix is no longer block-diagonal and prior information about the treatment can flow into the estimation of $\mathbf{\Sigma}$ via an optimal model-based experimental design \citep{chaloner1995bayesian}.
We highlight that due to the confounding between the treatment effect and the interaction among responses,
the prior knowledge on specificity of the treatment is necessary for such an optimal model-based experimental design. Future work could investigate the method of experimental design to best decode such networks and the theoretical properties of such designs.

\section{Open-source software}
We developed our algorithm in R 3.6.3 \citep{Rlang} and all the code as well as data used is available as an R package \texttt{CARlasso} hosted on 
\url{https://github.com/YunyiShen/CAR-LASSO}. All simulations and data analysies code are in the \texttt{dev} branch of the same repository.

\section*{Acknowledgements}

\ysrevision{
The authors thank the associate editor and two anonymous reviewers for greatly improving the manuscript with their feedback and suggestions.}
This work was supported by the National Institute of Food and Agriculture, United States Department of Agriculture, Hatch project 1023699.
This work was also supported by the Department of Energy [DE-SC0021016 to C.S.L.]. Y.S. would like to thank Xiang Li from the University of Hongkong for discussion on the Generalized Inverse Gaussian distribution.

\FloatBarrier

\bibliography{refs.bib}

\section*{Appendix}
\label{sec:appendix}
\appendix

\setcounter{figure}{0} 
\setcounter{equation}{0} 
\setcounter{table}{0} 
\setcounter{section}{0} 
\makeatletter 
\renewcommand{\thefigure}{A\@arabic\c@figure} 
\makeatother
\def\theequation{A\arabic{equation}}

\makeatletter 
\renewcommand{\thetable}{A\@arabic\c@table} %

\section{Derivation of the Gibbs sampling}
\label{Gibbs:Omega}
Let $\mathbf 1_n$ be the column vector of ones with dimension $n$, let $\mathbf{S} = \mathbf{Y}^T\mathbf{Y}\in \mathbb{R}^{k\times k}$ (here we have samples as row vectors in $\mathbf Y$),  let $\hat{\mu} = \mathbf{X}\mathbf{B}+\mathbf 1_n\mu^T$, and let $\mathbf{U} = \hat{\mu}^T\hat{\mu}\in \mathbb{R}^{k\times k}$. 
Equation \ref{full_Omega} shows the full conditional distribution of $\mathbf{\Omega}$ and $\eta$ (the hyperparameters in Equation \ref{eqn-prior}).

\begin{equation}
\label{full_Omega}
    \begin{aligned}
        p(\mathbf{\Omega},\eta|\mathbf{Y},\lambda_{\Omega},\hat{\mu})&\propto |\mathbf{\Omega}|^{\frac{n}{2}} \exp\left( -\frac{1}{2} \tr(\mathbf{S}\mathbf{\Omega})-\frac{1}{2}\tr(\mathbf{U}\mathbf{\Omega}^{-1}) \right) \prod_{q<q'}\left[\frac{1}{\sqrt{2\pi\eta_{qq'}}}\exp\left(-\frac{\omega_{qq'}^2}{2\eta_{qq'}}\right) \right]\\
        &\times \prod_{q=1}^{k}\left[\frac{\lambda_{\Omega}}{2}\exp\left(-\frac{\lambda_{\Omega}\omega_{qq}}{2}\right)\right]I_{\mathbf{\Omega}\in M^+}
    \end{aligned}
\end{equation}

We can update one row (column) at one iteration.
Let $\mathbf{H}$ be the symmetric matrix with $\mathbf{H}_{ml}=\mathbf{H}_{lm}=\eta_{ml}$ ($m<l$) on the off-diagonal entries and on the diagonal $\mathbf H_{mm}=0$. 
 We take one column out and partition $\mathbf\Omega$, $\mathbf{ S}$, $\mathbf{ U}$, and $\mathbf {H}$. Without lose of generality, we show the sampling scheme for the last row (column). Let $\mathbf\Omega_{11}\in \mathbb{R}^{(k-1)\times(k-1)}$, $\boldsymbol\omega_{12}\in\mathbb R^{k-1}$, and $\omega_{22}\in \mathbb R$. We partition $\mathbf S$, $\mathbf U$ and $\mathbf H$ in the same manner.
\[
\mathbf{\Omega}=\left[\begin{matrix}
\mathbf\Omega_{11} & \boldsymbol{\omega}_{12}\\
\boldsymbol{\omega}_{12}^T & \omega_{22}
\end{matrix}\right],
\mathbf{S}=\left[\begin{matrix}
\mathbf S_{11} & \boldsymbol s_{12}\\
\boldsymbol s_{12}^T & s_{22}
\end{matrix}\right],
\mathbf{U}=\left[\begin{matrix}
\mathbf U_{11} &  \boldsymbol u_{12}\\
\boldsymbol u_{12}^T & u_{22}
\end{matrix}\right],
\mathbf{H}=\left[\begin{matrix}
\mathbf H_{11} & \boldsymbol \eta_{12}\\
\boldsymbol \eta_{12}^T & 0
\end{matrix}\right].
\]

By setting 
\begin{equation}
\gamma =  \omega_{22}-\boldsymbol\omega_{12}^T\mathbf\Omega_{11}^{-1}\boldsymbol\omega_{12}\in \mathbb R, 
\label{omega22}
\end{equation}
$\mathbf{\Omega}^{-1}$ can be written in a block form: 

\[
\begin{aligned}
    \mathbf{\Omega}^{-1}
&=\left[\begin{matrix}
\mathbf\Omega_{11}^{-1}+\frac{1}{\gamma}\mathbf\Omega_{11}^{-1}\boldsymbol\omega_{12}\boldsymbol\omega_{12}^T\mathbf\Omega_{11}^{-1} & -\frac{1}{\gamma}\mathbf\Omega_{11}^{-1}\boldsymbol\omega_{12}\\
-\frac{1}{\gamma}\boldsymbol\omega_{12}^T\mathbf\Omega_{11}^{-1} & \frac{1}{\gamma}
\end{matrix}\right].
\end{aligned}
\]

Given
\[
\begin{aligned}
    \tr(\mathbf{U}\mathbf{\Omega}^{-1})
    &=\tr(\mathbf U_{11}\mathbf\Omega_{11}^{-1})+\frac{1}{\gamma}(\boldsymbol\omega_{12}^T\mathbf\Omega_{11}^{-1}\mathbf U_{11}\mathbf\Omega_{11}^{-1}\boldsymbol\omega_{12}-2\boldsymbol u_{12}^T\mathbf\Omega_{11}^{-1}\boldsymbol\omega_{12}+u_{22}),
\end{aligned}
\]

we have the full conditional distribution of $\boldsymbol\omega_{12}$ and $\gamma$:

\[
    \begin{aligned}
    p(\boldsymbol\omega_{12},\gamma|\mathbf\Omega_{11},\eta,\lambda_{\Omega})\propto& \gamma^{\frac{n}{2}}\exp\left(-\frac{1}{2}(s_{22}+\lambda_{\Omega})\gamma -\frac{u_{22}}{2\gamma}\right)\\
    &\times \exp\{-[\boldsymbol s_{12}-\frac{1}{\gamma}\mathbf\Omega_{11}^{-1}\boldsymbol u_{12}]^T\boldsymbol\omega_{12}\\
    &-\frac{1}{2}\boldsymbol\omega_{12}^T[D_{\eta}^{-1}+(s_{22}+\lambda_{\Omega})\mathbf\Omega_{11}^{-1}+\frac{1}{\gamma}\mathbf\Omega_{11}^{-1}\mathbf U_{11}\mathbf\Omega_{11}^{-1}]\boldsymbol\omega_{12}\}.
    \end{aligned}
\]

From the above equation, we get a closed form expression for the conditional distribution of $\gamma$:

\begin{equation}
    \begin{aligned}
    &p(\gamma|\boldsymbol\omega_{12},\mathbf\Omega_{11},\eta,\lambda_{\Omega})\propto \\
    &\gamma^{\frac{n}{2}}\exp\left(-\frac{1}{2}(s_{22}+\lambda_{\Omega})\gamma -\frac{u_{22}-2\boldsymbol u_{12}^T\mathbf\Omega_{11}^{-1}\boldsymbol\omega_{12}+\boldsymbol\omega_{12}^T\mathbf\Omega_{11}^{-1}\mathbf U_{11}\mathbf\Omega_{11}^{-1}\boldsymbol\omega_{12}}{2\gamma}\right)I_{\gamma \ge 0}\\
    \end{aligned}
    \label{gamma}
\end{equation}

which is a Generalized Inverse Gaussian (GIG) distribution \citep{hormann2014generating, jorgensen2012statistical} with parameters:
\[
\begin{aligned}
\lambda&=\frac{n}{2}+1\\
\psi&=s_{22}+\lambda_{\Omega}\\
\chi&=u_{22}-2\boldsymbol u_{12}^T\mathbf\Omega_{11}^{-1}\boldsymbol\omega_{12}+\boldsymbol\omega_{12}^T\mathbf\Omega_{11}^{-1}\mathbf U_{11}\mathbf\Omega_{11}^{-1}\boldsymbol\omega_{12}.
\end{aligned}
\]

GIG has a positive support. Thus, the determinant and the $k^{th}$ principal minor of the updated $\mathbf\Omega$ are positive, while the first $k-1$ principal minors remain unchanged and positive. In this manner, the updated $\mathbf\Omega$ always remains positive definite.

By denoting $\mathbf{D}_{\eta}=\diag(\boldsymbol \eta_{12})\in \mathbb R^{(k-1)\times (k-1)}$, the full conditional distribution of $\boldsymbol\omega_{12}$ is a Normal distribution:
\begin{equation}
    \begin{aligned}
    p(\boldsymbol\omega_{12}|\gamma,\mathbf\Omega_{11},\eta,\lambda_{\Omega})\propto& \exp\{-[\boldsymbol s_{12}-\frac{1}{\gamma}\mathbf\Omega_{11}^{-1}\boldsymbol u_{12}]^T\boldsymbol\omega_{12}\\
    &-\frac{1}{2}\boldsymbol\omega_{12}^T[\mathbf{D}_{\eta}^{-1}+(s_{22}+\lambda_{\Omega})\mathbf\Omega_{11}^{-1}+\frac{1}{\gamma}\mathbf\Omega_{11}^{-1}\mathbf U_{11}\mathbf\Omega_{11}^{-1}]\boldsymbol\omega_{12}\}
    \end{aligned}
    \label{omega12}
\end{equation}
with parameters:
\[
\begin{aligned}
\mathbf \Sigma_{\boldsymbol\omega_{12}}^{-1} &= \mathbf{D}_{\eta}^{-1}+(s_{22}+\lambda_{\Omega})\mathbf\Omega_{11}^{-1}+\frac{1}{\gamma}\mathbf\Omega_{11}^{-1}\mathbf U_{11}\mathbf\Omega_{11}^{-1}\\
\mu_{\boldsymbol\omega_{12}} &= -\mathbf\Sigma_{\boldsymbol\omega_{12}}[\boldsymbol s_{12}-\frac{1}{\gamma}\mathbf\Omega_{11}^{-1}\boldsymbol u_{12}].
\end{aligned}
\]

As in \citet{glasso}, the $z_{qq'}=1/\eta_{qq'}$ are independent Inverse Gaussians with parameters:
\[
\begin{aligned}
\mu_{z_{qq'}} &=\sqrt{\lambda_{\Omega}^2/\omega_{qq'}^2}\\
\lambda_{z_{qq'}} &=\lambda_{\Omega}^2
\end{aligned}
\]
and density:
\begin{equation}
\begin{aligned}
p(z_{qq'}|\mathbf{\Omega},\lambda_{\Omega})= \left(\frac{\lambda_{z_{qq'}}}{2\pi z_{qq'}^3}\right)^{1/2}\exp\left(\frac{-\lambda_{z_{qq'}}(z_{qq'}-\mu_{z_{qq'}})^2}{2(\mu_{z_{qq'}})^2z_{qq'}}\right) I_{z_{qq'}>0}.
\end{aligned}
\label{zij}
\end{equation}

The full conditional distribution of $\vect(\mathbf B)$ can be represented using tensor product \citep{Leng2014adaptivelasso}. Let $\mathbf D_{\tau^2}=\diag(\tau^2) \in \mathbb R ^{kp \times kp}$ for $\tau$ the scaling parameters in the prior density of $\mathbf{B}$ (Equation \ref{eqn-prior}). Then, the conditional distribution of $\vect(\mathbf{B})$ has the following form:

\begin{equation}
\begin{aligned}
    p(\vect(\mathbf{B})|\mathbf D_{\tau^2},\mathbf{\Omega},\mu,\mathbf X,\mathbf Y)
    \propto& \exp\{\mathbf X^T(\mathbf Y-\boldsymbol 1_n\mu^T\mathbf{\Omega}^{-1})\\
    &- \frac{1}{2} \vect(\mathbf{B})^T(\mathbf{\Omega}^{-1} \otimes \mathbf X^T\mathbf X + \mathbf D_{\tau^2}^{-1})\vect(\mathbf{B})\}.
\end{aligned}
\label{distB}
\end{equation}

Note that the information from data is encoded by $\mathbf \Omega^{-1}\otimes \mathbf X^T\mathbf X$ which differs from the canonical parameterization of the multiresponse linear regression model in which the information from data is encoded by $\mathbf \Omega\otimes \mathbf X^T\mathbf X$. This is because in the kernel of the likelihood, the term involving $\mathbf B$ is $\mathbf X_i \mathbf B \mathbf\Omega^{-1}\mathbf \Omega \mathbf\Omega^{-1} \mathbf B^T\mathbf X_i^T=\mathbf X_i \mathbf B \mathbf\Omega^{-1} \mathbf B^T\mathbf X_i^T$, instead of $\mathbf X_i \mathbf {\tilde{B}} \mathbf \Omega \mathbf {\tilde{B}}^T\mathbf X_i^T$ as in the canonical parametrization (see Section \ref{sec:cond_indp}).

Finally, we update $1/\tau_{jq}$ using an Inverse Gaussian distribution with parameters $\sqrt{\lambda_{\beta}^2/B_{jq}^2}$ and $\lambda_{\beta}^2$, and we update
$\mu$ using a Normal distribution with mean $(\mathbf Y\mathbf \Omega-\mathbf X \mathbf B)^T$ and variance $\mathbf \Omega/n$.

\section{Simulation settings}
\label{sec:graphical_structure}
Below we provide the details on the $\mathbf \Omega$ structure for the six graphical models:

\begin{itemize}
    \item Model 1 \ysrevision{(AR1)}: An AR(1) model with $\sigma_{qq'}=0.7^{|q-q'|}$
    \item Model 2 \ysrevision{(AR2)}: An AR(2) model with $\omega_{qq}=1$, $\omega_{q-1,q}=\omega_{q,q-1}=0.5$, $\omega_{q-2,q}=\omega_{q,q-2}=0.25$ for $i=1,\dots,k$
    \item Model 3 \ysrevision{(Block)}: A block model with $\sigma_{qq}= 1$  for $q=1,\dots,k$, $\sigma_{qq'}= 0.5$ for $1\le q\ne q'\le k/2$, $\sigma_{qq'}=0.5$ for $k/2 + 1\le q\ne q'\le 10$ and $\sigma_{qq'}=0$ otherwise.
    \item Model 4 \ysrevision{(Star)}: A star model with every node connected to the first node, with $\omega_{qq}=1$, $\omega_{1,q}=\omega_{q,1}= 0.1$ for $q=1,\dots,k$, and $\omega_{qq'}= 0$ otherwise.
    \item Model 5 \ysrevision{(Circle)}: A circle model with $\omega_{qq}= 2$, $\omega_{q-1,q}=\omega_{q,q-1}= 1$ for $q=1,\dots,k$, and $\omega_{1,q'}=\omega_{q',1}= 0.9$ for $q'=1,\dots,k$.
    \item Model 6 \ysrevision{(Dense)}: A full model with $\omega_{qq}= 2$ and $\omega_{qq'}= 1$ for $q\ne q' \in \{1,\dots,k\}$.
\end{itemize}

Note that model 1 and model 3 specify the entries of the covariance matrix $\mathbf \Sigma$ ($\sigma_{qq'}$) while the other models specify the entries of the precision matrix $\mathbf \Omega$ ($\omega_{qq'}$).

\section{Extension to other types of responses}
\label{app:nongaussian}
The model has been defined for continuous responses, yet there are different extensions for the case of binary data, counts and compositional data that we describe below.

\begin{itemize}
    \item \noindent \textit{Probit model for binary data.} 
For binary responses, we can use a Probit model with \ysrevision{CG} in the core of the dependence structure. We denote the \ysrevision{CG} latent variable as $\mathbf Z_i\in \mathbb R ^{k}$, and let $\Phi(Z_{ij})$ model the probability of observing a 1 where $\Phi$ is the cumulative distribution function of a standard Normal.

Equation \ref{eqn:probit} shows the alternative representation of the model:

\begin{equation}
\label{eqn:probit}
    \begin{aligned}
    \mathbf Z_{i}&\sim N(\mathbf{\Omega}^{-1}(\mathbf{B}^T\mathbf{X}_i^T+\mu),\mathbf{\Omega}^{-1})\\
     Y^*_{ij}&\sim N(Z_{ij},1)\\
     Y_{ij}&=\mathbf{1}_{Y^*_{ij}>0}
    \end{aligned}
\end{equation}

Then, the full conditional probability of $Y^*_{ij}$ is a truncated Normal with mean $Z_{ij}$ and variance 1. By denoting $\hat{\mathbf \mu}_i = (\mathbf{B}^T\mathbf{X}_i^T+\mu)$, we have the full conditional distribution of $\mathbf{Z}_i$:
\[
\mathbf{Z}_i|Y^*_{i},\hat{\mathbf\mu}_i,\mathbf{\Omega}\sim N([\mathbf{\Omega}+I]^{-1}(\hat{\mathbf \mu}_i+Y^*_i),[\mathbf{\Omega}+I]^{-1}).
\]

\item \textit{Log-normal Poisson model for counts.}
To model a response of multivariate counts, we use a Lognormal-Poisson model \citep{pois_lognorm}. Let $\mathbf Z_i\in \mathbb R ^{k}$ 
be the latent vector of log expected counts of the $i^{th}$ sample and let $\mathbf Y_i \in \mathbb N^{k}$ be the observed counts. We use $\mathbf{Z}_{i,-j}\in \mathbb R^{k-1}$ to denote the vector of log expected counts of the $i^{th}$ sample but without response $j$ and $Z_{ij}$ as the log expected counts of the $i^{th}$ sample and $j^{th}$ response.

The covariance matrix accounts for both over-dispersion and correlation of the counts:

\begin{equation}
\label{eqn:pois}
    \begin{aligned}
    \mathbf Z_{i}&\sim N(\mathbf{\Omega}^{-1}(\mathbf{B}^T\mathbf{X}^T_i+\mu),\mathbf{\Omega}^{-1})\\
    \lambda_{ij}&=\exp(Z_{ij})\\
     Y_{ij}&\sim \Pois(\lambda_{ij}).
    \end{aligned}
\end{equation}

Then, the density of $Y_{ij}$ is:
\[
p(Y_{ij}|Z_{ij})\propto \exp\{Y_{ij}Z_{ij}-e^{Z_{ij}}\}.
\]

Let $Z_{ij}|\mathbf Z_{i,-j} \sim N(\tilde{\mu}_{ij},\tilde{\sigma}^2_{ij})$ be the conditional prior so that the log full conditional is:
\[
\log[p(Z_{ij}|\mathbf Z_{i,-j},\hat{\mu},\Omega,Y)]=Y_{ij}Z_{ij}-\exp(Z_{ij})-\frac{1}{2\tilde{\sigma}^2_{ij}}(Z_{ij}-\tilde{\mu}_{ij})^2+C
\]
which is concave.
This means that we can sample the full conditional distribution of the latent variables using adaptive rejection sampling (ARS) \citep{gilks1992ars}, and this can be done in parallel to further speed up the sampling.

\item \textit{Normal-Logistic for multinomial data.}
As in \citet{Xia2013logistic}, we develop a Normal-Logistic model for multinomial compositional data. This type of data is very common in microbiome and ecology studies. 

Assume that we have $k+1$ responses in our sample and the last response serves as reference group. Let $\mathbf Z_i\in \mathbb R ^{k+1}$ denote the latent vector of logit transformed relative abundances for $i^{th}$ sample, and let $\mathbf Y_i \in \mathbb N^{k}$ be the observed species counts. Denote as $M$ the known total count (e.g. sequence depth in microbiome studies). Similarly we use $\mathbf Z_{i,-j}$ to denote the vector logit transformed relative abundance of the $i^{th}$ sample but without response $j$ and $Z_{ij}$ as the log expected counts of the $i^{th}$ sample and $j^{th}$ response.

The Normal-Logistic model has the following structure:

\begin{equation}
    \begin{aligned}
    \mathbf Z_{i}&\sim N(\mathbf{\Omega}^{-1}(\mathbf{B}^T\mathbf{X}^T_i+\mu),\mathbf{\Omega}^{-1})\\
    p_{ij}&=\frac{\exp(Z_{ij})}{\sum_{i=1}^{k}\exp(Z_{ij})+1}\\
    \mathbf Y_i&\sim \Multinomial(p_{i1},\dots,p_{ik},M)\\
    \end{aligned}
    \label{normal-logistic}
\end{equation}

 Note that the Normal latent variables take care of the over-dispersion. 

Then, the likelihood of $\mathbf{Y}_i$ is:
\begin{align*}
p(\mathbf Y_i|\mathbf Z_i)&= M!\prod_{j=1}^{k}\frac{1}{Y_{ij}!}\frac{\exp(Y_{ij}Z_{ij})}{\sum_{j=1}^{k}\exp(Z_{ij})+1}\\
&=\frac{M!}{\prod_{j=1}^{k}Y_{ij}!}\frac{\exp(\sum_{j=1}^{k}Y_{ij}Z_{ij})}{(\sum_{j=1}^{k}\exp(Z_{ij})+1)^M}\\
\end{align*}

Let $Z_{ij}|\mathbf Z_{i,-j}\sim N(\tilde{\mu}_{ij},\tilde{\sigma}^2_{ij})$ be 
the conditional prior so that the log full conditional is:
\[
    \begin{aligned}
    \log[p(Z_{ij}|\mathbf Z_{i,-j},\hat{\mu},\Omega,Y)]&=Y_{ij}Z_{ij}-M \log \left(\sum_{j=1}^{k} \exp(Z_{ij})+1 \right)-\frac{1}{2\tilde{\sigma}^2_{ij}}(Z_{ij}-\tilde{\mu}_{ij})^2+C
    \end{aligned}
\]

This function is concave because the first term is an affine, the second term is the negative log sum of exponential of an affine function, and the last term is a concave quadratic form. Thus, ARS \citep{gilks1992ars} can again be used during the Gibbs sampling, and this process can be parallelized for extra speed. 
\end{itemize}

\section{\ysrevision{Log concavity of posterior with fixed $\lambda$'s}}

\ysrevision{We here show the log-concavity of the posterior which make the Gibbs sampler efficient for any fixed $\lambda$.}

\ysrevision{The log-posterior is equivalent to a penalized likelihood. Let $\hat{\mu} = \mathbf{X}\mathbf{B}+\mathbf 1_n\mu^T$, and let $\mathbf{U} = \hat{\mu}^T\hat{\mu}\in \mathbb{R}^{k\times k}$. }
\begin{align*}
    \ysrevision{\log p(\mathbf{B},\mathbf{\Omega}|\mathbf{X},\mathbf{Y})}&\ysrevision{=C+\frac{n}{2}\log(|\mathbf{\Omega}|)-\frac{1}{2}\tr\left(\mathbf{Y^TY\Omega}\right)-\frac{1}{2}\tr\left(\mathbf{U\Omega}^{-1}\right)+\lambda_{\Omega}||\mathbf\Omega||_1+\lambda_\beta||\mathbf B||_1}
\end{align*}

\ysrevision{The last two terms are concave, so it remains to show that the first three terms are concave too. That is, we want to show that the log likelihood is concave as well. }

\ysrevision{This can be shown by calculating the Hessian. We observe that the random component $\mathbf Y$ is only involved in a linear term of $\mathbf\Omega$. Thus, the Hessian has no $\mathbf Y$ involved, i.e. the Hessian of the likelihood is itself the negative Fisher information of the chain graph model (expectation of constant is constant). Thus, the Hessian must be negative definite, thus we have the posterior being log-concave.}

\section{More simulation results}
\label{sec:simulation_results_more}

\begin{figure}[H]
	\centering
	\includegraphics[scale=0.5]{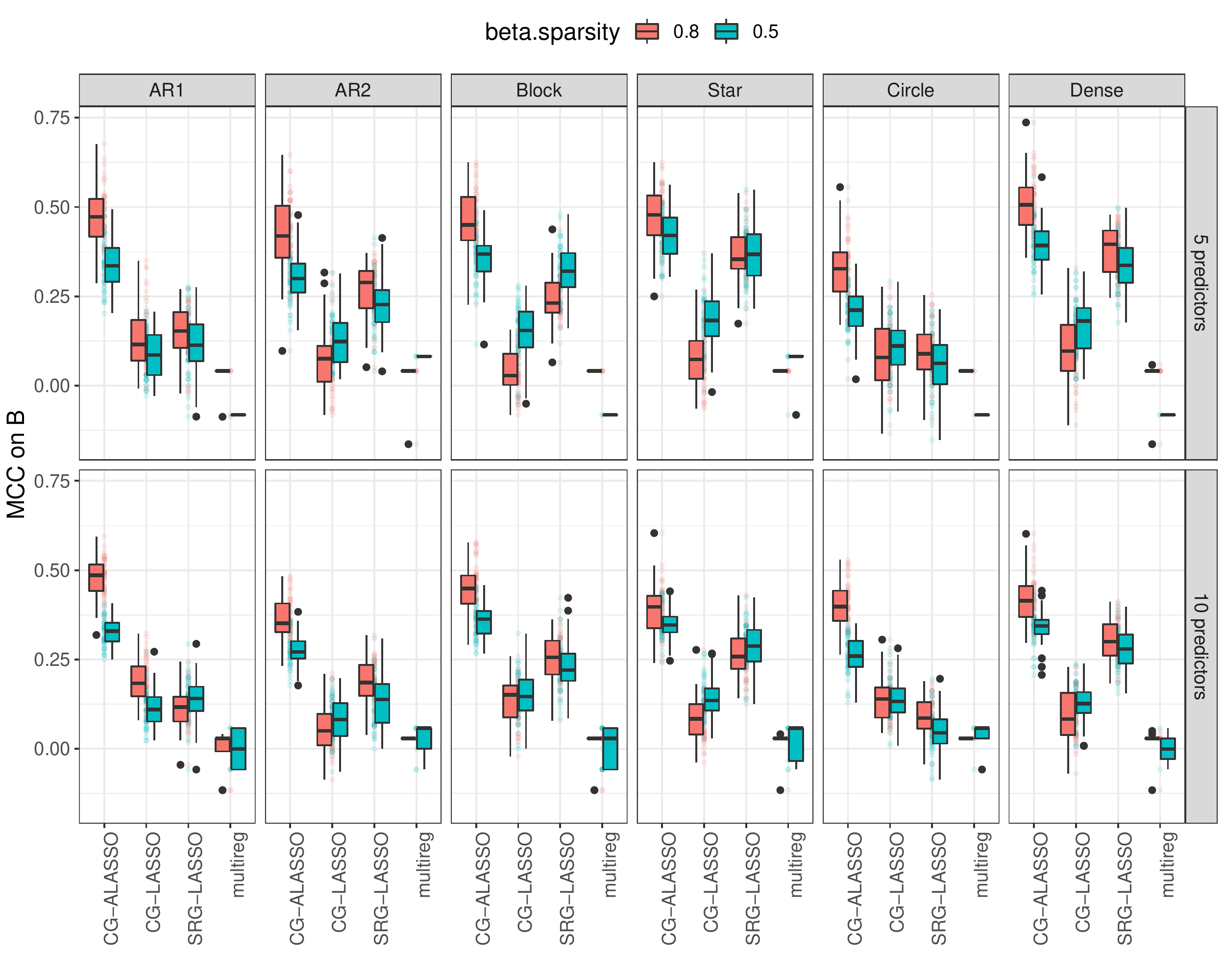}
	\caption{\textbf{Matthews Correlation Coefficients for} $\mathbf{B}$ for simulated datasets with 30 nodes and 50 samples under two levels of beta sparsity (red 0.8 and blue 0.5), two different number of predictors (10 in bottom row and 5 in top row) and six covariance models (columns, fully connected covariance model was omitted from $\mathbf \Omega$ result since MCC was not defined). X-axis corresponds to the models compared. MCC=1 means a perfect reconstruction. Our model Adaptive \ysrevision{CG}-LASSO gets the highest MCC in most cases. \ysrevision{We omit the \texttt{multireg\_mu0} model because it performs poorly across all cases (MCC close to 0).}}
	\label{fig:MCC_beta}
\end{figure}

\begin{figure}[H]
	\centering
	\includegraphics[scale=0.5]{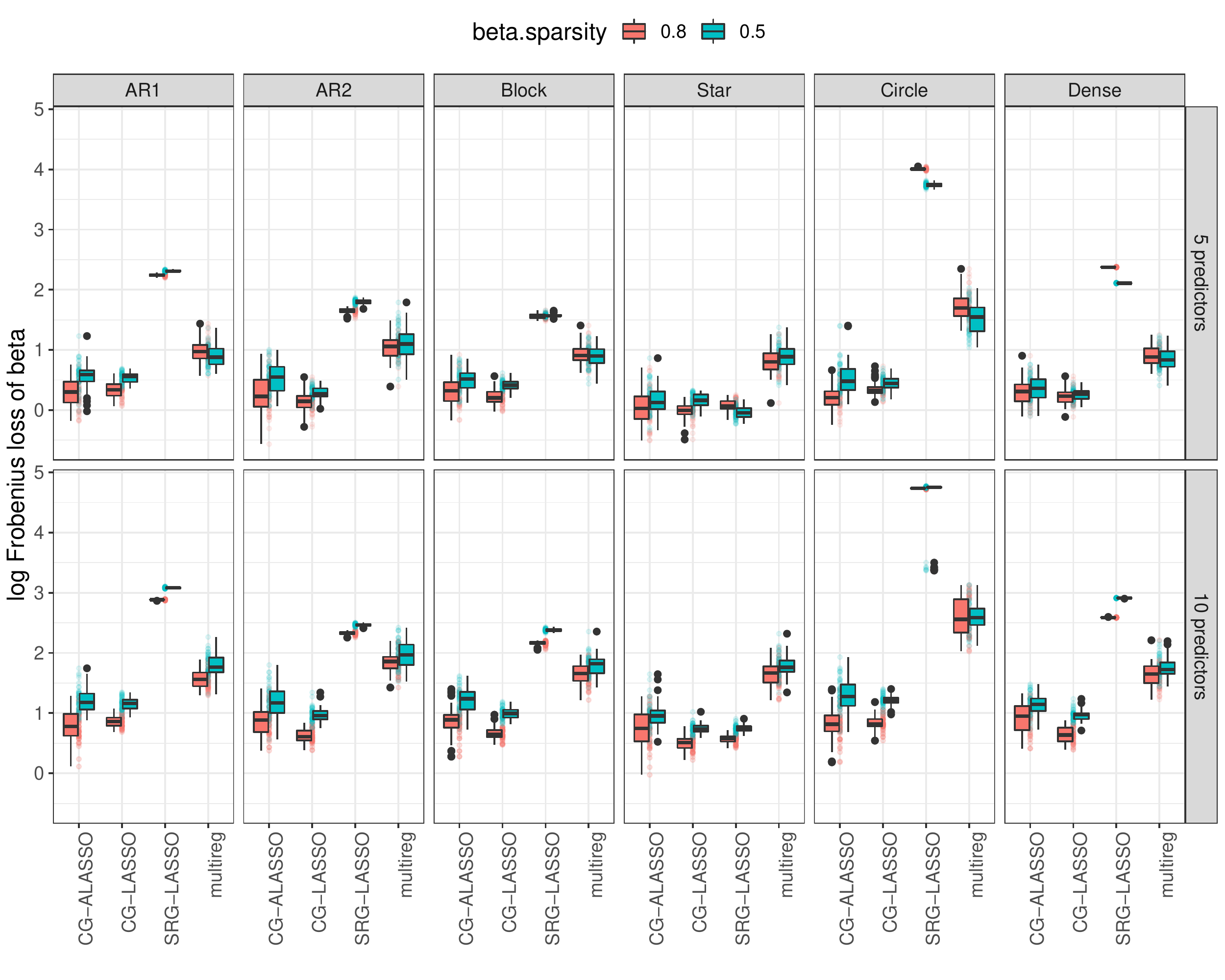}
	\caption[Estimation of $\mathbf B$]{\textbf{Frobenius Loss of} $\mathbf B$ (Y-axis in logarithmic scale) for simulated datasets with 10 nodes and 50 samples under two levels of beta sparsity (red 0.8 and blue 0.5), two different number of predictors (10 in bottom row and 5 in top row) and six covariance models (columns). X-axis corresponds to the models compared. Our models (Adaptive) \ysrevision{CG}-LASSO get the lowest loss in most cases.}
	\label{fig:beta_10}
\end{figure}

\begin{figure}[H]
	\centering
	\includegraphics[scale=0.5]{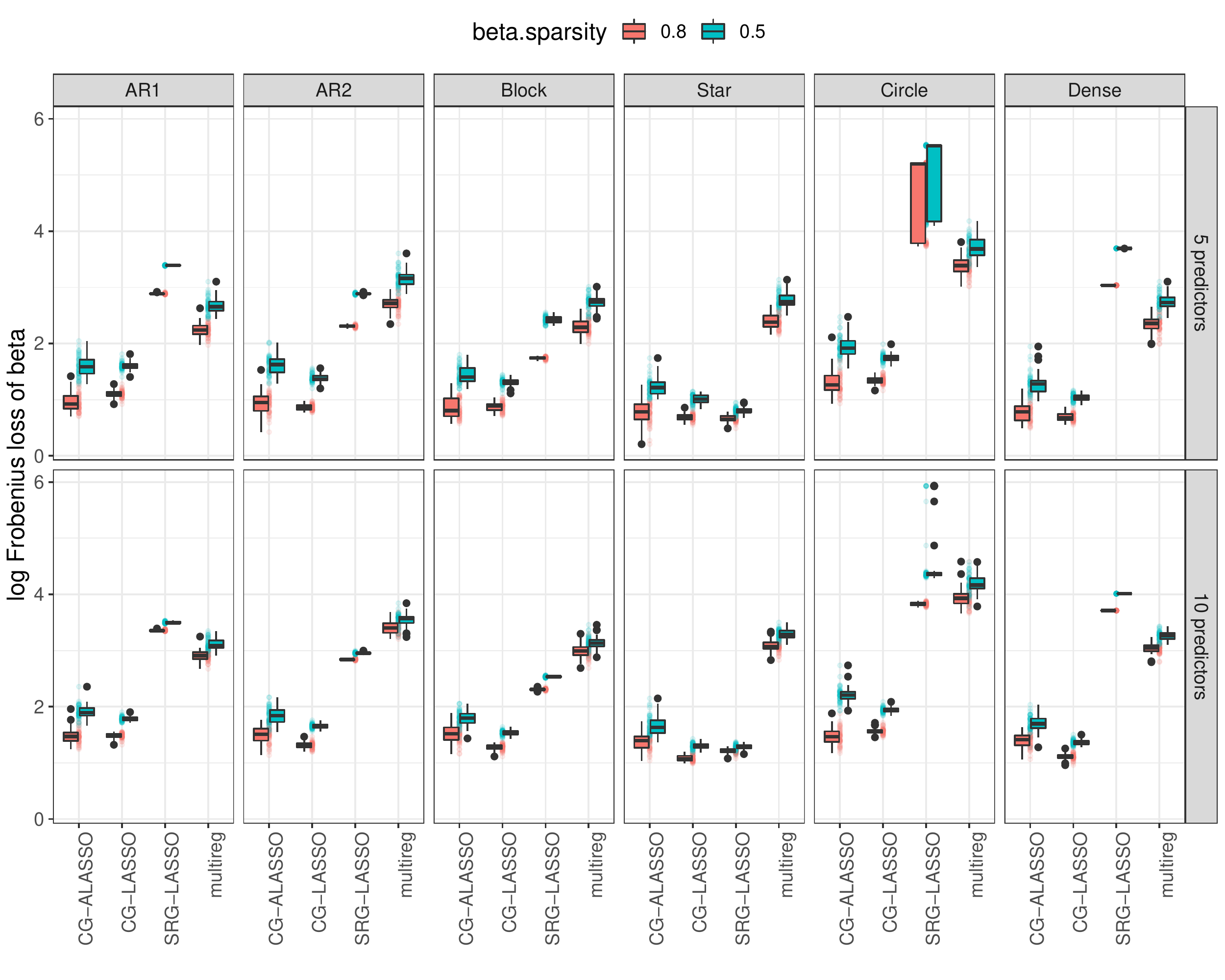}
	\caption[Estimation of $\mathbf B$]{\textbf{Frobenius Loss of} $\mathbf B$ (Y-axis in logarithmic scale) for simulated datasets with 30 nodes and 50 samples under two levels of beta sparsity (red 0.8 and blue 0.5), two different number of predictors (10 in bottom row and 5 in top row) and six covariance models (columns). X-axis corresponds to the models compared. Our models (Adaptive) \ysrevision{CG}-LASSO get the lowest loss in most cases.}
	\label{fig:beta_30}
\end{figure}

\begin{figure}[H]
    \centering
    \includegraphics[width=\linewidth]{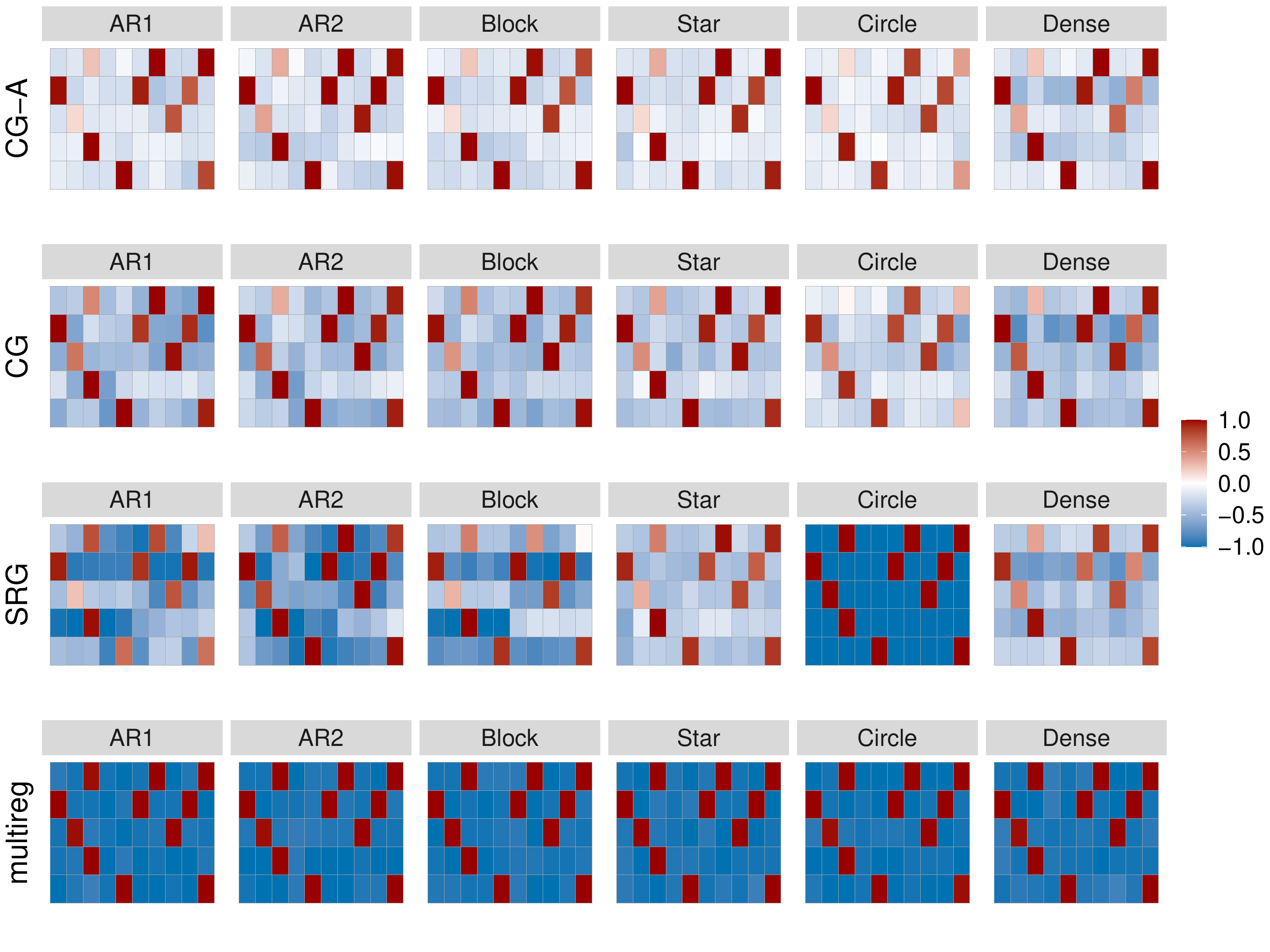}
    \caption{\textbf{Reconstruction accuracy of the graph between responses and predictors ($\mathbf B$)} for $k=10$ nodes, $p=5$ predictors and sparsity of $0.8$. Red entries correspond to true positive edges and blue entries correspond to false positive edges. Darker color means higher frequency of being estimated in 50 reconstructions. Our proposed method Adaptive \ysrevision{CG}-LASSO (\ysrevision{CG}-A) outperforms the other methods by displaying the lowest false positive rate (less blue).}
    \label{fig:visbeta_learnings.2_5}
\end{figure}

\begin{figure}[H]
    \centering
    \includegraphics[width=\linewidth]{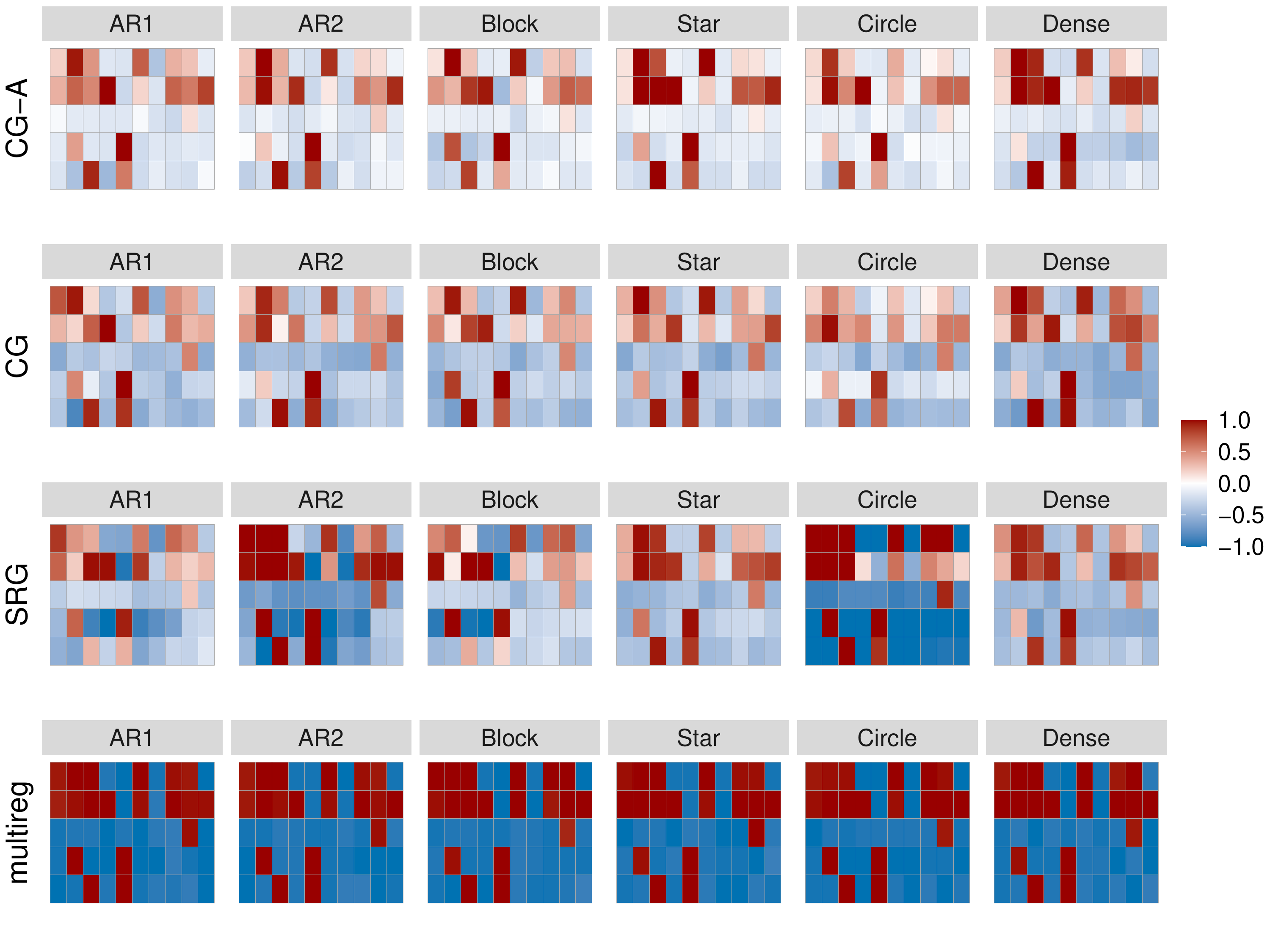}
    \caption{\textbf{Reconstruction accuracy of the graph between responses and predictors ($\mathbf B$)} for $k=10$ nodes, $p=5$ predictors and sparsity of $0.5$. Red entries correspond to true positive edges and blue entries correspond to false positive edges. Darker color means higher frequency of being estimated in 50 reconstructions. Our proposed method Adaptive \ysrevision{CG}-LASSO (\ysrevision{CG}-A) outperforms the other methods by displaying the lowest false positive rate (less blue).}
    \label{fig:visbeta_learnings.5_5}
\end{figure}

\begin{figure}[H]
    \centering
    \includegraphics[width=\linewidth]{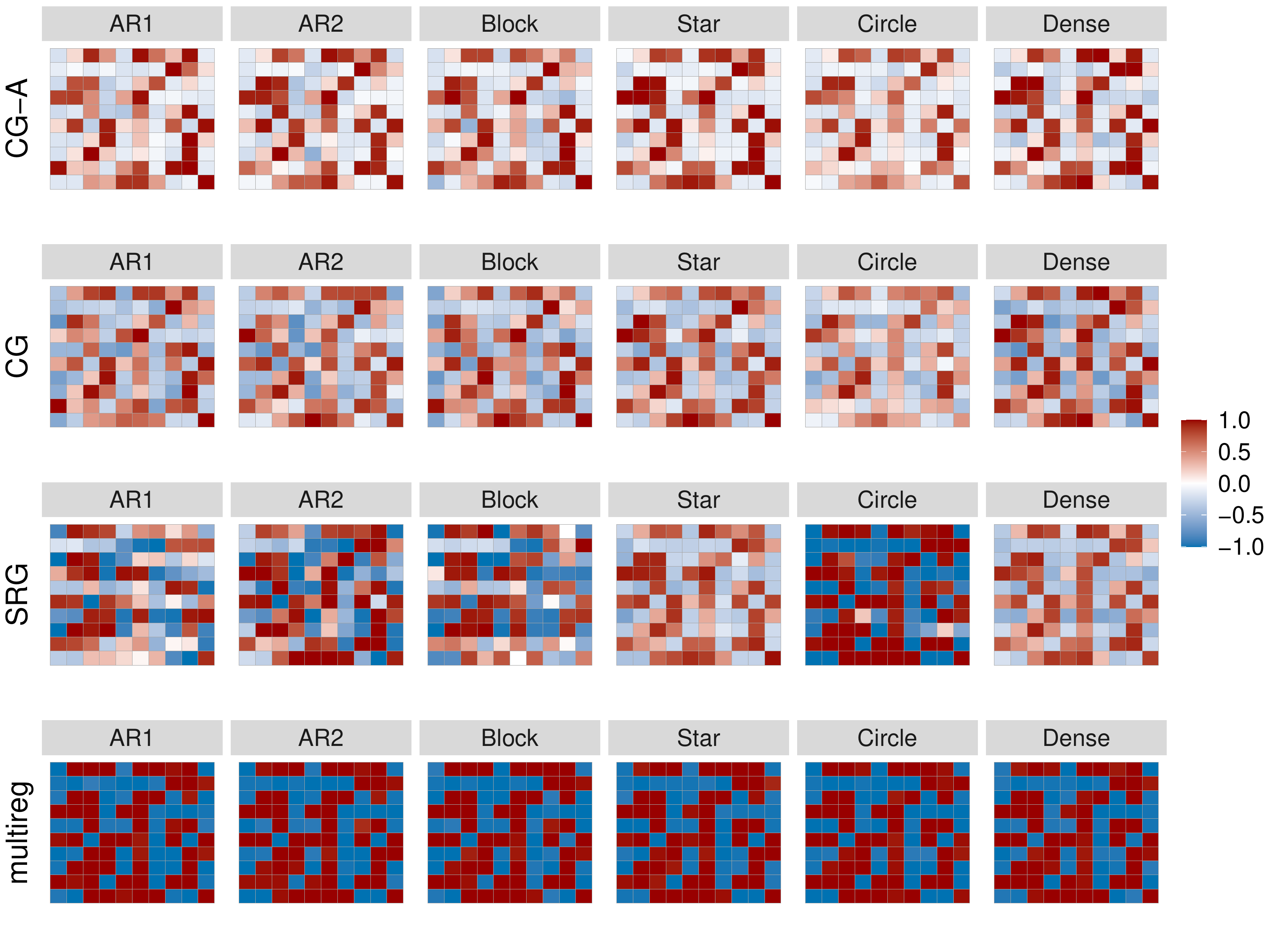}
    \caption{\textbf{Reconstruction accuracy of the graph between responses and predictors ($\mathbf B$)} for $k=10$ nodes, $p=10$ predictors and sparsity of $0.5$. Red entries correspond to true positive edges and blue entries correspond to false positive edges. Darker color means higher frequency of being estimated in 50 reconstructions. Our proposed method Adaptive \ysrevision{CG}-LASSO (\ysrevision{CG}-A) outperforms the other methods by displaying the lowest false positive rate (less blue).}
    \label{fig:visbeta_learnings.5_10}
\end{figure}

\begin{figure}[H]
	\centering
	\includegraphics[scale=0.5]{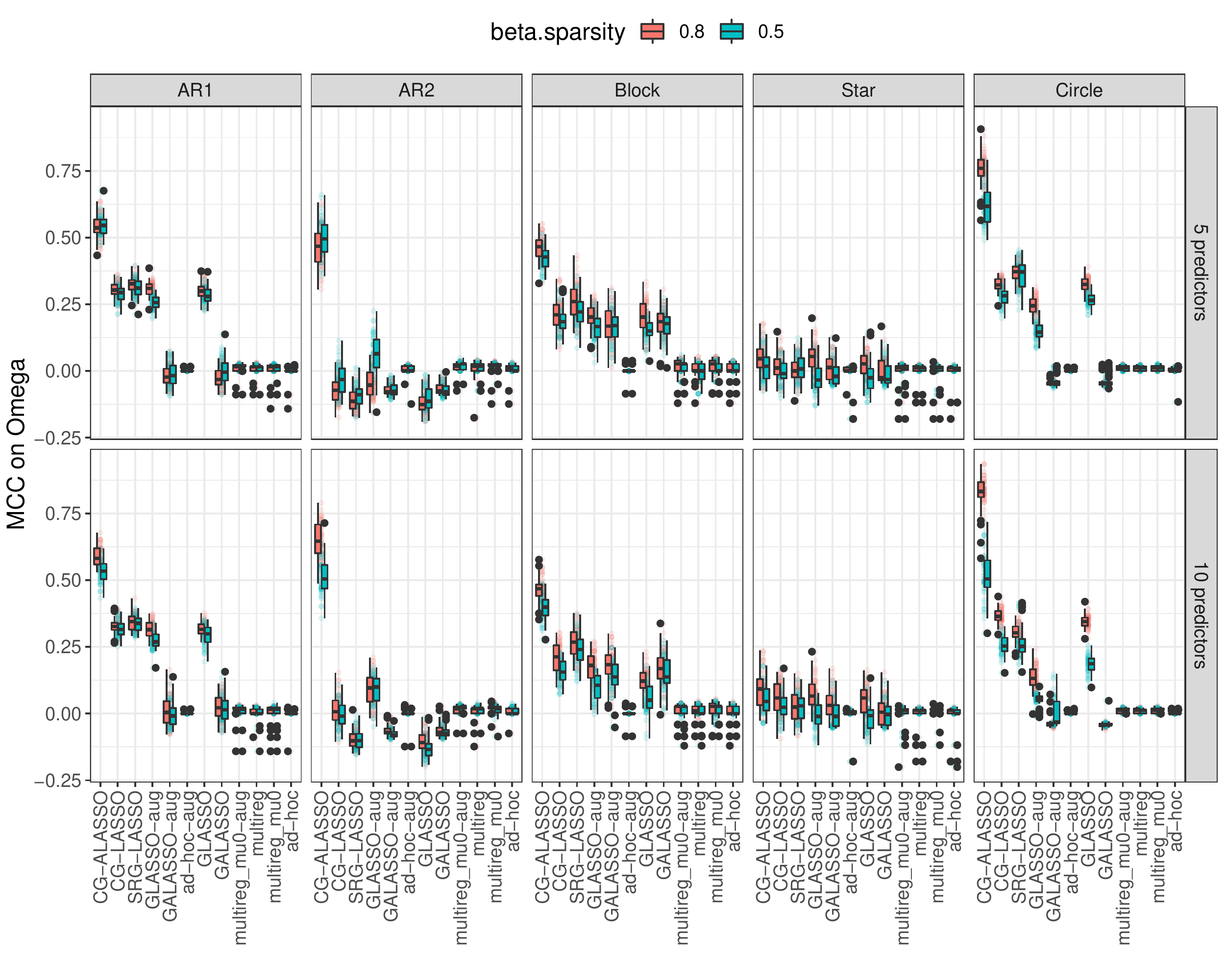}
	\caption{\textbf{Matthews Correlation Coefficients for} $\mathbf{\Omega}$ for simulated datasets with 30 nodes and 50 samples under two levels of beta sparsity (red 0.8 and blue 0.5), two different number of predictors (10 in bottom row and 5 in top row) and six covariance models (columns, fully connected covariance model was omitted from $\mathbf \Omega$ result since MCC was not defined). X-axis corresponds to the models compared. MCC=1 means a perfect reconstruction. Our model Adaptive \ysrevision{CG}-LASSO gets the highest MCC in most cases. We omit the dense model because MCC is not defined.}
	\label{fig:MCC_Omega}
\end{figure}

\begin{figure}[H]
	\centering
	\includegraphics[scale=0.5]{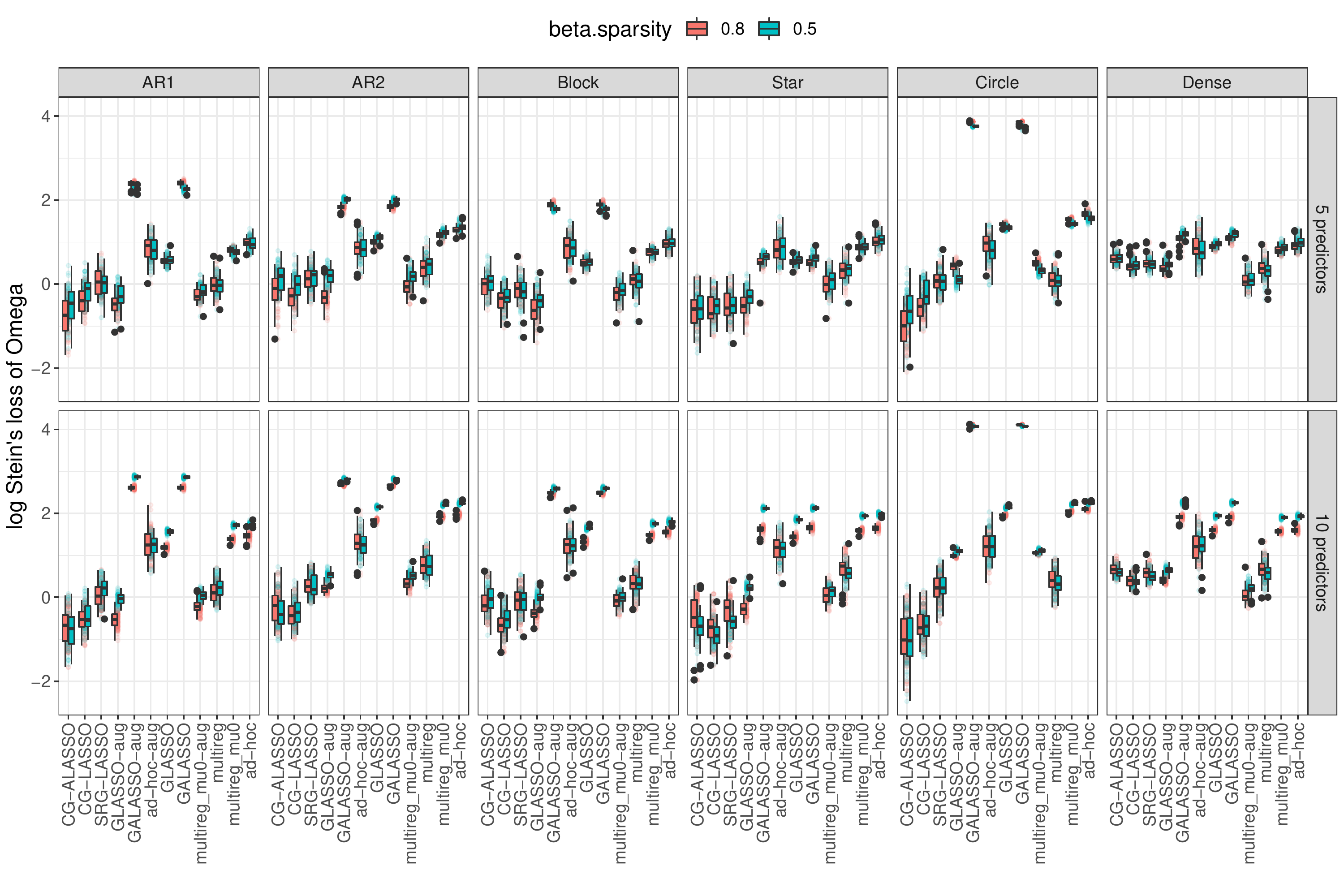}
	\caption[Estimation of $\mathbf\Omega$]{\textbf{Stein's Loss of }$\mathbf\Omega$, (Y-axis in logarithmic scale) for simulated datasets with 10 nodes and 50 samples under two levels of beta sparsity (red 0.8 and blue 0.5), two different number of predictors (10 in bottom row and 5 in top row) and six covariance models (columns). X-axis corresponds to the models compared. Our models (Adaptive) \ysrevision{CG}-LASSO get the lowest loss in most cases.}
	\label{fig:stein_10}
\end{figure}

\begin{figure}[H]
	\centering
	\includegraphics[scale=0.5]{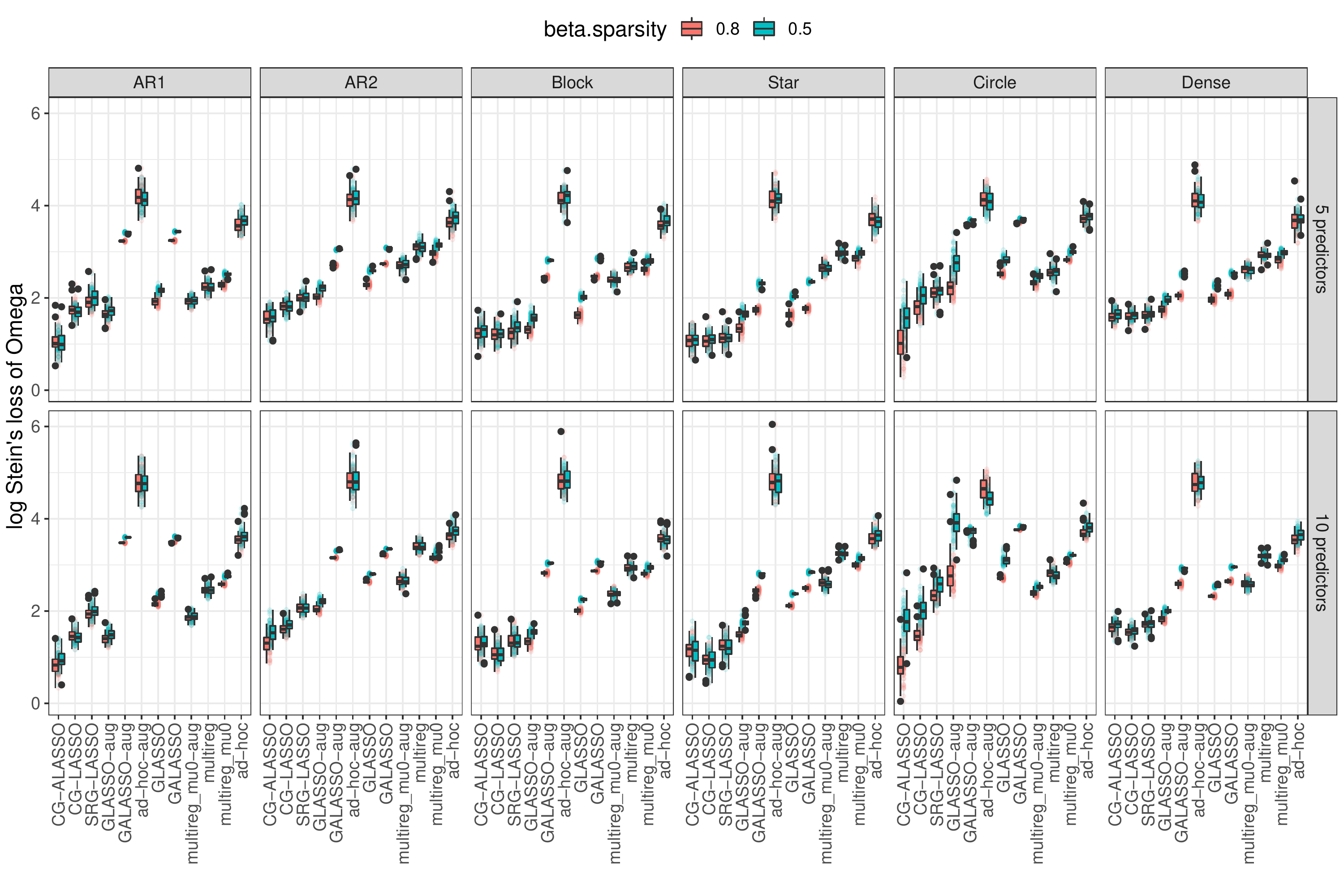}
	\caption[Estimation of $\mathbf\Omega$]{\textbf{Stein's Loss of }$\mathbf\Omega$, (Y-axis in logarithmic scale) for simulated datasets with 30 nodes and 50 samples under two levels of beta sparsity (red 0.8 and blue 0.5), two different number of predictors (10 in bottom row and 5 in top row) and six covariance models (columns). X-axis corresponds to the models compared. Our models (Adaptive) \ysrevision{CG}-LASSO get the lowest loss in most cases.}
	\label{fig:stein_30}
\end{figure}

\begin{figure}[H]
    \centering
    \includegraphics[width=\linewidth]{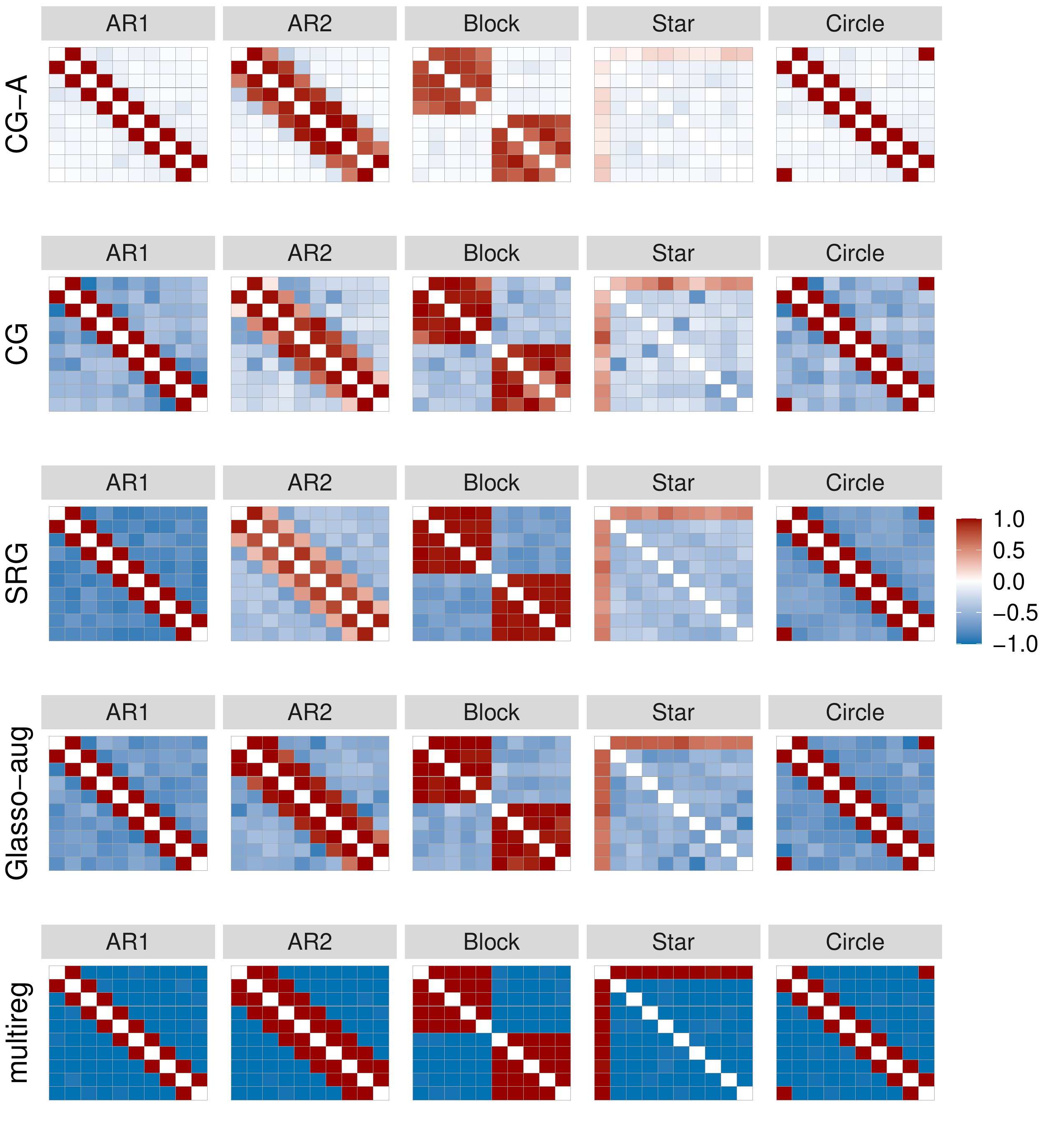}
    \caption{\textbf{Reconstruction accuracy of the graph among responses ($\mathbf \Omega$)} for $k=10$ nodes, $p=5$ predictors and sparsity of $0.8$. Red entries correspond to true positive edges and blue entries correspond to false positive edges. Darker color means higher frequency of being estimated in 50 reconstructions. Our proposed method Adaptive \ysrevision{CG}-LASSO (\ysrevision{CG}-A) outperforms the other methods by displaying the lowest false positive rate (less blue). We omit the dense model because it has no false positive or true negatives.}
    \label{fig:vis_learnings.2_5}
\end{figure}

\begin{figure}[H]
    \centering
    \includegraphics[width=\linewidth]{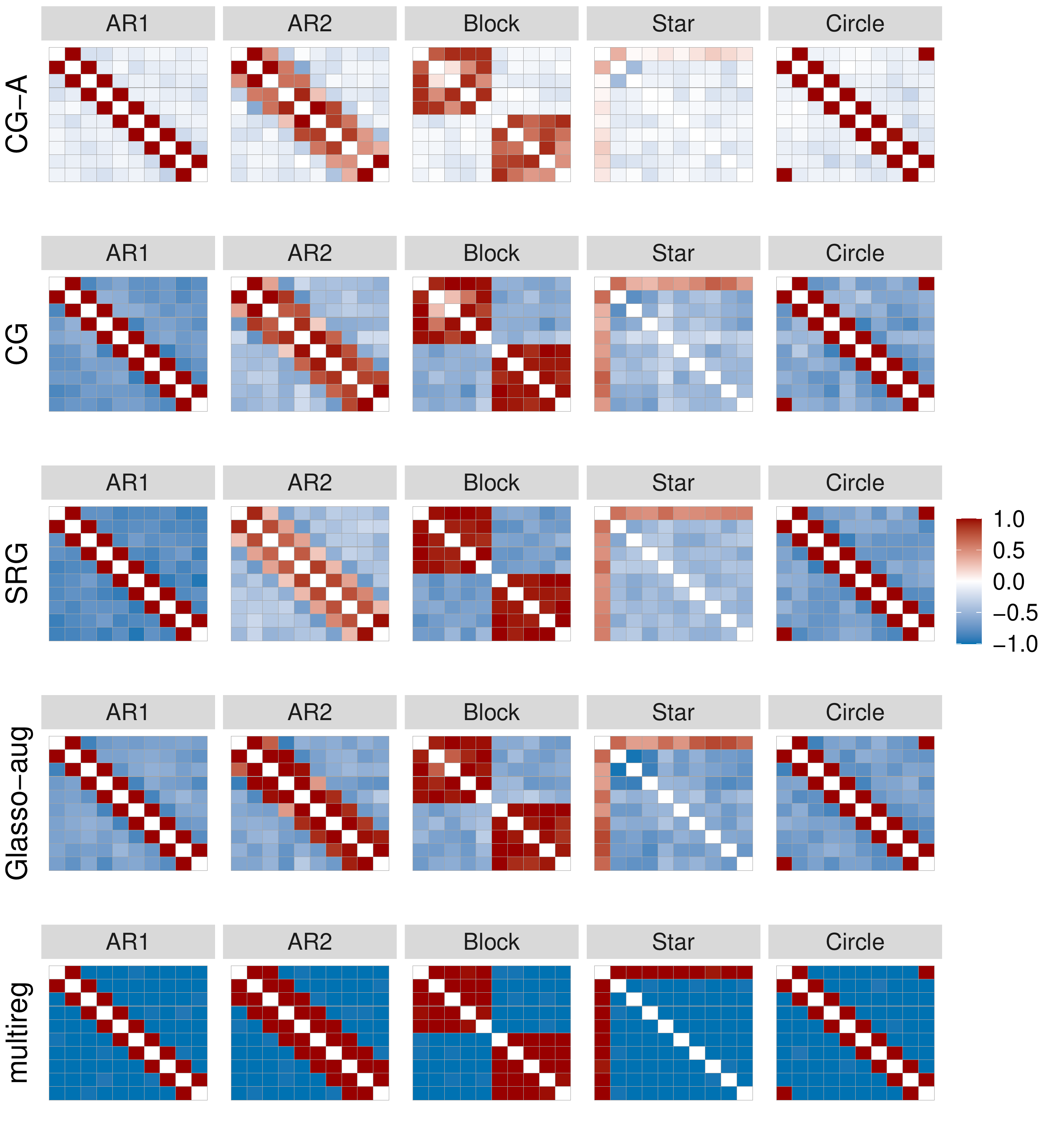}
    \caption{\textbf{Reconstruction accuracy of the graph among responses ($\mathbf \Omega$)} for $k=10$ nodes, $p=5$ predictors and sparsity of $0.5$. Red entries correspond to true positive edges and blue entries correspond to false positive edges. Darker color means higher frequency of being estimated in 50 reconstructions. Our proposed method Adaptive \ysrevision{CG}-LASSO (\ysrevision{CG}-A) outperforms the other methods by displaying the lowest false positive rate (less blue). We omit the dense model because it has no false positive or true negatives.}
    \label{fig:vis_learnings.5_5}
\end{figure}

\begin{figure}[H]
    \centering
    \includegraphics[width=\linewidth]{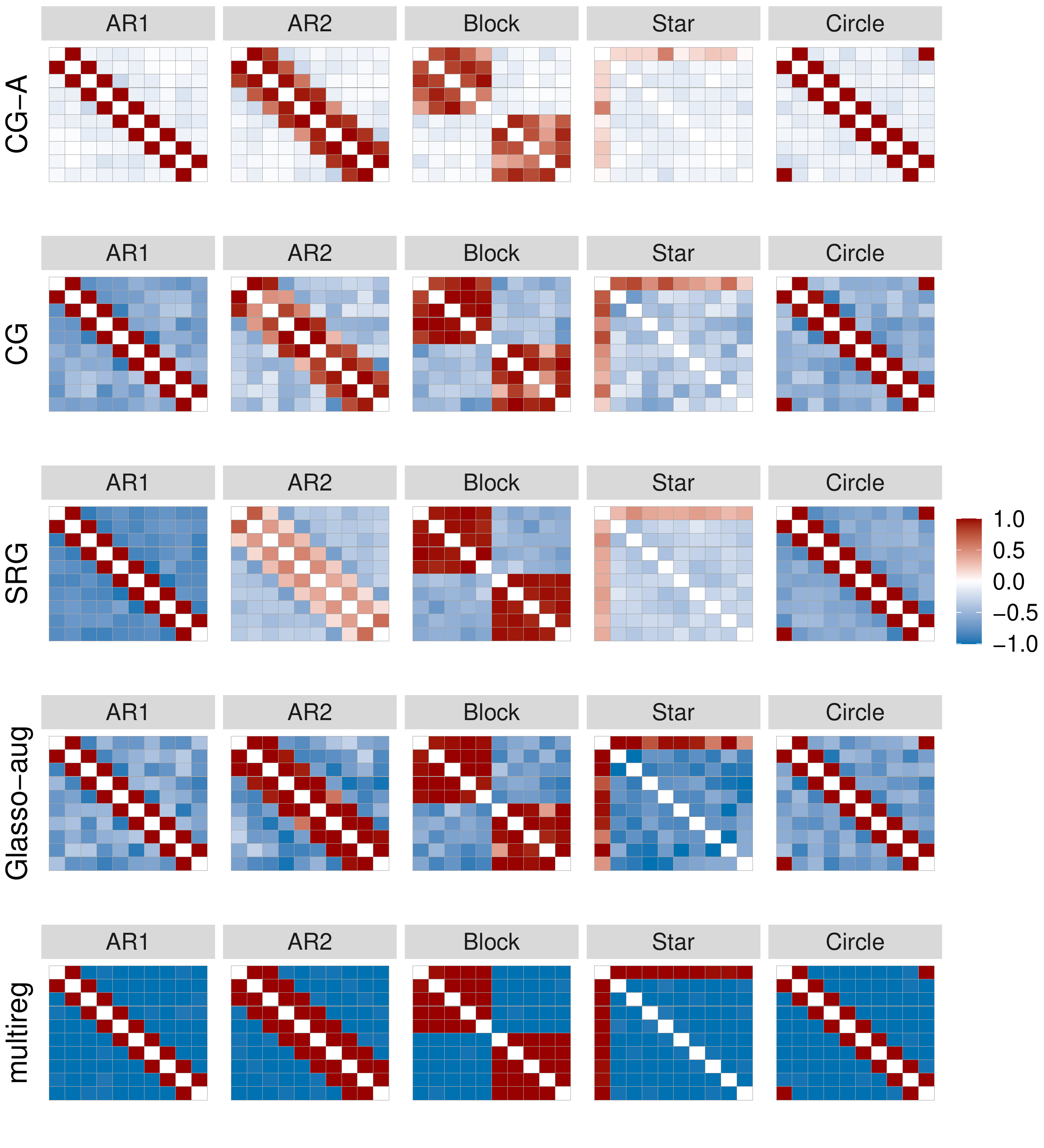}
    \caption{\textbf{Reconstruction accuracy of the graph among responses ($\mathbf \Omega$)} for $k=10$ nodes, $p=10$ predictors and sparsity of $0.5$. Red entries correspond to true positive edges and blue entries correspond to false positive edges. Darker color means higher frequency of being estimated in 50 reconstructions. Our proposed method Adaptive \ysrevision{CG}-LASSO (\ysrevision{CG}-A) outperforms the other methods by displaying the lowest false positive rate (less blue). We omit the dense model because it has no false positive or true negatives.}
    \label{fig:vis_learnings.5_10}
\end{figure}

\section{Computational speed and scaling}
\label{sec:scaling}

\begin{figure}[H]
	\centering
	\includegraphics[scale=0.6]{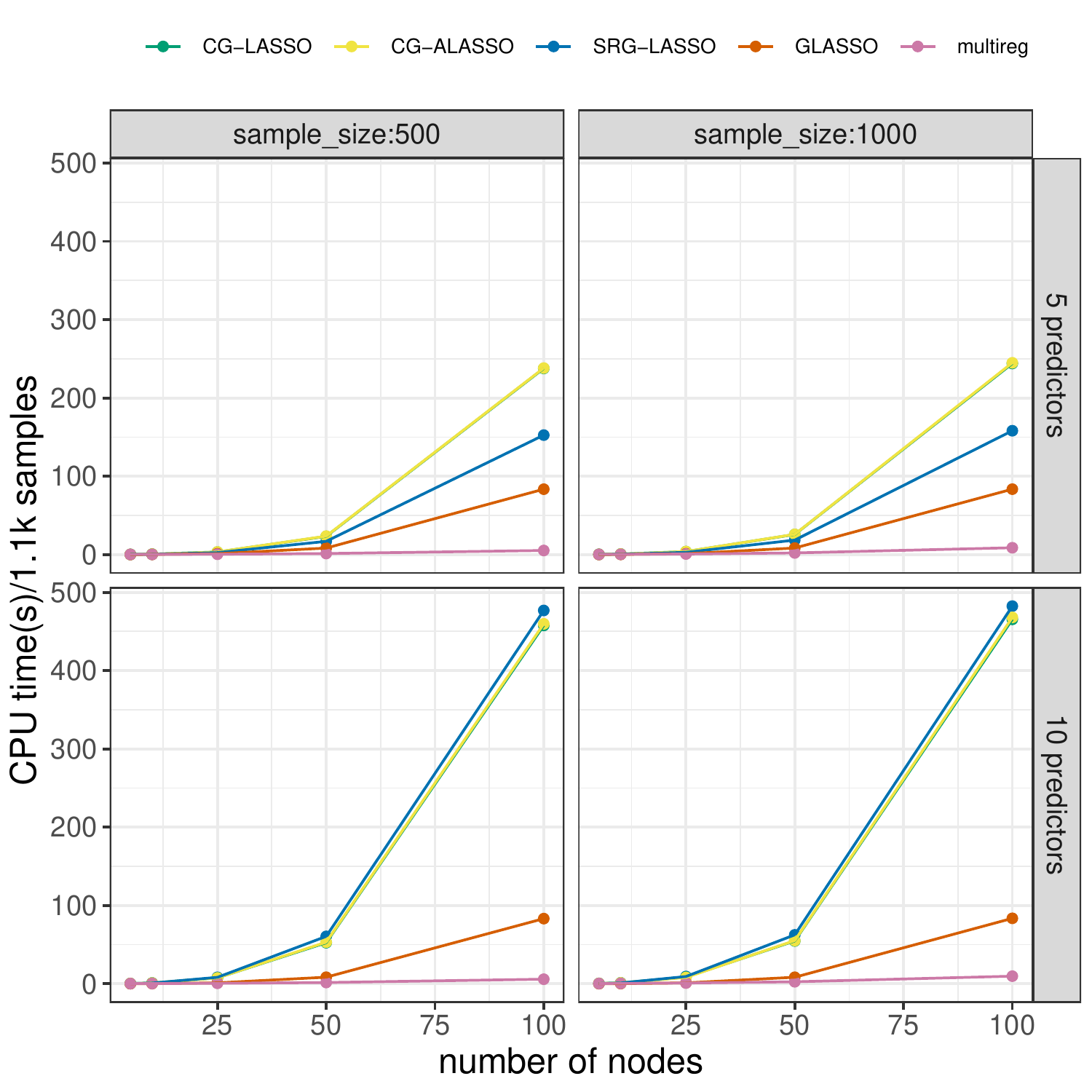}
	\caption[Scalability test]{\textbf{Scalability test.} Computational time for each algorithm in CPU seconds as a function of the number of nodes, the number of predictors, and sample size. Speed depends on the number of nodes and number of predictors, but not on sample size. Our proposed method is efficient, yet slower than Graphical LASSO. }
	\label{fig:scaling}
\end{figure}

\section{\ysrevision{Transformation of marginal effects into conditional effect in real data}}

\begin{figure}
    \centering
    \includegraphics[scale=0.6]{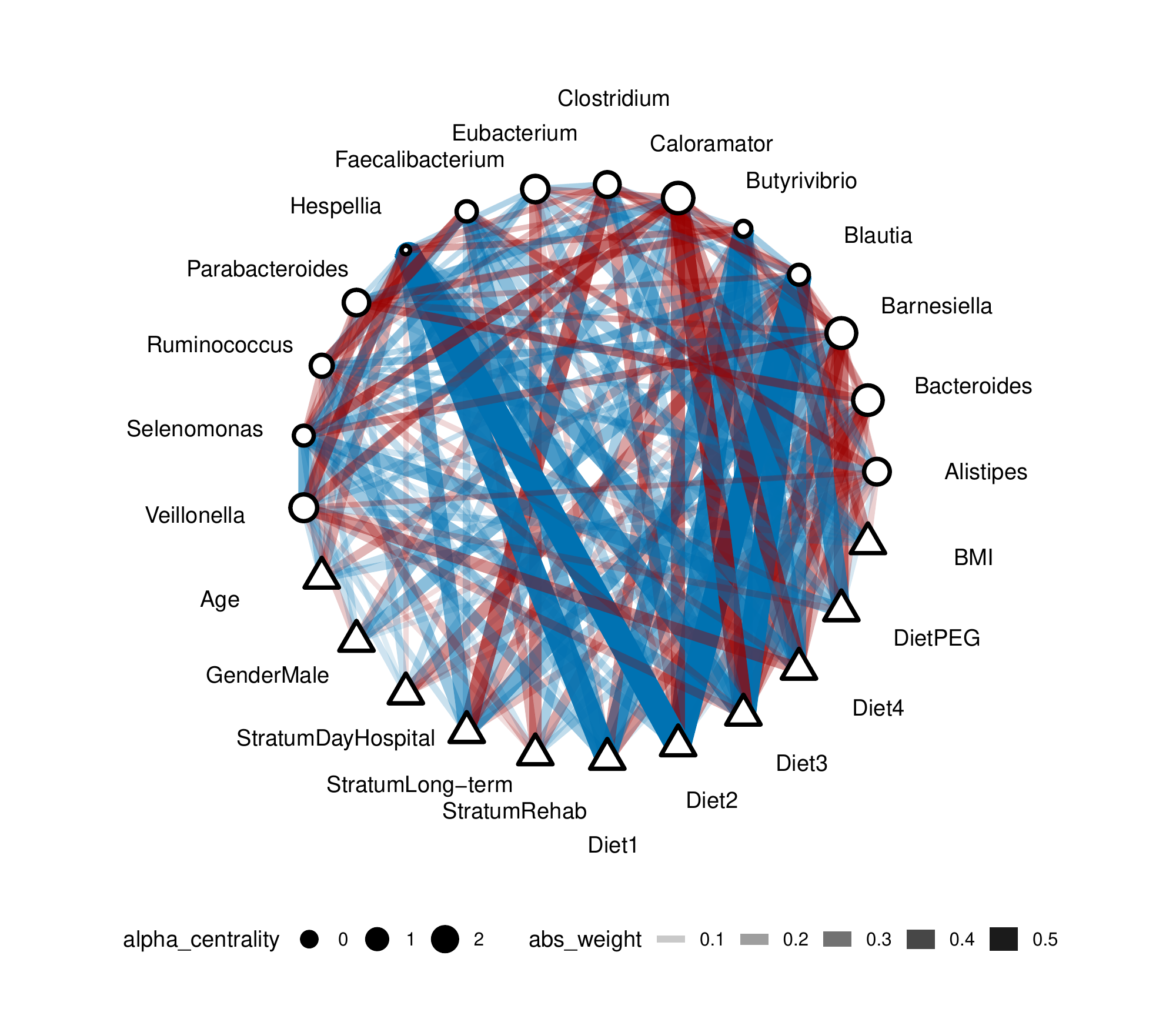}
    \caption{\ysrevision{\textbf{Conditional network of human gut network via multiresponse regression and transformation into chain graph parameterization}. Triangle nodes correspond to predictors and circle nodes correspond to relative abundances of genus. The node size on the circle nodes correspond to the $\alpha-$centrality values \citep{bonacich2001eigenvector}. The width of the edges correspond to the absolute weight, and the color to the type of interaction (red positive, blue negative).}}
    \label{fig:human_cond_from_marg}
\end{figure}

\begin{figure}
    \centering
    \includegraphics[scale=0.6]{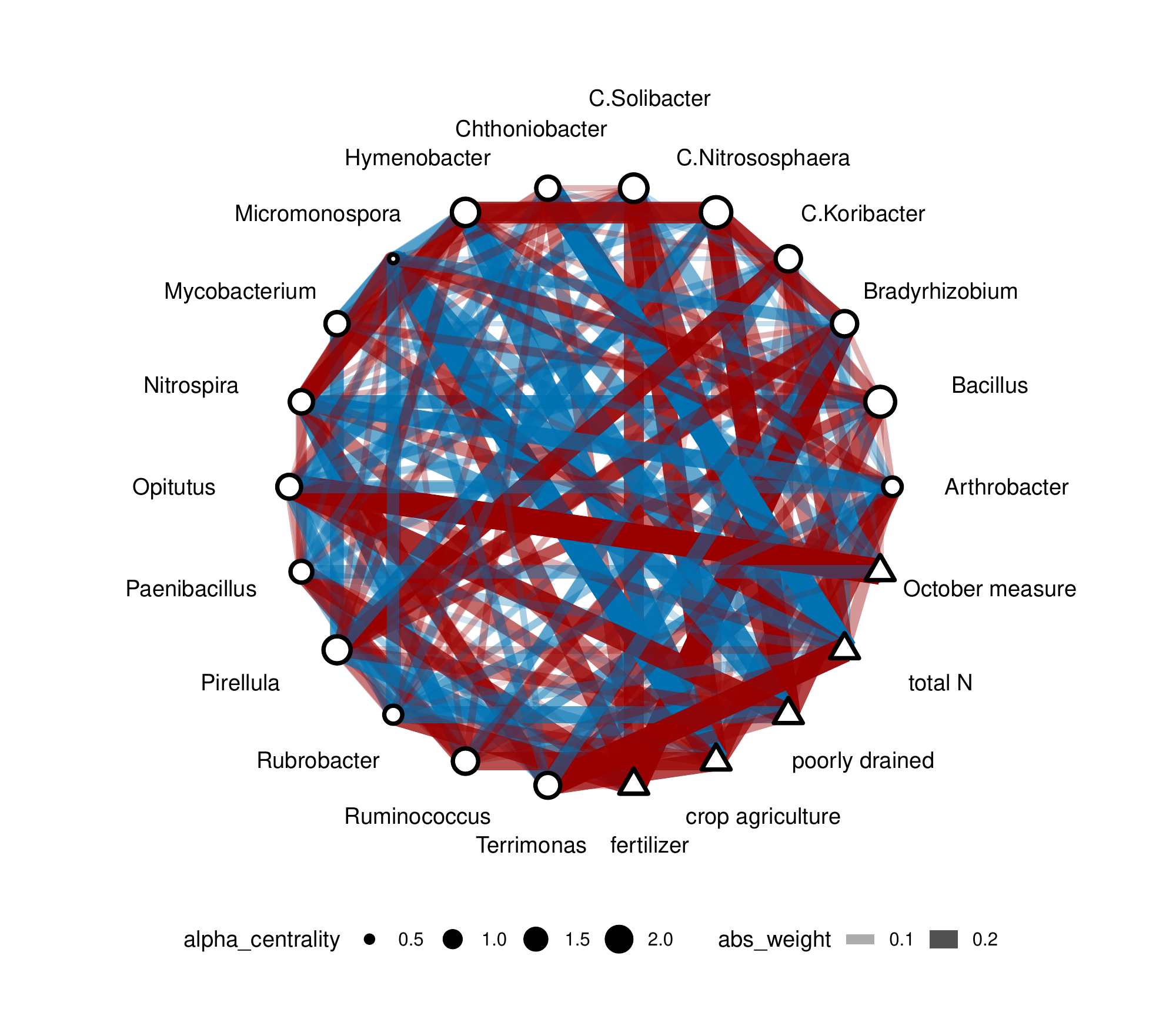}
    \caption{\ysrevision{\textbf{Conditional network of soil network via multiresponse regression and transformation into chain graph parameterization}. Triangle nodes correspond to predictors and circle nodes correspond to relative abundances of genus. The node size on the circle nodes correspond to the $\alpha-$centrality values \citep{bonacich2001eigenvector}. The width of the edges correspond to the absolute weight, and the color to the type of interaction (red positive, blue negative).}}
    \label{fig:soil_cond_from_marg}
\end{figure}

\ysrevision{We fit a hierarchical model with multiresponse linear regression as the core and the same logit-multinomial sampling distribution (Equation \ref{normal-logistic} with $\mathbf Z_i$ as multiresponse regression model). 
The density of the networks show that the sparsity assumption on the chain graph model has a strong impact on the estimated network (Figures \ref{fig:human_cond_from_marg} and \ref{fig:soil_cond_from_marg}).}

\end{document}